\documentclass[11pt,aps,nofootinbib,superscriptaddress,notitlepage,longbibliography,tightenlines]{revtex4-2}

\usepackage[T1]{fontenc}
\usepackage{stix}
\usepackage{epsfig}
\usepackage{amsmath,bm,bbm}
\usepackage{graphicx}
\usepackage{verbatim}
\usepackage{simplewick}
\usepackage{appendix}
\usepackage[usenames,dvipsnames]{xcolor}
\usepackage{enumitem}
\usepackage{ascmac}

\usepackage{tikz}
\usetikzlibrary{arrows.meta, positioning, shadows}

\usepackage{ragged2e}

\numberwithin{equation}{section}
\def\dd{\mathrm{d}}
\def\ee{\mathrm{e}}

\newcommand{\beal}{\begin{equation}\begin{aligned}}
\newcommand{\enal}{\end{aligned}\end{equation}}

\begin{document}

\title{
Exact WKB Formulation of Quantization and Particle Production in Time-Dependent Backgrounds}
\author{Ryo Namba}
\affiliation{RIKEN Center for Interdisciplinary Theoretical and Mathematical Sciences (iTHEMS), Wako, Saitama 351-0198, Japan}
\author{Motoo Suzuki}
\affiliation{SISSA, Via Bonomea 265, Trieste 34136, Italy}
\affiliation{INFN, Sezione di Trieste, Via Valerio 2, 34127, Italy}
\affiliation{IFPU, Via Beirut 2, 34014 Trieste, Italy}
\date{\today}

\begin{abstract}
Divergence in perturbative expansions is where interesting physics takes place. Particle production on time-dependent backgrounds, as one such example, is interpreted as transition from one vacuum to another. Vacuum is typically defined as an asymptotic state in which the WKB approximation is valid. The use of the WKB method, however, poses several conceptual and computational issues, as the WKB series is divergent in general, quantization is insensitive to higher orders in the series, and the global behavior of solutions cannot be captured.
Exact WKB analysis is a powerful resummation technology that provides an analytical tool for a global structure of exact solutions to overcome these problems.
In this paper, we establish quantization by fully employing the exact WKB solutions as mode functions and by defining the vacua with respect to them.
We provide a self-contained exact WKB formulation to obtain evolution matrices without resorting to the use of known special functions and without approximations.
We find that the quantity called Voros coefficient plays an important role to re-normalize the exact WKB solutions compatible with asymptotic states. 
We show that the ambiguity that coexists with nontrivial Voros coefficients is eliminated by requiring physical quantization conditions.
Our formalism provides a conceptual as well as practical framework to upgrade our treatment of quantization and particle production. Combined with other approximating techniques, it can form a basis to tackle a broad class of problems that are beyond technical ability of the existing formulations.
\end{abstract}

\maketitle
\tableofcontents

\section{Introduction \& Conventional Approach to Particle Production}
\label{sec:intro}

Particle production is ubiquitous in physical systems. 
It is intrinsically a non-perturbative process in the field-theoretical language, and a common interpretation of particle production is a change of vacuum states due to a 
nontrivial background. 
A realistic system typically consists of many, sometimes infinite, degrees of freedom (dof's); yet, its global nature can often be characterized by a few components. Under a high degree of symmetry, one can reduce a problem down to a one-variable system. 
The equation of motion then has a universal form given by
\begin{align}
    \frac{\dd^2}{\dd x^2} \tilde\psi(x; \bm{\vartheta}) + A(x; \bm{\vartheta}) \, \frac{\dd}{\dd x} \tilde\psi(x; \bm{\vartheta}) + B(x; \bm{\vartheta}) \, \tilde\psi(x; \bm{\vartheta}) = 0 \; , 
\end{align}
where $\psi$ denotes the target degree of freedom that depends on a coordinate variable $x$ and a collection of parameters, $\bm{\vartheta}$, and $A$ and $B$ are coefficients that are to be fixed once a model is specified.
This equation can always be further simplified to that without a first-order derivative term, i.e.,
\begin{align}
\label{eq:schrodinger_canonical}
    \frac{\dd^2}{\dd x^2} \psi(x; \bm{\vartheta}) + \left[ B(x; \bm{\vartheta}) - \frac{A^2(x; \bm{\vartheta})}{4} - \frac{1}{2} \, \frac{\dd A(x; \bm{\vartheta})}{\dd x} \right] \psi(x; \bm{\vartheta}) = 0
\end{align}
where the target function is redefined via $\psi(x;\bm{\vartheta}) \equiv \exp\left[ \frac{1}{2} \int^x A(x'; \bm{\vartheta}) \, \dd x' \right] \tilde\psi(x; \bm{\vartheta})$.
This implies that the problem is reduced to solving a one-dimensional Schr\"{o}dinger-type equation. Let us emphasize that any one-dof, one-variable, linear equation can be transformed into \eqref{eq:schrodinger_canonical}.
To extend it to a multi-dof case, one can promote $\psi$ to an array of dof's and $A$ and $B$ to coefficient matrices.

Examples in which the governing equation takes the form \eqref{eq:schrodinger_canonical} are ample. Schr\"{o}dinger equations of quantum mechanics, condensed matter systems in materials, perturbations in curved spacetime count a few. 
Let us give some concrete examples of field interactions that are often employed in cosmology, which we have in mind as applications of the methodology we develop in this paper.
In a field-theoretical description of cosmology, one can classify model contents in terms of scalar, vector, tensor, and fermions and consider interactions among them. For instance, a scalar field $\varphi$ can couple to itself, to another scalar $\chi$, to a vector/gauge field $A_\mu$, to a tensor $g_{\mu\nu}$, and to a fermion $\psi$ through interaction terms at dimension-$4$
\begin{align}
  \label{eq:interactions_renorm}
    - \frac{\lambda}{4} \, \varphi^4 \; , \qquad
    -g \varphi^2 \chi^2 \; , \qquad
    e A_\mu \left( \varphi \, \partial^\mu \varphi^\dagger - \varphi^\dagger \partial^\mu \varphi \right) \; , \qquad
    \frac{\xi}{6} \, \varphi^2 R \; , \qquad
    y \, \varphi \bar\psi \psi \; ,
\end{align}
where $R$ is the Ricci scalar associated with $g_{\mu\nu}$, and $\lambda, \, g, \, e , \, \xi , \, y$ are dimensionless coupling constants. Note that $\varphi$ is implicitly assumed to be a real scalar except for the above coupling to $A_\mu$.
Higher-dimension effective operators also play important roles in cosmology, such as 
\begin{align}
  \label{eq:interactions_eft}
    - \frac{I^2(\varphi)}{4} F_{\mu\nu} F^{\mu\nu} \; , \quad
    \frac{\varphi}{\Lambda} \left( R^2 - 4 R_{\mu\nu} R^{\mu\nu} + R_{\mu\nu\rho\sigma} R^{\mu\nu\rho\sigma} \right) \; , \quad
    - \frac{\varphi}{f} F_{\mu\nu} \tilde{F}^{\mu\nu} \; , \quad
    \frac{\varphi}{f} R_{\mu\nu\rho\sigma} \tilde{R}^{\mu\nu\rho\sigma} \; , \quad
    \frac{\partial_\mu \varphi}{f} \, \bar\psi \gamma^\mu \gamma_5 \psi \; ,
\end{align}
where $I(\varphi)$ is some function of $\varphi$, $R_{\mu\nu\rho\sigma}$ and $R_{\mu\nu}$ are Riemann and Ricci tensor of $g_{\mu\nu}$, respectively, $F_{\mu\nu}$ and $\tilde{F}^{\mu\nu}$ are the field-strength tensor of $A_\mu$ and its dual, respectively, $\tilde{R}^{\mu\nu\rho\sigma}$ is the dual of $R_{\mu\nu\rho\sigma}$, and $\Lambda$ and $f$ are coupling constants of mass dimension-$1$. In the last three interactions, $\varphi$ is regarded as a pseudo-scalar.

Nontrivial backgrounds arise in various physical environments. As a leading application we have in mind, the cosmological background breaks the temporal symmetries, i.e.~boost and time translation. In the language of effective field theory, this is realized when a scalar field acquires a timelike vacuum expectation value (vev). Assuming that the $\varphi$ field in \eqref{eq:interactions_renorm} and \eqref{eq:interactions_eft} is responsible for the symmetry breaking with $\partial_\mu \varphi \ne 0$ being timelike, perturbations around such a background follow equations of motion of the form \eqref{eq:schrodinger_canonical} after proper diagonalization and canonical normalization. The perturbative analysis is often done in the Fourier space, in which case parameters $\bm{\vartheta}$ contain not only the coupling constants, as in \eqref{eq:interactions_renorm} and \eqref{eq:interactions_eft}, but also the wavenumber $\bm{k}$ associated with the Fourier transformation.

Particle production is captured along the time evolution governed by \eqref{eq:schrodinger_canonical}. As implied earlier, determination of vacuum states is part of the essential ingredients of the phenomenon. The canonical quantization is based on the mode functions obtained as the solutions to the linear equation \eqref{eq:schrodinger_canonical}. We review the standard procedure with respect to the adiabatic approximation of the solutions in the following subsection, proceeding then to the more exact method we propose in this work.

\subsection{Quantization \& Defining Vacuum States}

The Lagrangian that leads to the equation of motion in \eqref{eq:schrodinger_canonical} takes the form
\begin{align}
    \label{eq:lagr_quantum}
    \hat{L} = \frac{1}{2} \left[ \partial_x \hat\psi \partial_x \hat\psi - V(x) \, \hat\psi \hat\psi \right] \; ,
\end{align}
where the real function $\psi$ is now promoted to a quantum operator, denoted by the hat symbol, and $V = B - A^2/4 - \partial_x A /2$ in comparison with \eqref{eq:schrodinger_canonical}. For a closed system with a Hermitian Hamiltonian, $V(x)$ is assumed to be real.%
\footnote{The value of $V(x)$ can be either positive or negative. In the exact WKB analysis, we often introduce small imaginary parameter to $V(x)$, in order to ensure the Borel summability. We take the limit where this parameter vanishes at the end of calculation.}
In our consideration, $x$ is the time variable.
The conjugate momentum of $\hat\psi$ is then
\begin{align}
    \label{eq:def_pi}
    \hat\pi \equiv \frac{\partial \hat{L}}{\partial \left( \partial_x \hat\psi \right)} = \partial_x \hat\psi \; ,
\end{align}
and the corresponding Hamiltonian is obtained by the Legendre transformation of the Lagrangian, i.e.,
\begin{align}
    \label{eq:hamil_quantum}
    \hat{H} \equiv \hat\pi \, \partial_x \hat\psi - \hat{L} = \frac{1}{2} \left[ \hat\pi \hat\pi + V(x) \, \hat\psi \hat\psi \right] \; .
\end{align}
Here we assume real fields, that is $\hat\psi^\dagger = \hat\psi$ and $\hat\pi^\dagger = \hat\pi$, and impose the standard condition of second quantization on the conjugate pair, that is
\begin{align}
    \label{eq:condition_quantum}
    \left[ \hat\psi , \hat\pi \right] = i \; , \qquad
    \left[ \hat\psi , \hat\psi \right] = \left[ \hat\pi , \hat\pi \right] = 0 \; ,
\end{align}
whereas their time evolution is determined by
\begin{align}
    \label{eq:eom_conjugatepair}
    \partial_x \hat\psi = -i \left[ \hat\psi , \hat H \right] = \hat\pi \; , \qquad
    \partial_x \hat\pi = -i \left[ \hat\pi , \hat H \right] = -V(x) \, \hat\psi \; .
\end{align}
These equations give the Schr\"{o}dinger-like equation in our system,
\begin{align}
  \label{eq:diffeq_psi}
  \partial_x^2\hat\psi + V(x) \, \hat \psi=0\ .
\end{align}
Note that the quantization condition \eqref{eq:condition_quantum} is preserved along time evolution, i.e.~$\partial_x \left[ \hat\psi , \hat\pi \right] = 0$. 

Quantization involves an appropriate identification of positive and negative frequency modes. For this purpose we first define the inner product of arbitrary functions $f(x)$ and $g(x)$ by
\begin{align}
    \label{eq:innerproduct}
    \left( f , \, g \right) \equiv - i \left( f \, \partial_x \bar g - \bar g \, \partial_x f \right) \; ,
\end{align}
where complex conjugation of a complex function $f(z)$ on $z\in \mathbb{C}$, denoted by $\overline{f(z)}$, takes the overall value $f(z)$ to its complex conjugate value for input $z$.%
\footnote{This is rather a standard notion of complex conjugation of a function. We make it explicit here just to avoid confusing ourselves and to distinguish it from $f(\bar{z})$ and $f_R(z) - i f_I(z)$, where $f(z) = f_R(z) + i f_I(z)$ with $f_R$ and $f_I$ real-valued functions for real $z$.}
We then define a set of functions $u_\pm(x)$ of respectively positive and negative frequency solutions (in an asymptotic state) such that they respect the conditions
\begin{align}
    \label{eq:modes}
    \left( u_+ , u_+ \right) = 1 \; , \qquad
    \left( u_- , u_- \right) = - 1 \; , \qquad
    \left( u_+ , u_- \right) = \left( u_- , u_+ \right) = 0 \; .
\end{align}
Now, we decompose the conjugate variables in terms of time-independent creation and annihilation operators, $\hat a^\dagger$ and $\hat a$, respectively,
\begin{align}
    \label{eq:decompose}
    \hat\psi(x) = u_+(x) \, \hat a + u_-(x) \, \hat a^\dagger \; , \qquad
    \hat\pi(x) = v_+(x) \, \hat a + v_-(x) \, \hat a^\dagger \; ,
\end{align}
where the reality condition of the variables enforces $\overline{u_+} = u_-$ and $\overline{v_+} = v_-$.%
\footnote{Note that by inverting the relations we can express $u_+ = \left[ \hat\psi , \hat a^\dagger \right]$, $v_+ = \left[ \hat\pi , \hat a^\dagger \right]$, and $\hat a = v_- \hat\psi - u_- \hat\pi$.}
In order to respect the quantization condition \eqref{eq:condition_quantum}, it suffices to impose
\begin{align}
    \label{eq:commutation_a}
    \left[ \hat a , \hat a^\dagger \right] = 1 \; , \qquad
    \left[ \hat a , \hat a \right] = \left[ \hat a^\dagger , \hat a^\dagger \right] = 0 \; ,
\end{align}
and
\begin{align}
    \label{eq:condition_uv}
    u_+ v_- - u_- v_+ = i \; .
\end{align}
With $\hat{a}$ and $\hat{a}^\dagger$ time-independent, the first equation in \eqref{eq:eom_conjugatepair} enforces $v_\pm = \partial_x u_\pm$.
In regard to the expansion \eqref{eq:decompose}, the Hamiltonian is expressed in terms of $\hat a$ and $\hat a^\dagger$,

\begin{align}
    \label{eq:H_by_a}
    \hat{H} =
    \left( \hat a^\dagger , \, \hat a \right)
    \mathscr{H}
    \left(
    \begin{array}{c}
      \hat a \vspace{1mm} \\ \hat a^\dagger
    \end{array}
    \right)
    \; , \qquad
    \mathscr{H} \equiv 
    \frac{1}{2}
    \left(
    \begin{array}{cc}
      v_+ v_- + V u_+ u_- \quad & v_-^2 + V u_-^2 \vspace{1mm} \\
      v_+^2 + V u_+^2 \quad & v_+ v_- + V u_+ u_-
    \end{array}
    \right) \; .
\end{align}
Note that, if we impose that $\hat a$ is a time-independent operator, all the time dependence is carried by $u_\pm$ and $v_\pm$. The operator evolution equations \eqref{eq:eom_conjugatepair} translates to a set of first-order equations
\begin{align}
  \label{eq:eom_uv}
  \partial_x u_\pm = v_\pm \; , \qquad
  \partial_x v_\pm = - V(x) \, u_\pm \; .
\end{align}
This reduction implicitly assumes the reality of $V(x)$ for all $x$ and the commutation relations \eqref{eq:commutation_a}.

Quantization involves diagonalization of an Hermitian operator to define its eigenstates as a basis of the Hilbert space that spans the quantum system. A natural choice for such an operator in our setup is the Hamiltonian, and it is in general not in a diagonal form in terms of $\hat a$ and $\hat a^\dagger$.
An exception is the case of $V = {\rm const.} \equiv V_0 > 0$, in which one can take the solutions to be 
$u_\pm \propto \exp\left( \mp i \sqrt{V_0} \, x \right)$
and 
$v_\pm = \mp i \sqrt{V_0} \, u_\pm$, 
and the off-diagonal components of $\mathscr{H}$ automatically vanish.
However, this happens thanks to the fact that the eigenvalues of $\mathscr{H}$ are degenerate under this special condition, and we need extra steps to diagonalize the Hamiltonian for a generic time-dependent $V$.

The eigenvalues of $\mathscr{H}$ are
\begin{align}
  \label{eq:eigenvalues_Hamiltonian}
    \omega_\pm = \frac{1}{2}\left(v_+ v_- + V u_+ u_- \pm \vert v_+^2 + V u_+^2 \vert \right)\; ,
\end{align}
where the condition \eqref{eq:condition_uv} has been used. Clearly $\omega_\pm$ are real provided $V$ is real. Using the triangle inequality $\left\vert \vert v_+ \vert^2 - \vert V \vert \vert u_+ \vert^2 \right\vert \le \vert v_+^2 + V u_+^2 \vert \le \vert v_+ \vert^2 + \vert V \vert \vert u_+ \vert^2$, we observe
\begin{subequations}
\begin{align}
    \omega_\pm \ge 0 \; , \qquad & \mbox{ for } \; V \ge 0 \; , \\
    \omega_- < 0 < \omega_+ \; , \qquad & \mbox{ for } \; V < 0 \; .
\end{align}    
\end{subequations}
Thus there exists a tachyonic mode for $V < 0$, and this region is unitarily disconnected from the $V > 0$ region, in the sense that they cannot go back and forth by an operation of the form $U^\dagger \mathscr{H} U$ with an arbitrary matrix $U$, since such a transformation does not change the sign of the determinant.

The ground state, as the lowest energy state at a classical fixed point in the (real) phase space spanned by $(\psi, \pi)$, makes sense only for $V>0$,%
\footnote{We exclude the case $V=0$ from our discussion, as the system in \eqref{eq:lagr_quantum} would contain the kinetic term alone without interesting dynamics. Moreover, in the field-theoretical language, $V=0$ would typically correspond to a vanishing sound speed, $c_s = 0$, which would be outside the validity range of effective action, see e.g.~\cite{Cheung:2007st}.}
as otherwise an arbitrary negative energy could be quantum-mechanically accessible.
Thus we hereafter assume $V>0$ whenever we attempt canonical quantization and define a Hilbert space at some asymptotic points.
In order to quantize the system with respect to the Hamiltonian \eqref{eq:H_by_a}, we still need a further step in general. If a vacuum $\vert 0 \rangle$ were defined with respect to $\hat{a}$, i.e.~ $\hat{a} \, \vert 0 \rangle = 0$, one would find $: \!\!\! \hat{H} \!\!\! : \!\! \vert 0 \rangle = \left( v_-^2 + V u_-^2 \right) \big( \hat{a}^\dagger \big)^2 \vert 0 \rangle \ne 0$, where $: \!\! \mdblkcircle \!\! :$ denotes normal ordering, which would imply that energy could be extracted from the ``vacuum.'' Hence we understand that we need to complete the diagonalization of the Hamiltonian matrix $\mathscr{H}$ before identifying the true vacuum state.%
\footnote{The unitary matrix constructed from the eigenvectors of $\mathscr{H}$ would not preserve the Hermitian conjugation between the creation and annihilation operators or the complex conjugation between the two mode functions. This is ultimately due to the time dependence of $V$. The exception would be the case with time-independent $V > 0$, in which one could trivially choose $u_\pm$ and $v_\pm$ such that $\mathscr{H}$ be diagonal, and $\omega_\pm = \omega = \sqrt{V}/2$.}

In the following subsections, we first describe particle production as change of vacuum states, under the assumption that these states are properly defined. We then summarize the conventional method of the diagonalization using the WKB (or adiabatic) approximation and point out its potential issues to motivate our subsequent construction based on the exact WKB formulation.

\subsection{Particle production}
\label{subsec:part_prod}

In general, the vacuum is not uniquely defined in a time-dependent background, due to the fact that a time-dependent Hamiltonian is inherently intertwined to the absence of globally defined timelike Killing vector and leads to the frame-dependent notion of energy, see e.g.~\cite{Birrell:1982ix}.
Moreover, each of such vacua is defined as an asymptotic state. Generically the Hamiltonian \eqref{eq:H_by_a} cannot be diagonalized while keeping the commutation relation \eqref{eq:condition_quantum}, or equivalently \eqref{eq:commutation_a} with \eqref{eq:condition_uv}, at an arbitrary time, but can be (approximately) done so only at some asymptotic time. 

Let us suppose that the $\big( \hat a , \hat a^\dagger \big)$ basis in \eqref{eq:decompose} (approximately) diagonalizes the Hamiltonian, i.e.~$v_+^2 + V u_+^2 = \big( v_-^2 + V u_-^2 \big)^* \simeq 0$ at some asymptotic time, which means that the Hamiltonian matrix $\mathscr{H}$ in \eqref{eq:H_by_a} is approximately diagonal and $\hat{H} \simeq \big( v_+ v_- + V u_+ u_- \big) \big( \hat a^\dagger \hat a + \hat a \hat a^\dagger \big)/2$. This allows us to define a vacuum state $\vert 0\rangle$ such that
\begin{align}
  \hat{a} \, \vert 0\rangle =0 \; ,
\end{align}
at this time.
This condition and the commutation algebra in \eqref{eq:commutation_a} are sufficient to generate a tower of orthonormal states, forming a complete Fock space as
\begin{align}
  \vert 1 \rangle = \hat a^\dagger \, \vert 0 \rangle \; , \qquad
  \vert 2 \rangle 
  = \frac{\big( \hat a^\dagger \big)^2}{\sqrt{2}} \, \vert 0 \rangle  \; , \quad \dots\dots \; , \qquad
  \vert n \rangle 
  = \frac{\big( a^\dagger \big)^n}{\sqrt{n!}} \, \vert 0 \rangle \; .
\end{align}
Thanks to the diagonalized form of the Hamiltonian, these states form its orthonormal eigenbasis at the asymptotic time, and the system is properly quantized.

Now we further assume the existence of another asymptotic time in which we can define a new vacuum state $\vert \tilde 0 \rangle$ with a corresponding (again time-independent) annihilation operator $\hat{\tilde a}$, such that
\begin{align}
  \hat{\tilde a} \, \vert \tilde 0\rangle =0\ ,
\end{align}
leading to an alternative decomposition of the same $\hat\psi(x)$ as,
\begin{align}
    \hat\psi(x) = \tilde u_+(x) \, \hat {\tilde a} + \tilde u_-(x) \, \hat {\tilde a}^\dagger\ .
\end{align}
Here, the new mode functions $\tilde u_\pm(x)$ are respectively positive and negative frequency modes for a different asymptotic state, satisfying the inner product conditions in \eqref{eq:modes}.
The quantization procedure identically follows the untilded case above.
Provided that each set of $u_\pm$ and $\tilde u_\pm$ forms the basis of the two independent solutions to the second-order differential equation \eqref{eq:diffeq_psi},
these new mode functions $\tilde u_\pm(x)$ (or the new vacuum) can be related to the original mode functions $u_\pm(x)$ as follows,
\begin{align}
\label{eq:decomposition}
    \tilde u_+=\alpha\, u_++\beta\, u_-\; , \qquad
    \tilde u_-=\bar\beta\, u_++\bar\alpha\, u_- \; .
\end{align}
The coefficients $\alpha$ and $\beta$, known as Bogoliubov coefficients,
are invertedly determined by the inner product between the mode functions corresponding to different vacua,
\begin{align}
  \alpha=(\tilde u_+,u_+) \; , \qquad
  \beta=-(\tilde u_+, u_-) \; ,
\end{align}
and they satisfy the consistency condition,
\begin{align}
    |\alpha|^2-|\beta|^2=1 
\end{align}
to preserve the 
commutation relation structure, namely the counterpart of \eqref{eq:condition_uv} for the tilded mode functions. Particle production is characterized by the non-zero expectation value of the number operator of one vacuum with respect to another vacuum,
\begin{align}
  \label{eq:number_operator}
  \langle \tilde 0 \vert \, \hat a^\dagger \hat a \, \vert \tilde 0\rangle = 
  \langle 0 | \, \hat {\tilde a}^\dagger \hat {\tilde a} \, | 0\rangle =|\beta|^2 \; .
\end{align}
The evaluation of the number operator by the vacuum state in a different frame is the essential part of particle production, the feature not only of cosmological background but also of the Hawking radiation, the Unruh effect, etc.. The existence of multiple vacua, or equivalently multiple pairs of positive and negative frequency modes, results from a nontrivial background, and the occupation number in \eqref{eq:number_operator} in a way measures the ``mismatch'' between the two different vacua. This interpretation of particle production therefore makes sense only if the Hamiltonian can be diagonalized, at least approximately, at different asymptotic times each of which the vacuum is well defined at.

We here summarize the algebraic structure of particle production, or rather that of the Bogoliubov transformation, in the hope of extending our formulation to a system of multiple degrees of freedom.
The connection formulae of the mode functions in \eqref{eq:decomposition} can be written in a matrix form as
\begin{align}
\label{eq:connection_upm}
\begin{pmatrix}
\tilde{u}_+ \\
\tilde{u}_-
\end{pmatrix}
=
\begin{pmatrix}
\alpha & \beta \\
\bar\beta & \bar\alpha
\end{pmatrix}
\begin{pmatrix}
u_+ \\
u_-
\end{pmatrix} \; , 
\end{align}
or in terms of the creation/annihilation operators as
\begin{align}
  \label{eq:8}
  \begin{pmatrix}
    \hat{\tilde a} \\
    \hat{\tilde a}^\dagger
  \end{pmatrix}
  =
  \begin{pmatrix}
    \bar\alpha & - \bar\beta \\
    - \beta & \alpha
  \end{pmatrix}
  \begin{pmatrix}
    \hat a \\
    \hat a^\dagger
  \end{pmatrix} \; .
\end{align}
We refer to the matrix that relates the mode functions in \eqref{eq:connection_upm} as the connection matrix. 
The determinant of this matrix is unity, implying that the connection matrix is an element of the ${\rm SU}(1,1)$ group,
\begin{align}
    \begin{pmatrix}
\alpha & \beta \\
\bar\beta & \bar\alpha
    \end{pmatrix}\; ,
  \quad |\alpha|^2-|\beta|^2=1
  \quad \in \; {\rm SU}(1,1)\ .
\end{align}
In a more general case, a coupled system with $N$ bosonic degrees of freedom in a time-evolving background is described with a connection matrix belonging to ${\rm U}(N,N) \cap {\rm Sp}(2N,\mathbb{C})$.%
\footnote{See the appendix \ref{app:n_system}.}
In the absence of particle production, i.e., when $\beta=0$, the connection matrix is diagonal and corresponds to an element of U(1)
\begin{align}
\begin{pmatrix}
\alpha & 0 \\
0 & \bar\alpha
\end{pmatrix}
=
\begin{pmatrix}
{\rm e}^{i\theta} & 0 \\
0 & {\rm e}^{-i\theta}
\end{pmatrix}\; , \quad
  \theta\in [0,2\pi)
\quad \in \; {\rm U}(1)\ .
\end{align}
This indicates that the state $(u_+,u_-)$ differing only by a U(1) element does not lead to any physical differences, i.e.
\begin{align}
\label{eq:redundancy}
\begin{pmatrix}
\tilde{u}_+ \\
\tilde{u}_-
\end{pmatrix}
  \cong
\begin{pmatrix}
{\rm e}^{i\theta} & 0 \\
0 & {\rm e}^{-i\theta}
\end{pmatrix}
\begin{pmatrix}
\tilde{u}_+ \\
\tilde{u}_-
\end{pmatrix}\ ,
\end{align}
where 
$\cong$ denotes the left and right-hand expressions are physically equivalent, i.e. they give the same number density.
Consequently, the evolution of the Bogoliubov coefficients is described by an ${\rm SU}(1,1)$ matrix, while this description has a redundancy under left and right multiplications by arbitrary diagonal phases in ${\rm U}(1)$,
\begin{align}
        \begin{pmatrix}
\alpha & \beta \\
\bar\beta & \bar\alpha
\end{pmatrix}
  \cong
\begin{pmatrix}
\ee^{i\theta_L} & 0 \\
0 & \ee^{-i\theta_L}
\end{pmatrix}
\begin{pmatrix}
\alpha & \beta \\
\bar\beta & \bar\alpha
\end{pmatrix}
\begin{pmatrix}
\ee^{i\theta_R} & 0 \\
0 & \ee^{-i\theta_R}
\end{pmatrix}\ ,
\end{align}
where $\theta_{L,R}$ denote the phases.
This reflects the fact that the Bogolyubov transformation is defined up to phase rotations associated with the choice of basis in the asymptotic vacuum states. 

Particle production is formulated above as a change of vacuum states, which are defined as (approximate) eigenstates of the Hamiltonian at asymptotic temporal regions. While the vacuum is not globally unique in a time-dependent system, one can define a unique vacuum state at each asymptotic region, at least in principle, provided $V>0$ and that the adiabatic condition $\vert \partial_x V \vert / V^{3/2} \ll 1$ is respected.%
\footnote{In fact, this uniqueness relies on the frame of an observer. If an observer is in a non-inertial frame, for example, even the Minkowski vacuum may be perceived as a thermalized state as in the Unruh effect. Thus we here implicitly assume fixing the coordinate system first.}
In this sense, particle production is to be understood only as a relational statement between two states.
The remaining task is how to diagonalize the Hamiltonian while keeping the quantum commutation relation \eqref{eq:condition_quantum}. We review the conventional procedure in the following subsection, followed then by our proposal using the exact WKB solutions.
We emphasize that, in general, diagonalization of a time-dependent Hamiltonian can be done only approximately, and its asymptotic points correspond to singular points of the differential equation of the system.

\subsection{Conventional Method of Quantization and its Issues}
\label{subsec:issues}

The conventional scheme for diagonalization of $\mathscr{H}$ in \eqref{eq:H_by_a} relies on the intuition developed from the harmonic oscillator case and uses the leading-order WKB solutions as the mode functions in quantization. 
Concretely, it assumes that the WKB approximate solutions determine the positive/negative frequency mode functions $u_\pm$ at some asymptotic time $x = \sigma_i$. Then one picks the asymptotic form of the mode functions as
\begin{align}
  \label{eq:upm_conventional}
  u_{\pm}(x) \to \frac{1}{\sqrt{2 \omega}} \exp \left( \mp i\int^x \omega \, \dd x \right) \; , \quad
  v_\pm(x) \to \mp i \sqrt{\frac{\omega}{2}} \exp \left( \mp i\int^x \omega \, \dd x \right) \; , \quad
  \text{for}~x\to \sigma_i \; ,
\end{align}
where $\omega \equiv \sqrt{V}$ at $x \to \sigma_i$. Provided $\omega \in \Re$ and $\omega > 0$, $u_\pm$ respect the quantization condition \eqref{eq:condition_uv}, derived from the commutation relation \eqref{eq:condition_quantum} of the conjugate pair.
In order to consider production of \textit{particles}, one assumes another asymptotic time, say $x = \sigma_f$, at which a different vacuum state is defined and the corresponding mode functions $\tilde{u}_\pm$ at $\sigma_f$ are written as
\begin{align}
  \label{eq:upm_conventional_tilded}
  \tilde{u}_{\pm}(x) \to \frac{1}{\sqrt{2 \tilde\omega}}\exp \left( \mp i\int^x \tilde\omega \, \dd x \right) \; , \quad
  \tilde{v}_\pm(x) \to \mp i \sqrt{\frac{\tilde\omega}{2}} \exp \left( \mp i\int^x \tilde\omega \, \dd x \right) \; , \quad
  \text{for}~x\to \sigma_f \; ,
\end{align}
where $\tilde\omega \equiv \sqrt{V}$ at $x \to \sigma_f$. In general $\omega \ne \tilde\omega$, and hence $u_\pm \ne \tilde{u}_\pm$ and $v_\pm \ne \tilde v_\pm$.
As long as the mode functions $u_\pm$ and $\tilde{u}_\pm$ are proper asymptotic solutions to the equation of motion, they are related to each other by a linear combination, at least approximately. The coefficients of this relation play the role of the Bogoliubov coefficients $\alpha$ and $\beta$, and $\beta \ne 0$ is interpreted as particle production in one of the vacuum states viewed from the other.

This conventional approach inherits several conceptual issues, some of which are intertwined to each other.
\begin{enumerate}
  \item \textit{Breakdown of the WKB approximation:}
  
    As is clear from the expression \eqref{eq:upm_conventional} (or \eqref{eq:upm_conventional_tilded}), these solutions diverge at $V = 0$. Moreover, since $u_\pm$ and $v_\pm$ (or $\tilde u_\pm$ and $\tilde v_\pm$) of the above form do not satisfy the equation of motion in \eqref{eq:eom_uv}, the true solutions may be expanded as $\alpha(x) \, u_+(x) + \beta(x) \, u_-(x)$ and $\alpha(x) \, v_+(x) + \beta(x) \, v_-(x)$, where the Bogoliubov coefficients are now promoted to time-dependent quantities.%
    \footnote{These functions $\alpha(x)$ and $\beta(x)$ should be identified as the conventional Bogoliubov coefficients that are constants at the asymptotic time $\sigma_i$. In this paper we abuse the use of the term ``Bogoliubov coefficients'' to call both the time-independent and time-dependent quantities as long as no confusion is invoked.}
    Eq.~\eqref{eq:eom_uv} can then be rewritten as the evolution equations for $\alpha$ and $\beta$, reading
    \begin{align}
      \label{eq:eom_aphabeta}
      \partial_x \alpha = \frac{\partial_x \omega}{\omega} \, \beta \, \exp \left( 2 i \int^x \omega \, \dd x \right) \; , \qquad
      \partial_x \beta = \frac{\partial_x \omega}{\omega} \, \alpha \, \exp \left( - 2 i \int^x \omega \, \dd x \right) \; .
    \end{align}
    While the equations in \eqref{eq:eom_aphabeta} clearly indicate that $\partial_x \omega \ne 0$ is the source of mixing $\alpha$ and $\beta$, they are also ill-defined at $\omega = 0$. Therefore, the conventional formulation based on the WKB approximate solutions is obscure in its interpretation for systems that admit zero crossing of $\omega$.

\item \emph{Ambiguity of the vacuum with respect to higher orders of the WKB series:}

The determination of ``positive'' and ``negative'' frequency modes is purely based on the leading order of the WKB series in the conventional formulation. This approach is thus agnostic about higher orders in the series, which in turn implies that the decomposition of creation and annihilation operators has the same amount of ambiguity, and so does the vacuum state. For example, it is insensitive to exponentially suppressed contributions, if any, in choosing the frequency modes. Even if suppressed in the asymptotic region, they may switch to a growing function in the course of time evolution. In such a case, whether to include such contributions or not may change the results significantly.

\item \emph{Incompatibility with the exact solutions:}
\label{item:incompatible}

In general, it is not guaranteed that the leading-order WKB solutions \eqref{eq:upm_conventional} act as the asymptotic expansions of the exact solutions. For example, in the cosmological perturbation theory, the exact solutions for the scalar and tensor modes are known in terms of the Hankel functions under the slow roll approximation. While each mode behaves as a free massless field in de Sitter in the asymptotic past, in which the Bunch-Davies vacuum provides the proper initial conditions, the leading-order WKB poorly approximates the exact modes in superhorizon scales.

\item \emph{Divergent WKB series expansion:}

The WKB solutions are expressed as a divergent series in general, which we discuss more details in the next section. Because of this, the WKB approximation makes sense only in the asymptotic regions. This divergence is a generic feature of the WKB expansion with a zero convergence radius, occuring even away from $\omega = 0$. This fact does not immediately undermine the conventional approach and rather means that the WKB series is the asymptotic expansion of the exact solution. Nevertheless, these approximate solutions are valid only locally in the time domain, and the divergent behavior is a crucial obstacle for a global analysis of the system.

\item \emph{Lack of knowledge on the global behavior: connection formulae between two asymptotic solutions:}

Even if asymptotic solutions are known as WKB approximations, the correct behavior of solutions in intermediate regions needs to be known somehow. 
In simplest scenarios, there may exist analytic solutions in full regions. This is, however, only in limited cases when using specific potentials under some approximations.
The conventional method to overcome the above issue is to go back to the original variable (wave function) and solve its equation of motion around the point where such a pathological behavior occurs. This appears to work often times, but there are at least $3$ issues: (i) in the analytical approach this method often requires an expansion around the saddle point and connects the obtained solution with some junction conditions to the full solution, but whether the performed connection is mathematically well justified or not is obscure,
(ii) quantization is done for the asymptotic states, and how to connect the full solution to an asymptotic one, which is essentially the WKB 
solution, where the initial and final states are defined, is not as obvious as people typically assume (related to point \ref{item:incompatible} above), and (iii) in the presence of the potential singular behavior of the Bogoliubov coefficients, the interpretation of particle production is not necessarily straightforward.

\item \emph{Diagonalization of the Hamiltonian is approximate:}

From the expressions of the Hamiltonian in  \eqref{eq:H_by_a} or of its eigenvalues in \eqref{eq:eigenvalues_Hamiltonian}, it is observed that, if the WKB mode functions \eqref{eq:upm_conventional} or \eqref{eq:upm_conventional_tilded} were the exact solutions, the off-diagonal components of the Hamiltonian would indeed vanish, but generically this is not the case with a realistic potential. One might think that this would not cause a problem as long as the off-diagonal components are much smaller than the on-diagonal ones, i.e.~$\vert v_\pm^2 + V u_\pm^2 \vert \ll v_+ v_- + V u_+ u_-$ in \eqref{eq:H_by_a}. It is true that the \emph{eigenvalues} of such a Hamiltonian are dominated by its on-diagonal components; however, the corresponding \emph{eigenvectors} are in general not at all close to the standardized form $(1,0)$ and $(0,1)$.%
\footnote{As a simple example, if the Hamiltonian takes the form, e.g.,
\begin{align*}
  \begin{pmatrix}
    \omega^2 & \epsilon \\
    \epsilon & \omega^2
  \end{pmatrix}
\end{align*}
with $\vert \epsilon \vert \ll \omega^2$, then its eigenvalues are $\omega^2 \pm \epsilon \simeq \omega^2$, but its eigenvectors are $(1,\pm 1)/\sqrt{2}$. Notably, the eigenvectors are independent of $\epsilon$.}
This implies that, even with tiny off-diagonal terms in the Hamiltonian, while further diagonalization procedure is not mandatory to obtain the energy spectrum, determination of the ground state cannot be done until the Hamiltonian is fully diagonalized.

\item \emph{Failure in non-adiabatic backgrounds (tachyonic region, quenches):}

The validity of the WKB asymptotic approximation in \eqref{eq:upm_conventional} and \eqref{eq:upm_conventional_tilded} relies on two basic assumptions: (i) $\omega^2 = V > 0$ and (ii) $\vert \partial_x \omega / \omega^2 \vert \ll 1$ in the asymptotic regions. If (i) were violate, then the WKB solutions could not meet the quantization condition \eqref{eq:condition_uv} while respecting the conjugate relation between $u_\pm$ or $v_\pm$ even approximately. If (ii) were violated, then \eqref{eq:eom_aphabeta} would tell us that the Bogoliubov coefficients evolve rapidly and thus the adiabatic vacuum would be poorly defined. A typical example of the first violation is the presence of tachyonic modes, which actually includes gradient instabilities in this context. An example of the second one is sudden quenches, in which $V(x)$ changes rapidly for a short amount of time.

\item \emph{Ultraviolate divergence:}

In the conventional canonical formulation of quantum field theory in curved spacetime, one often encounters ultraviolet divergences in local observables, such as in the expectation value of the energy-momentum tensor. These divergences originate from the high-momentum behavior of the quantum modes and reflect the fact that the energy density and pressure are composite operators requiring proper regularization and renormalization.
The leading-order WKB approximation does not provide a sufficiently accurate representation of the high-frequency modes to yield a finite result. As a consequence, higher-order adiabatic expansions are required to consistently subtract the divergent contributions through adiabatic regularization or other renormalization schemes such as point-splitting or dimensional regularization. While this issue is not a failure of the WKB approach \emph{per se}, it points to the need to better account for the short-distance structure of quantum fields, which cannot be captured by low-order approximations alone.

\end{enumerate}

The structure of this paper is as follows. Section~\ref{sec:OurFormulation} develops our formulation for describing particle production using exact WKB solutions. Section~\ref{sec:review_exact_WKB} provides a concise review of the foundations of exact WKB analysis, while Section~\ref{sec:voros_coefficient} offers a detailed account of the Voros coefficients and their concrete calculation. In Section~\ref{sec:particle_production_exact_WKB}, we outline a general procedure to compute the number density and illustrate it in a concrete potential. Finally, Section~\ref{sec:conclusion} summarizes our conclusions.

\section{Exact WKB approach to quantization and particle production \lowercase{\textit{\`{a} la}} Bogoliubov 
}
\label{sec:OurFormulation}

As discussed in Sec.~\ref{subsec:issues}, the conventional method to describe particle production has several conceptual shortcomings that originate from the limitation due to the use of the WKB approximated solutions. To overcome some of these issues, we adopt so-called \textit{exact WKB solutions} instead of their approximated counterparts to define the mode functions and describe our approach of quantization with these solutions.
The exact WKB method is in short a resummation of full WKB series to obtain a globally well-defined solution to a given second-order differential equation. Properly set up, it guarantees that the resummed function has the same asymptotic behavior as the WKB leading approximation, while being a solution in the non-asymptotic regions without divergent behaviors. With this methodology, we are equipped to analyze the global structure of the exact solutions. We formulate how to use the exact WKB formulation to understand quantization and particle production in this section. Please refer to Sec.~\ref{sec:review_exact_WKB} for details of the exact WKB method and the corresponding notations. 
For recent works applying the exact WKB approach or Stokes phenomenon to 
cosmological particle production, see 
Refs.~\cite{Kim:2013jca,Li:2019ves,Enomoto:2020xlf,Hashiba:2020rsi,Hashiba:2021npn,Enomoto:2021hfv,Yamada:2021kqw,Hashiba:2022bzi,Enomoto:2022nuj}.
For a broader set of applications of the exact WKB method 
in high-energy physics and mathematical physics, 
see, for a partial list, 
Refs.~\cite{Nekrasov:2009rc,Sueishi:2020rug,Taya:2020dco,Yan:2020kkb,Ito:2021boh,Grassi:2021wpw,Kamata:2021jrs,Enomoto:2022mti,Misumi:2024gtf,Fujimori:2025kkc,Hao:2025azt,Matsuda:2025nwp,Morikawa:2025grx,Morikawa:2025xjq,Morikawa:2025vvs,Meynig:2025lnk,Kamata:2025dkk,
Basar:2015xna,Dunne:2016qix,Sueishi:2019xcj,Cavusoglu:2023bai,Ture:2024nbi,Pazarbasi:2021fey,
Kashani-Poor:2015pca,Kashani-Poor:2016edc, Ashok:2016yxz,Gaiotto:2012rg,Dorey:2001uw, Dorey:2007zx,Ito:2018eon,Ito:2019jio,Imaizumi:2020fxf,Emery:2020qqu,Ito:2024nlt,Grassi:2014cla, Codesido:2017dns,Codesido:2017jwp,Hollands:2019wbr,Ashok:2019gee,Coman:2020qgf,Iwaki:2023cek,Imaizumi:2022dgj,vanSpaendonck:2022kit,Kamata:2023opn,Bucciotti:2023trp,Kamata:2024tyb}.

We start from the mode function $u_+(x)$ given as a linear combination of the exact WKB solutions $\Psi_\pm(x)$.
We are in general only able to define vacuum states at some asymptotic time, say $\sigma$, and thus it is appropriate to take a function that is properly normalizable at $x=\sigma$. Denoting the exact WKB solutions normalized at $x=\sigma$ by $\Psi_\pm^{(\sigma)}(x)$, we write $u_+(x)$ as
\begin{align}
\label{eq:modefunction_exactWKB}
    & u_+(x)=\frac{1}{\sqrt{2}}\left[\alpha\, \Psi_-^{(\sigma)}(x)+\beta\, \Psi_+^{(\sigma)}(x)\right]\ , 
\end{align}
where the coefficients $\alpha,\beta\in \mathbb{C}$ are determined from boundary conditions discussed later.
There is, however, difficulty to find the mode functions with the exact WKB solutions due to the operation of complex conjugation. In fact, since complex conjugation of a complex function is in general not analytic, the complex conjugates of the Borel summed solutions
are not analytic generically.
Since the exact WKB method applies to analytic (or more precisely meromorphic) functions, this poses an additional complication.
Also, finding the complex conjugation in some given $x$ is difficult since the exact WKB solutions often lack explicit analytic forms.
Complex conjugation becomes a non-trivial task. To avoid these problems, we consider the complex conjugation only in the asymptotic region, always keeping in mind that we take the limit ${\rm Im} \, (x) \to 0$ in the same region at the end of the calculation.
Note that, 
when an exact WKB solution satisfies the Schr{\"o}dinger-like equation,
\begin{align}
    \left[
    - \frac{\dd^2}{\dd x^2} - 
    V(x)
\right]
\Psi(x) = 0\ ,
\end{align}
its complex conjugate $\overline{\Psi(x)}$ also satisfies
\begin{align}
\label{eq:schrodinger_cc}
\left[
    - \frac{\dd^2}{\dd x^2} - 
    V(x)
\right]
\overline{\Psi(x)}
= 0 \ , 
\end{align}
provided $V(x) \in \mathbb{R}$ with $x \in \mathbb{R}$, which is a typical variable domain of physical interests.
That is, the complex conjugation of the exact WKB solution 
is also a solution of the original equation in $x\in \mathbb{R}$ and with real potential.

Recovering the analyticity, we are to (again) complexify $x$. Then $\overline{\Psi(x)}$ is to be analytically continued as an analytic solution to the original Schr\"{o}dinger-like equation \eqref{eq:schrodinger_cc} and can thus be identified as a linear combination of $\Psi_\pm(x)$.
Since the standard WKB solution should serve as an asymptotic series of the exact solution in terms of the expansion parameter, we construct the series such that its leading-order approximation well describes the exact solution in the asymptotic region $x \to \sigma$, i.e.,%
\footnote{
The existence of the asymptotic expansion is guaranteed when the Watson's lemma is applicable as discussed later.}
\begin{align}
    \label{eq:exactWKB_leading_expansion}
    \Psi^{(\sigma)}_{\pm}(x) \sim
    \frac{1}{\left( - V \right)^{1/4}} \exp\left( \pm \int^x \sqrt{-V} \, \dd x \right) \; , \qquad
    x \to \sigma \; .
\end{align}
Then $\overline{\Psi_\pm^{(\sigma)}(x)}$ should be approximated by the complex conjugate of the above expression in the limit $x \to \sigma$. However, how the analytical continuation into the complex $x$ domain is done is in general not unique. In other words, a general exact WKB solution $\Psi(x)$ with complex $x$, which is analytic in the vicinity of $x=\sigma$, may take different values depending on which direction $x$ and $V(x)$ approach the asymptotic point $\sigma$ on the complex plane. In the language of exact WKB analysis, this difference is closely related to the quantity called \textit{Voros coefficients}.
In short, the properly normalized exact WKB solution that has the asymptotic behavior of \eqref{eq:exactWKB_leading_expansion} would not be realized without taking into account the effect of Voros coefficients; moreover, they contribute to the amount of particle production significantly. 
In Sec.~\ref{sec:voros_coefficient}, we carefully discuss how to compute the Voros coefficients and how to construct the exact WKB solutions free from the ambiguity they may cause.%
\footnote{In order to avoid the ambiguity associated with complex conjugation, taking the so-called \textit{median resummation} \cite{ecalle1981fonctions,ecalle1984singularites,ecalle1994weighted} with respect to the Voros coefficients may be considered to be an option, as it commutes with complex conjugation \cite{Delabaere_10.1063/1.532206,Delabaere_AIHPA_1999__71_1_1_0}. While not necessarily introducing a fundamental issue, the median resummed exact WKB solutions appear to have non-trivial normalization, and its leading-order amplitude may not reproduce the WKB approximation. For this reason, we refrain from using the median resummation in this paper.}

Determination of a vacuum, i.e. the determination of the coefficients $\alpha$ and $\beta$ in \eqref{eq:modefunction_exactWKB}, requires boundary conditions.
In the context of particle production, these conditions are deduced from the quantization condition at an asymptotic point $x=\sigma$, which is also a singular point of the given differential equation. 
It is also important to specify the direction along which the asymptotics is approached; for the ease of finding the complex conjugation in the above prescription, it appears appropriate for our purpose to take this path along the \textit{anti-Stokes curve} that flows into $\sigma$.%
\footnote{The anti-Stokes curve is properly defined in Sec.~\ref{sec:review_exact_WKB}. 
For the purpose of the current discussion, it is sufficient to define it by
\begin{align*}
    {\rm Re} \int^x_\tau \sqrt{-V} \, \dd x = 0
\end{align*}
where $\tau$ is an appropriate turning point $V(\tau) = 0$.}
Since the leading order of the asymptotic expansion of the exact WKB solution as in \eqref{eq:exactWKB_leading_expansion} is purely oscillatory on this curve, we can straightforwardly identify the positive and negative frequency modes for $x \to \sigma$.
We then choose $u_\pm$ without mixing those two modes.%
\footnote{This boundary condition can make the Hamiltonian approximately diagonal in the asymptotic limit. See 
\eqref{eq:H_components}.}
In practice, we consider complex conjugates of the exact WKB solutions $\Psi_\pm^{(\sigma)}$ only along the anti-Stokes curves, analytically continue them to identify them with the opposite modes, and perform quantization only at the asymptotic limit $x \to \sigma$.

What is the physical meaning to take positive or negative frequency modes as the boundary condition?
To proceed, let us first summarize our formulation of canonical quantization.
We have defined the mode functions $u_{\pm}$ with the exact WKB solution normalized at some singular point $x=\sigma$, denoted by $\Psi^{(\sigma)}_\pm$,  introducing constant coefficients $\alpha$ and $\beta$,
\begin{align}
\label{eq:uplus_eWKB}
    u_+(x) 
    =\frac{1}{\sqrt{2}}
    \left[\alpha\, \Psi^{(\sigma)}_-(x)+\beta\, \Psi^{(\sigma)}_+(x)\right] \; .
\end{align}
Notice that the above prescription of complex conjugation practically identify $\overline{\Psi_\pm^{(\sigma)}(x)} \leftrightarrow 
i \, \Psi_\mp^{(\sigma)}(x)$, where 
the additional factor $i$ comes from the phase of $1/\sqrt{S_{\rm odd}} \sim (-V)^{1/4}$ for $V \in \mathbb{R}$.%
\footnote{This phase also depends on the branch choice of $(-V)^{1/4}$ at $x \to \sigma$. If we choose the other branch, the only effect is that the roles of $\alpha$ and $\beta$ flip.}
Hence we choose the minus mode as
\begin{align}
\label{eq:uminus_eWKB}
    u_-(x) = \frac{
    i}{\sqrt{2}} \left[\overline{\alpha} \, \Psi_+^{(\sigma)}(x) + \overline{\beta} \, \Psi_-^{(\sigma)}(x) \right] \; .
\end{align}
The corresponding conjugate momenta are $v_\pm = \partial_x u_\pm$.
As we show in Sec.~\ref{sec:review_exact_WKB}
, the Wronskian of the exact WKB solutions is given by,%
\footnote{Strictly speaking, this relation in \eqref{eq:wronskians} holds for the exact WKB solutions that are normalized at some turning point. Yet, it is valid even for those that are normalized at a singular point.}
\begin{align}
    \mathcal{W}\left\{ \Psi_-(x) , \, \Psi_+(x) \right\} \equiv \Psi_- \, \frac{\dd \Psi_+}{\dd x} - \frac{\dd \Psi_-}{\dd x} \, \Psi_+ = 2 \; .
\end{align}
Using this property, one can show that the commutation condition \eqref{eq:condition_uv} results in the equivalent condition on $\alpha$ and $\beta$ introduced in \eqref{eq:uplus_eWKB}, reading
\begin{align}
\label{eq:commutation_condition_eWKB}
    u_+ v_- - u_- v_+ = i \qquad \Longleftrightarrow \qquad \vert \alpha \vert^2 - \vert \beta \vert^2 = 1 \; .
\end{align}
This shows that the mode functions using the exact WKB solutions satisfy the necessary commutation relation and are thus eligible for quantization.

The quantization is not fully demonstrated yet, as it is to be done by quantizing the energy eigenvalues in terms of the Fock space. This amounts to showing that the mode functions defined in \eqref{eq:uplus_eWKB} and \eqref{eq:uminus_eWKB} and their conjugate momenta properly diagonalize the Hamiltonian in \eqref{eq:H_by_a} in the limit $x\to \sigma$. Using the leading-order asymptotic expansion \eqref{eq:exactWKB_leading_expansion}, we find the on- and off-diagonal components of $\mathscr{H}$ as
\begin{subequations}
\label{eq:H_components}
\begin{align}
    v_+ v_- + V u_+ u_-
    & \to \left( \vert \alpha \vert^2 + \vert \beta \vert^2 \right) \sqrt{V} + \dots \; ,
    \\
    v_+^2 + V u_+^2 
    & \to -2 i \alpha \beta \sqrt{V} + \dots \; ,
\end{align}
\end{subequations}
where $\dots$ denotes the higher-order terms near $x=\sigma$, which are of negative powers of $V$ depending on the details of the behavior of the potential around $\sigma$.
It is evident from the above expression that the off-diagonal components of $\mathscr{H}$ cancels in the limit of $x\to \sigma$ only if either $\alpha=0$ or $\beta=0$; due to the condition \eqref{eq:commutation_condition_eWKB}, the only choice is $\vert \alpha \vert = 1$ and $\beta = 0$. The phase of $\alpha$ is then an overall phase and thus redundant as we have discussed in Sec.~\ref{subsec:part_prod}.
This is the boundary conditions we impose at the asymptotic point for quantization.
Therefore, we have shown that our boundary condition can make the Hamiltonian diagonal in the corresponding asymptotic region.
Let us emphasize that the diagonalization is only approximate near the singular points of the equation of motion. This implies that quantization makes sense only in an asymptotic respect. 
Once a Stokes curve is crossed, the Stokes phenomenon occurs, and the asymptotic expansion changes discontinuously. Then the identification of vacuum changes subsequently. It is in this sense that particle production is associated with a transition between different vacua and is a result of non-uniqueness of vacuum.

We now comment on how the quantization prescription described above can resolve the first $5$ of the issues in the conventional prescription raised in Sec.~\ref{subsec:issues}. The numbering corresponds to that of the enumerated list there.

\begin{enumerate}
\item \textit{Breakdown of the WKB approximation:}

Even though constructed based on the WKB series, the exact WKB solutions are resummed, exact solutions to the original Schr\"{o}dinger-type differential equation, provided the resummation is well defined and convergent (the \textit{Borel summability}). 
Thus they do not encounter any issue around $V=0$.

\item \emph{Ambiguity of the vacuum with respect to higher orders of the WKB series:}

We employ the exact WKB solutions in defining the positive and negative frequency modes. Since these solutions are obtained by resummation of the full infinite WKB series, all the higher orders are taken into account. Hence there is no ambiguity in the decomposition of creation and annihilation operators associated with the mode functions.

\item \emph{Incompatibility with the exact solutions:}

The exact WKB solutions are exact solutions to the original Schr\"{o}dinger-type equation by construction, as long as the Borel summability is assured, while they may not necessarily reproduce the known special functions. In fact, the usefulness of using the exact WKB method resides in the fact that we do not have to rely on those special functions. Even in the cases where the explicit form of exact solutions is unknown, the exact WKB provides a tool to analyze the global behavior of exact solutions.

\item \emph{Divergent WKB series expansion:}

While the original WKB series is divergent in general, its resummed version, that is the exact WKB solution, is convergent and well defined, given the Borel summability. One may wonder the correspondence between a divergent series and the exact counterpart. The standard WKB series is a formal, asymptotic series solution to the original differential equation but is divergent. The corresponding exact WKB solution is a convergent completion of the divergent one, while being a legitimate solution to the equation and keeping the asymptotic behaviors coincide.

\item \emph{Lack of knowledge on the global behavior: connection formulae between two asymptotic solutions:}

We employ the exact WKB analysis in order precisely to overcome this issue. The standard WKB series is resummed to give exact solutions while their asymptotic behaviors are kept to coincide. The connection formulae between different asymptotic regions are obtained without approximations, as given in \eqref{eq:connection_formula_1} and \eqref{eq:connection_formula_2}. Hence the method resolves the difficulties raised in Sec.~\ref{subsec:issues}: (i) no ambiguity in connecting two different asymptotic regions arises since the analysis does not rely on special functions as solutions or connect an approximate solution around a saddle point to the asymptotic one, (ii) quantization, done in an asymptotic limit, can be formally conducted in accordance with the exact solutions as mode functions, and (iii) the Bogoliubov coefficients are computed as constant coefficients of the exact mode functions, and the interpretation of particle production as a change of vacua is transparent.

\item \emph{Diagonalization of the Hamiltonian is approximate:}

This issue still remains. As discussed around \eqref{eq:H_components}, the off-diagonal components of the Hamiltonian vanishes only at the leading order in the asymptotic expansion. Thus a vacuum and the corresponding Fock space are unambiguously defined only at (each) asymptotic limit.

\item \emph{Failure in non-adiabatic backgrounds (tachyonic region, quenches):}

In our formulation, we assume positive potential $V(x) > 0$ in the asymptotic regions. This is to ensure that the energy eigenvalues are positive definite, see \eqref{eq:eigenvalues_Hamiltonian}, and thus the energy spectrum is bounded from below, or in other words, the ground state exists. We do not attempt quantization in a tachyonic region $V(x) < 0$ in this paper.%
\footnote{Note that we require the positiveness of $V(x)$ only in the asymptotic limits. Taking negative values in the intermediate regions is perfectly fine. Such a tachyonic region can in fact contribute to particle production significantly, see e.g.~Sec.~\ref{subsec:example_x2}.}
An ``asymptotic limit'' is identified with a singular point $\sigma$ of the differential equation in the context of quantization, around which the potential can be 
locally expanded as $V(x) \sim (x-\sigma)^{-\nu}$ with $\nu>0$. Then we observe $\partial_x \omega / \omega^2 \sim \partial_x V / V^{3/2} \sim (x-\sigma)^{(\nu-2)/2}$.%
\footnote{In the case where $x \to \infty$ is a singular point, $V(x)$ reduces to this form after changing variables $x = 1/y$ and corresponding redefinition of $\Psi$.}
Therefore the adiabaticity condition $\vert \partial_x \omega / \omega^2 \vert \ll 1$ is automatic for $\nu > 2$. Indeed, in the exact WKB formulation, the pole $\nu = 1$ is considered to play a role similar to a turning point rather than a singular one [CITE]. The case with $\nu = 2$ is marginal, and the adiabaticity condition depends on the coefficients of $\omega$. In this paper, we assume the adiabaticity condition when performing quantization.

\item \emph{Ultraviolate divergence:}

Subtraction of divergent UV Fourier modes in a framework of quantum field theory in curved spacetime inherits multiple conceptual and technical issues, and we do not attempt to address them in this paper. At the very least, since the exact WKB method is a resummation of the original infinite WKB series with correct asymptotic behaviors, it is in principle a straightforward task to include up to an appropriate WKB order to subtract divergent contributions.%
\footnote{See \cite{Corba:2022ugu} for an attempt along this line.}
Yet, whether this approach gives a self-consistent, unambiguous result seems to require further considerations. We would like to come back to this issue in the future publication.

\end{enumerate}

We have thus far postulated how the difficulties in the approach based on the standard WKB series can be resolved by promoting the series to the exact WKB result. In the following sections, we show the technical side of the formulation and its concrete implementation in the context of particle production.

\section{Practical Review of Exact WKB Formulation}
\label{sec:review_exact_WKB}

In this section, we provide a concise and practical summary of the relevant notions and quantities needed to apply the exact WKB analysis to particle production. 
For more detailed and complementary discussions, see Appendix~\ref{app:basics_eWKB} or a textbook~\cite{Kawai:1998book}. 
We also include some topics that are often not covered in the physics literature on exact WKB analysis.
In Fig.~\ref{fig:summary_basics}, We collect some of the key definitions and properties of the exact WKB analysis that are found useful in this paper.

Our starting point is the one-dimensional Schr\"odinger(-like) equation,
\begin{align}
\label{eq:Sheq}
    \left[ -\frac{\dd^2}{\dd x^2}-\eta^2 V(x,\eta) \right] \psi(x,\eta)=0\ ,
\end{align}
where $\eta$ is the expansion parameter for the WKB analysis,%
\footnote{Note that $\eta$ is regarded as $1/\hbar$ in the WKB method when finding approximate solutions in quantum mechanics.
However, in broader contexts solving the Schrodinger-like equation, $\eta$ is not necessarily corresponding to $1/\hbar$. Rather, $\eta$ is a parameter giving a formal power series of a formal solution of the Schrodinger-like equation, i.e.~$\eta$ is an auxiliary parameter to utilize the exact WKB analysis. Notably, when working in natural units, the original equation may lack suitable large or small parameters. In such cases, $\eta$ must be introduced manually into the potential. We will later provide guidelines on how to incorporate $\eta$, which proves useful in studying particle production.}
and $V(x,\eta)$ denotes a potential expanded as
\begin{align}
    V(x,\eta)
    =\sum_{n=0}^N \eta^{-n} V_n(x)
    =V_0(x)+\eta^{-1} V_1(x)+\eta^{-2} V_2(x)+\cdots\ ,
\end{align}
with $N$ a non-negative integer.%
\footnote{Changing the variable $x$ to
$z(x)$ can be useful in certain cases, which also introduces a friction-like term.
In order to recast the equation into a form like \eqref{eq:Sheq}, one can also redefine the wave function.
The resulting Schr\"odinger-like equation then takes the form of \eqref{eq:Sheq}, with the potential proportional to the original one, plus higher-order terms of $\eta^{-2}$ that are proportional to the so-called Schwarzian derivative. A brief summary of this discussion can be found in~\cite{Iwaki2016}.}
The two independent formal WKB solutions take the form
\begin{align}
\label{eq:WKB_2}
    \psi_\pm(x,\eta)=\frac{1}{\sqrt{S_{\rm odd}(x,\eta)}}\,
    \exp\!\left[ \pm\int_{\tau_0}^x S_{\rm odd}(x',\eta)\, \dd x'\right]\ ,
\end{align}
where the lower integration bound is customarily taken at 
a turning point $\tau_0$ (see Fig.~\ref{fig:summary_basics} for definition) and
\begin{align}
 S_{\rm odd}(x,\eta)\equiv \sum_{j\geq 0}S_{2j-1}(x)\,\eta^{1-2j} \; ,
\end{align}
and $S_j(x)$ are determined recursively,
\begin{subequations}
\label{eq:Sj_recursion_example0}
\begin{align}
\label{eq:Sm1}
 &S_{-1}^2=-V_0 \; , \qquad
 S_0 = - \frac{1}{2 S_{-1}} \left( \frac{\dd S_{-1}}{\dd x} + V_1 \right) \; , \\
 \label{eq:Sj}
 &S_{j+1}=-\frac{1}{2S_{-1}}
 \left(
 \frac{\dd S_j}{\dd x}
 +
 \sum_{k=0}^j S_{j-k} S_k
 +V_{j+2}
 \right)\ ,
 \qquad j=0,\, 1,\, 2,\dots \; .
\end{align}
\end{subequations}
The $\pm$ sign in \eqref{eq:WKB_2} corresponds to the branch choice when solving $S_{-1}^2 = - V_0$.
In general, the WKB series obtained this way is divergent.  This motivates the use of Borel resummation, which can provide well-defined solutions from divergent asymptotic expansions.  
Formally, the WKB solutions \eqref{eq:WKB_2} can be reduced to
\begin{align}
\label{eq:asym_series}
    \psi(x, \eta)=\ee^{\zeta_0(x) \, \eta}\sum_{n=0}^\infty \eta^{-n-\alpha}f_n(x)\ ,
\end{align}
where $\zeta_0(x) \equiv \int^x_{\tau_0} \sqrt{- V_0(x')} \, \dd x'$ and $f_n(x)$ are complex 
functions, while $\alpha\in\mathbb{R}$ with $\alpha\notin \{0,-1,-2,\dots\}$.
Eq.~\eqref{eq:asym_series} serves as an asymptotic expansion of the true solution to \eqref{eq:Sheq}.
Its Borel transform is defined by
\begin{align}
\label{eq:Boreltransdef}
    \psi_B(x, \zeta)=\sum_{n=0}^\infty \frac{f_n(x)}{\Gamma(\alpha+n)}\left[ \zeta+\zeta_0(x) \right]^{\alpha+n-1}\ ,
\end{align}
where $\Gamma(x)$ is the gamma function, and the corresponding Borel sum is the Laplace transform of $\psi_B$,
\begin{align}
\label{eq:Borelsumdef}
    \Psi(\eta)=\int_{-\zeta_0}^\infty \ee^{-\eta \zeta}\,
    \psi_B(\zeta)\, \dd\zeta\ ,
\end{align}
with the integration contour taken parallel to the real axis for real and positive $\eta$ unless otherwise noted.
Some properties of the Borel-summed solutions are in order.

\begin{figure}[t]
\begin{screen}
\begin{itemize}[label=$\ast$, align=parleft, left=0pt]
    \item {\bf A turning point $\bm{\tau_0}$} $\Leftrightarrow$ a zero point of $V_0$ ($\tau_0\in \mathbb{C}$)
    \item {\bf A simple turning point}
    $\Leftrightarrow$ $\tau_0$ is a turning point with $\frac{\dd V_0(\tau_0)}{\dd x}\neq 0$
    \item {\bf Stokes curves} ({\bf Stokes lines})
    $\Leftrightarrow$
    Curves on the complex plane defined by ${\rm Im}\int_{\tau_0}^x\sqrt{-V_0(\chi)} \, \dd\chi=0$
    \item {\bf A Stokes region}
    $\Leftrightarrow$
    A region surrounded by Stokes curves
    \item {\bf Dominance relation} 
    $\Leftrightarrow$
    $\mathop{\rm sign} \left( {\rm Re}\int_{\tau_0}^x\sqrt{-V_0(\chi)} \, \dd\chi \right)$
    on the Stokes curve emanating from $\tau_i$
    \item {\bf Formal WKB solutions normalized at 
     $\bm{\tau_0}$}  $\Leftrightarrow$ $\psi_{\pm,\tau_0}=\frac{1}{\sqrt{S_{\rm odd}}}\exp\left(\pm\int_{\tau_0}^x S_{\rm odd}(\chi) \, \dd \chi \right)$
    \item {\bf Branch cuts:} cuts due to two-valueness of $S_{\rm odd} \sim \sqrt{-V_0}$, and due to four-valueness of $1/\sqrt{S_{\rm odd}} \sim (-V_0)^{-1/4}$
    \item {\bf Borel sum:} $\psi(\eta)
    \; \xrightarrow[\text{transform}]{\text{Borel}} \;
    \psi_B(\zeta) 
    \; \xrightarrow[\text{transform}]{\text{Laplace}} \;
    \Psi(\eta) 
    $
    \item {\bf Functional product:} $\psi_1(\eta)\,\psi_2(\eta) \xrightarrow[\text{sum}]{\text{Borel}} \Psi_1(\eta)\Psi_2(\eta)$
    \item {\bf Wronskian}:
    $\mathcal{W} \big\{ \psi_{+,\tau_0}(x),\psi_{-,\tau_0}(x) \big\} = -2\ , \mathcal{W} \big\{ \psi_{+,\tau_0}(x),\psi_{+,\tau_0}(x) \big\} = \mathcal{W} \big\{ \psi_{-,\tau_0}(x),\psi_{-,\tau_0}(x) \big\} =0$
    \item {\bf Voros coefficients:} Integration of regularized $S_{\rm odd}$ for a proper normalization of mode functions, see Sec.~\ref{sec:voros_coefficient}
\end{itemize}
\end{screen}
\caption{Summary of exact WKB basics.}
\label{fig:summary_basics}
\end{figure}

\begin{description}
    \item[\it Borel summation and products] ${}$\\
    The Borel sum and functional product commute, that is, the Borel sum of the product is obtained as the product of the Borel sum of individual asymptotic series, and vice versa. This is proven by using the property that the Borel transformation of the product of two formal asymptotic power series $\psi_1(\eta)$ and $\psi_2(\eta)$ is obtained by a convolution of the Borel transform of the original series, and the Laplace transformation of a convolution of two functions is equal to the product of the Laplace transforms of these individual functions. Explicitly, denoting the Borel transform of $\psi_i(\eta)=\ee^{\eta\phi_i}\sum_{n=0}^\infty f_{i,n} \eta^{-n+\alpha_i}$ by $\psi_{i , B}(\zeta)$, the Borel sum of the product is,%
    \footnote{To the integral form of the Borel transform, it may useful to note
    \begin{align*}
        \int_0^{\zeta} \eta^{n - 1} \left( \zeta - \eta \right)^{m - 1} \dd\eta 
        = \frac{\Gamma(n) \, \Gamma(m)}{\Gamma(n+m)} \, \zeta^{n+m-1} \; ,
    \end{align*}
    provided the integral converges.}
    \begin{align}
    \psi_1(\eta)\,\psi_2(\eta)\, 
    & \xrightarrow[\text{transform}]{\text{Borel}} \int_{-\phi_1}^{\zeta+\phi_2} \dd\tilde\zeta\, \psi_{1,B}(\tilde \zeta)\, \psi_{2,B}(\zeta-\tilde\zeta)\, \nonumber \\
    & \xrightarrow[\text{transform}]{\text{Laplace}}
    \int_{-\phi_1-\phi_2}^\infty \dd\zeta \, \ee^{-\zeta \eta}\int_{-\phi_1}^{\zeta+\phi_2} \dd\tilde\zeta\, \psi_{1,B}(\tilde \zeta)\, \psi_{2,B}(\zeta-\tilde\zeta)\ .
    \label{eq:product_series}
    \end{align}
    On the other hand, the product of the Borel summed series is also obtained as
    \begin{align}
    \Psi_1(\eta)\,\Psi_2(\eta)&=
    \int_{-\phi_1}^\infty \ee^{-\tilde\zeta \eta} \, \psi_{1,B}(\tilde\zeta ) \, \dd\tilde\zeta  
    \int_{-\phi_2}^\infty \ee^{-z \eta} \, \psi_{2,B}(z) \, \dd z \nonumber \\
    &=\int_{-\phi_1-\phi_2}^\infty \dd\zeta \, \ee^{-\zeta\eta} \int_{-\phi_1}^{\zeta+\phi_2} \dd\tilde\zeta  \, \psi_{1,B}(\tilde\zeta )\,\psi_{2,B}(\zeta-\tilde\zeta )\ ,
    \label{eq:product_Borelsummed}
    \end{align}
    where $\zeta=\tilde \zeta+z$.
    Comparing \eqref{eq:product_series} and \eqref{eq:product_Borelsummed}, we observe that the Borel sum of $\psi_1 \psi_2$ is equal to $\Psi_1 \Psi_2$, provided that the Borel summability is assured.
    Therefore, the Borel resummation commutes with multiplication.

    \item[\it Wronskian] ${}$\\ 
    The Wronskian of two WKB solutions is defined as
    \begin{align}
    \mathcal{W}\big\{\psi_k(x),\psi_m(x) \big\} \equiv  \psi_k\frac{\dd\psi_m}{\dd x}-\frac{\dd\psi_k}{\dd x}\psi_m\ ,
    \end{align}
    where $\psi_i(x)$ may denote either the formal or the Borel-summed WKB solutions. Even at the level of the formal WKB series, one easily finds
    \begin{subequations}
    \label{eq:wronskians}
    \begin{align}
    &\mathcal{W}\big\{\psi_{+,\tau_0}(x),\,\psi_{-,\tau_0}(x)\big\}=-2\ ,\\
    &\mathcal{W}\big\{\psi_{+,\tau_0}(x),\,\psi_{+,\tau_0}(x)\big\}
    =\mathcal{W}\big\{\psi_{-,\tau_0}(x),\,\psi_{-,\tau_0}(x)\big\}=0\ .
    \end{align}
    \end{subequations}
    Thanks to the property that the Borel sum commutes with multiplication, discussed above, the above Wronskian relations hold for the Borel-summed exact WKB solutions.
    These relations will be used later in computing the particle number density. 

    \item[\it Watson's lemma: Borel summability and asymptotic expansion] ${}$\\ 
    The original Watson's lemma~\cite{Hardy:1949book} gives sufficient conditions that a (analytic) function $f(z)$ is recovered by the Borel summation of its asymptotic expansion, i.e.~the lemma ensures existence and uniqueness of the asymptotic series of a class of the Borel summable functions.
    An improved Waston's lemma (Nevanlinna-Sokal theorem)~\cite{Sokal:1980ey} weakens the conditions: let $f(z)$ be an analytic function in ${\rm Re} \, (z^{-1})>R^{-1}$, and also assume $f(z)$ has an asymptotic expansion,
    \begin{align}
    f(z)=\sum_{k=0}^{N-1}a_k z^k+R_N(z)\ , \qquad |R_N(z)|\leq A B^N N! \, \vert z \vert^N\ ,
    \end{align}
    for some positive real constants $A$ and $B$, uniformly in $N$ and in ${\rm Re} \, (z^{-1})>R^{-1}$. 
    Then, the Borel sum of the asymptotic series is the same as $f(z)$ in ${\rm Re}\, (z^{-1})>R^{-1}$.
    The original papers can be referred to for more details.
    The authors in~\cite{Aoki:2019} also argue that the Borel summation of the formal exact WKB solution recovers the asymptotic expansion of the original formal solution when the path of the integration $S_{\rm odd}$ in~\eqref{eq:WKB_2} never cross any Stokes lines connecting turning points. 
    In later sections, we utilize Watson's lemma to connect the behavior of the exact WKB solutions in asymptotic regions with the WKB approximated solutions; in other words, the exact WKB solutions in asymptotic regions, which define the vacua for particle production, are usually approximated well by the WKB approximated solutions
owing to the existence of the asymptotic expansion ensured by this lemma.
\end{description}

%%%%
\begin{figure}[th]
\hspace{1cm}
\includegraphics[width=0.4\linewidth]{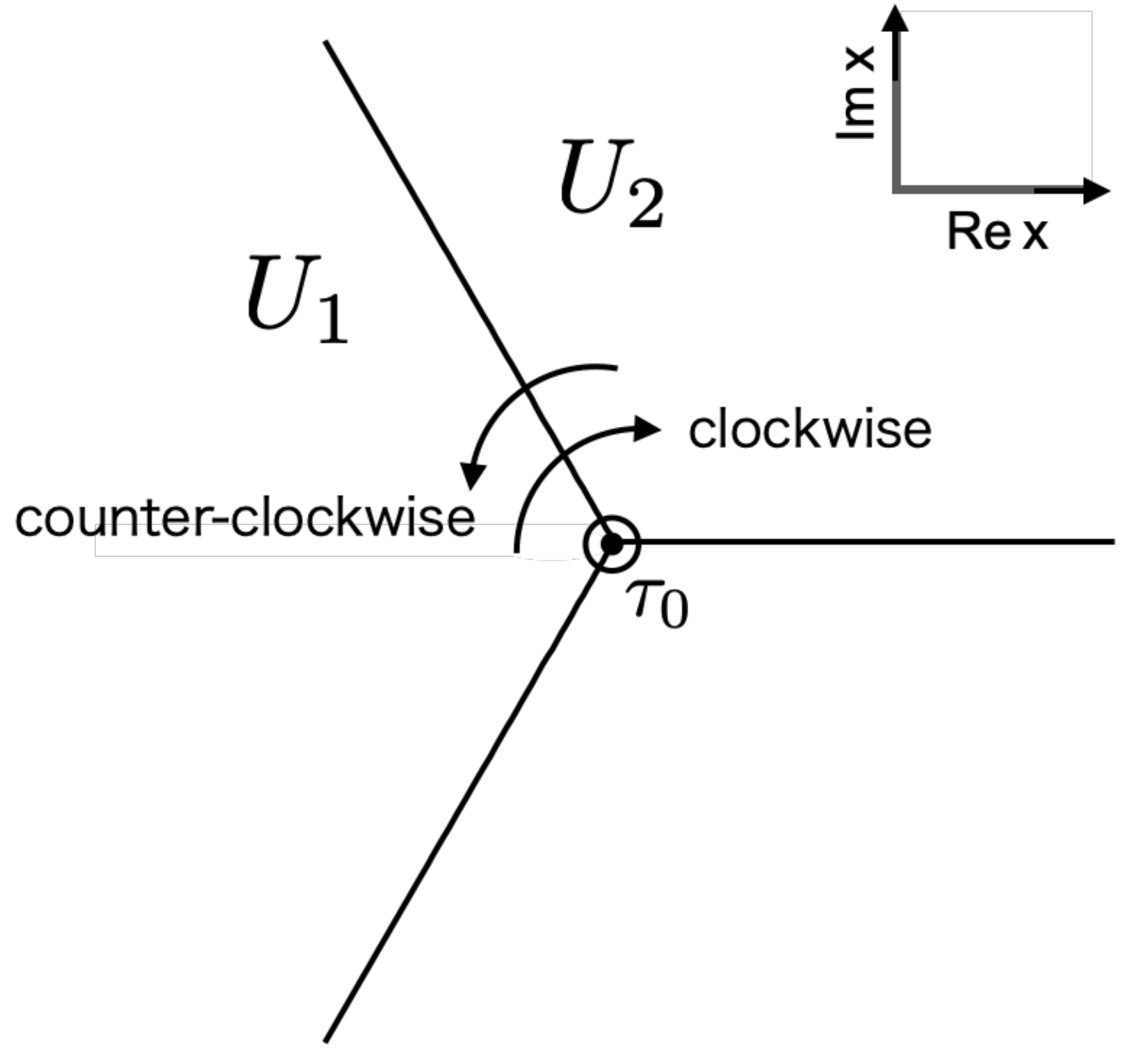}
   \vspace{0.2cm}
\caption{A schematic figure for Stokes curves, Stokes regions, a turning point, and clockwise or counter-clockwise paths about the turning point.
The turning point $\tau_0$ is denoted by $\odot$, the three lines emanating form the turning point are the Stokes curves, and $U_{1,2}$ label the two Stokes regions.
} 
\label{fig:schematic_stokes_regions}
\end{figure}
%%%% 

\begin{figure}[h!]
\begin{screen}
\justifying

\begin{center}
{\bf Connection formula}
\end{center}

\noindent
{\it Assumption}\\
All turning points are simple, and the turning points are never connected by Stokes curves.
\\

\noindent
{\it Statement}\\
Suppose two Stokes regions $U_1$ and $U_2$ that have a Stokes curve $\Gamma$ as a common boundary, and this curve is extended via a turning point $\tau_0$ (see Fig.~\ref{fig:schematic_stokes_regions} for a schematic picture of the setup).
Let $\Psi_{\pm,\tau_0}^i (i=1,2)$ be the Borel sums of the WKB solutions,
$\psi_{\pm,\tau_0}=\frac{1}{\sqrt{S_{\rm odd}}}\exp\left(\pm \int_{\tau_0}^x S_{\rm odd} \, \dd x \right)$, in the Stokes regions $U_{1,2}$, respectively.
Then, $\Psi^i_{\pm,\tau_0}$ is analytically continued to $U_2$ by one of the following equations,
\begin{align}
\label{eq:connection_formula_1}
\begin{cases}
    \displaystyle
    \; \Psi_{+,\tau_0}^1=\Psi_{+,\tau_0}^2 \vspace{2mm}\\
    \displaystyle
    \; \Psi_{-,\tau_0}^1=\Psi^2_{-,\tau_0}\pm i\Psi^2_{+,\tau_0}
\end{cases} \; ,
\end{align}
\begin{align}
\label{eq:connection_formula_2}
\begin{cases}
    \displaystyle
    \; \Psi_{+,\tau_0}^1=\Psi_{+,\tau_0}^2\pm i\Psi^2_{-,\tau_0} \vspace{2mm}\\
    \displaystyle
    \; \Psi_{-,\tau_0}^1=\Psi^2_{-,\tau_0}
\end{cases} \; .
\end{align}
Here, \eqref{eq:connection_formula_1} is for the case of ${\rm Re}\int_{\tau_0}^x\sqrt{-V_0(x')} \, \dd x'<0$ on $\Gamma$, and \eqref{eq:connection_formula_2} is for the case of ${\rm Re}\int_{\tau_0}^x\sqrt{-V_0(x')} \, \dd x'>0$ on $\Gamma$.
The signs $\pm$, respectively, correspond to the counter-clockwise and clockwise  crossing of $\Gamma$ of the path of analytic continuation from $U_1$ to $U_2$ about the turning point (see also Fig.~\ref{fig:schematic_stokes_regions}).
\end{screen}
\caption{Summary of the connection formulas for the Stokes phenomena between adjacent Stokes regions.}
\label{fig:connection_formula}
\end{figure}

The connection formula between exact WKB solutions in different Stokes regions plays a crucial role in analyzing particle production. 
A theorem~\cite{Voros:1983t} provides the connection relations (connection formulas) between Borel-summed WKB solutions in adjacent Stokes regions. See Fig.~\ref{fig:connection_formula} for the summary of the connection formulas.%
\footnote{For a simple pole, one also obtains a connection formula~\cite{Ko2} that differs from a connection formula shown in the main text.}
See also \cite{Kawai:1998book} for a proof of the theorem.

The connection formula is saying, e.g., the positive mode solution $\Psi_+$ in a Stokes region can be described by the mixing of the positive and negative mode solutions in a different Stokes region. We note that there is no jump of the value of the wave function $\Psi^1_{\pm,\tau_0}$ when $x$ changes from a point in $U_1$ to another point in $U_2$ of Fig.~\ref{fig:schematic_stokes_regions}, i.e., the value of the wave function is continuously changed because the connection formula is the relation obtained by the analytic continuation.
It is the form of an asymptotic expansion that changes discontinuously.
We also note that $\Psi^{1,2}_{\pm,\tau_0}$ are obtained from one formal WKB solution with the turning point $\tau_0$, i.e., the Borel sum and analytic continuation between Stokes regions make the difference of $\Psi^{1,2}$ while the original formal WKB solution is the same.

Regarding the last point of Fig.~\ref{fig:summary_basics} we find an extensive discussion useful. The Voros coefficients are one of the crucial ingredients for estimating the amount of particle production, as well as for ensuring the conservation of the Bogoliubov coefficients in \eqref{eq:commutation_condition_eWKB}, which allows an important consistency check. We thus devote the following section for this subject.

\section{Voros Coefficients}
\label{sec:voros_coefficient}

%%%%
\begin{figure}[h]
\centering
\hspace{1cm}
   \includegraphics[width=0.4\linewidth]{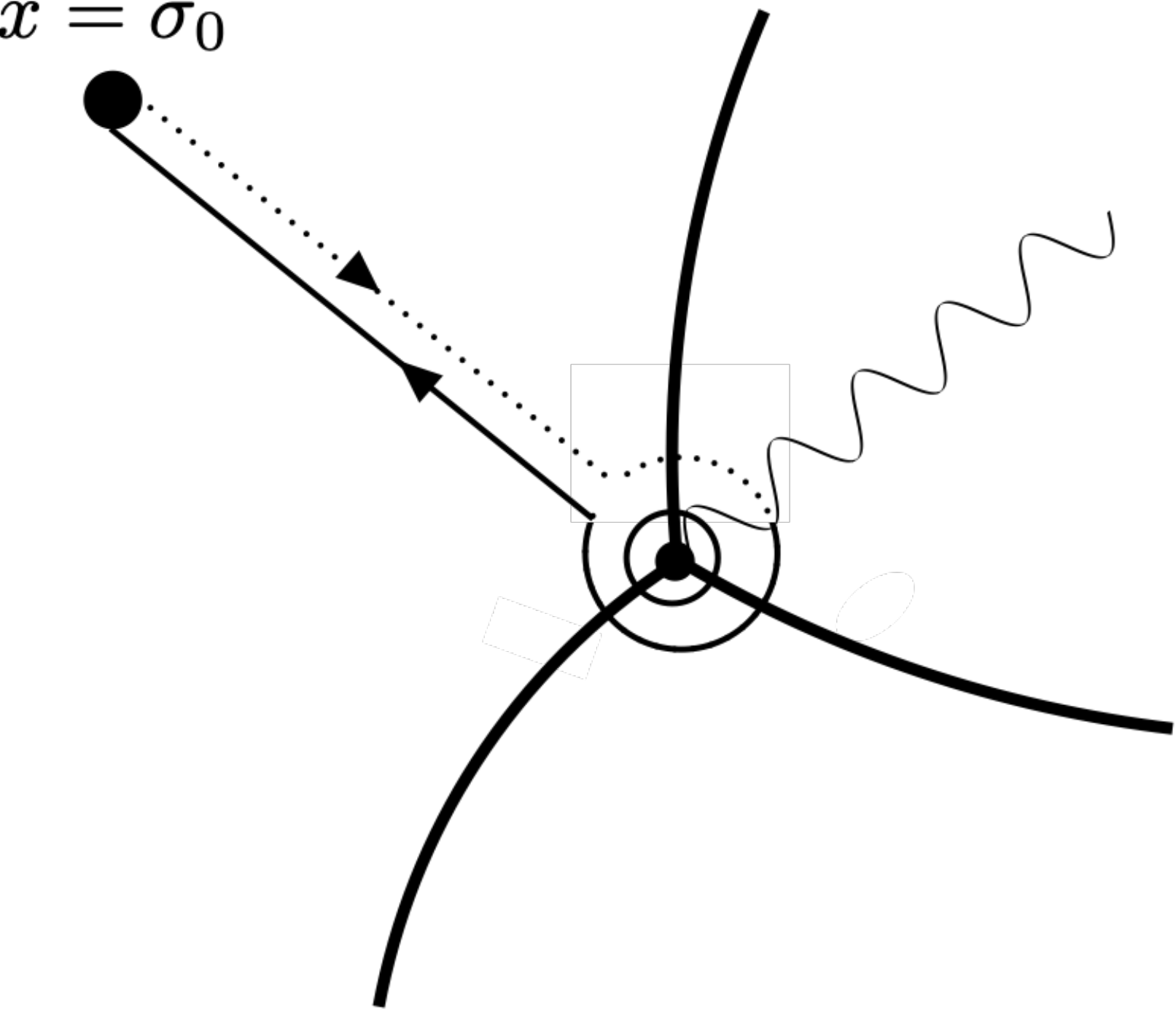}
   \caption{A schematic figure of the integration path of the Voros coefficient: The thick black solid curves represent the three Stokes curves emanating from the turning point denoted by a black dot surrounded by a circle. The waving line denotes the branch cut. The larger black dot corresponds to the singular point that is the starting point of the integration. The solid and dashed lines show the integration path in the first and second Riemann sheets, respectively.
   } 
\label{fig:voros_path}
\end{figure}
%%%% 

\subsection{Definition of Voros coefficients}

The Voros coefficient is defined as the integration of ``properly regularized $S_{\rm odd}$'' along the path connecting singular points and passing through a turning point by eliminating the terms $S_{j}$ having singularities. Notably, let us consider that the potential $V(x)$ has a regular or irregular singularity at $x=\sigma_0$,%
\footnote{
Considering a second-order differential equation,
$
    \left(\frac{\dd^2}{\dd x^2}+p_1(x) \frac{\dd}{\dd x}+p_2(x)\right)f(x)=0\ ,
$
we have a regular singular point $x=\sigma_0$ when either
$p_1(x)$ has a pole up to order $1$, like $p_1(x) \sim \frac{1}{x-x_0}$, or $p_2(x)$ has a pole up to order $2$, like $p_1(x)\sim \frac{1}{(x-\sigma_0)^2}$.
If either $p_{1,2}(x)$ has a higher pole at $x=\sigma_0$, it is a irregular singular point. 
We will see some concrete examples later in this section.
}
and $S_{2j-1}$ $(j\geq 1)$ are integrable around the singularities.
In that case, we define the Voros coefficient $\mathcal{V}_{\rm voros}$ as
\begin{align}
\label{eq:Voros_1}
    \mathcal{V}_{\rm voros}\equiv 
    \frac{1}{2}\int_{\gamma_{\sigma_0,\tau_0}} \left( S_{\rm odd}-\eta S_{-1} \right) \dd x \sim\int_{\tau_0}^{\sigma_0} \left( S_{\rm odd}-\eta S_{-1} \right) \dd x \ ,
\end{align}
where $\gamma_{\sigma_0,\tau_0}$ denotes the path of the contour integral whose contour is not closed in general. When the Stokes regions are described with two Riemann sheets, the path starts from $x=\sigma_0$ in the second Riemann sheet, passing around the turning point $\tau_0$, and ends at $x=\sigma_0$ in the first Riemann sheet (see Fig.~\ref{fig:voros_path}). 
We may use the abbreviation expressed in the second equation of \eqref{eq:Voros_1} to write the integration for simplicity of the notation. 
Note that the path from the turning point to $x=\sigma_0$ is assumed to never cross the Stokes curves connecting turning points.
In the examples we consider in this paper with $V(x)$ a rational function, $S_{2j}~(j\in \mathbb{Z})$ (and so $S_{\rm even} \equiv \sum_{j \ge 0} S_{2j} \eta^{-2j}$) are single-valued functions, and hence the integrals $\int_{\gamma_{\sigma_0,\tau_0}} S_{2j} \, \dd x$ are given as a contour integral for which the path $\gamma_{\sigma_0,\tau_0}$ is closed.  
In some cases, $S_{\rm even}$ contains poles, not necessarily only the singular points $\sigma_0$ of $V(x)$ but also the turning points $\tau_0$. In the cases of our interest in this paper, $S_0$ is the only term in $S_{\rm even}$ that contains poles.%
\footnote{If no terms in $S_{\rm even}$ contain poles, the following discussion still holds, as $\oint_{\gamma_{\sigma_0,\tau_0}} S_{\rm even} \, \dd x = \oint_{\gamma_{\sigma_0,\tau_0}} S_0 \, \dd x = 0$.}
Then the Voros coefficient \eqref{eq:Voros_1} can be rewritten as
\begin{align}
    \mathcal{V}_{\rm voros}
    = \frac{1}{2} \int_{\gamma_{\sigma_0,\tau_0}} \left( S^{(+)}-\eta S_{-1}-S_0 \right) \dd x
    \sim
    \int_{\tau_0}^{\sigma_0} \left( S^{(+)}-\eta S_{-1}-S_0 \right) \dd x \ .
\end{align}
This procedure is straightforwardly generalized to more complex cases by removing poles from $S^{(+)}$ through subtraction of the corresponding $S_j$ components.
Here, we define $S^{(\pm)}$ for later convenience,
\begin{align}
     S^{(\pm)}\equiv \pm S_{\rm odd} + S_{\rm even}\ .
\end{align}
This form of the Voros coefficient with $S^{(+)}$ (or $S^{(-)}$ in some cases) is utilized for the explicit calculation of the Voros coefficient as explained later.

Using the Voros coefficients, the WKB solutions normalized at the turning point in \eqref{psi_turning} are (formally) decomposed to
\begin{align}
    \psi_{\pm,\tau_0}=\exp \left( \pm \mathcal{V}_{\rm voros} \right) \psi^{(\sigma_0)}_{\pm}\ .
\end{align}
On the right-hand side of the equation, the explicit form of the latter function $\psi^{(\sigma_0)}_{\pm}$ is 
\begin{align}
\label{eq:WKB_asymptotic}
    \psi^{(\sigma_0)}_{\pm}=\frac{1}{\sqrt{S_{\rm odd}}} 
    \exp \left( \pm\eta \int_{\tau_0}^x S_{-1} \, \dd x \right)\,
    \exp \left[ \pm \int_{\sigma_0}^x \left( S_{\rm odd}-\eta S_{-1} \right) \dd x \right] \ ,
\end{align}
where the Voros coefficient is given by~\eqref{eq:Voros_1}.
The integration $\int_{\sigma_0}^x \left( S_{\rm odd}-\eta\,S_{-1} \right) \dd x$ on the right-hand side of the equation is integrable and approaches zero in the limit of $x\to \sigma_0$. Consequently, $\psi^{(\sigma_0)}_{\pm} \sim \frac{1}{\sqrt{S_{\rm odd}}} \exp \left( \pm \eta \int_{\tau_0}^x S_{-1} \dd x \right)$ give the properly normalized asymptotic solutions around $x=\sigma_0$. In other words, the Voros coefficients relate the WKB solutions normalized at $x=\sigma_0$ to those normalized at the turning point $x=\tau_0$.
As we show in the later sections, the Voros coefficients are relevant to describe the particle production properly, i.e., without the coefficients, we may wrongly obtain the connection matrix that does not belong to the elements of ${\rm SU}(1,1)$ discussed in Sec.~\ref{subsec:part_prod}.%
\footnote{In some limited cases, the Voros coefficients become trivial, i.e.~zero.}

\subsection{A generic flow of Voros coefficient calculation}

An overview of the calculation of the Voros coefficient is given as follows. See later sections and appendices for more detailed calculations with concrete examples.
Consider a potential $V(\alpha,\eta)$ where $\alpha$ denotes some constant parameter in the potential. For instance, we explicitly study the potential with the form of $V=-(\alpha-x^2)$ in Sec.~\ref{subsec:Voros_ex1}.
To perform a Borel resummation of a Voros coefficient,
it is critical to find the following difference relation of the Voros coefficient in terms of the WKB parameter $\eta$,
\begin{align}
\label{eq:del_eta,voros}
    \mit{\Delta}_\eta \mathcal{V}_{\rm voros}(\alpha,\eta)\equiv 
    \mathcal{V}_{\rm voros}(\alpha,\eta+\tilde c)-\mathcal{V}_{\rm voros}(\alpha,\eta)=f(\alpha,\eta)\ ,
\end{align}
where $\tilde c$ is a constant and $f(\alpha,\eta)$ is some concrete function. The Borel transformation of both sides of the equation is computed as
\begin{align}
\label{eq:voros_B}
    \left( \ee^{-\tilde c\,\zeta}-1 \right) \mathcal{V}_{{\rm voros},B}(\alpha,\zeta)=f_B(\alpha,\zeta)\ ,
\end{align}
where the subscript $B$ denotes the Borel transform of the function. 
This equation can be solved for $\mathcal{V}_{{\rm voros},B}$ algebraically, and one obtains the Borel summed Voros coefficient when the Laplace integration of $\mathcal{V}_{{\rm voros},B}$ from this relation can be evaluated.%
\footnote{
The formal series of the Voros coefficients for the the confluent family of the Gauss hypergeometric equations are described with Bernoulli numbers~\cite{Iwaki:2018I,Iwaki:2018II}.}

In order to obtain the relation \eqref{eq:del_eta,voros}, 
our first crucial step is to find the ladder operators $\mathcal{L}_{\pm}$ that relate the formal solution $\psi(x,\alpha,\eta)$ for the (original) potential $V(x,\alpha,\eta)$ to the formal solution for the $\alpha$-shifted potential $V(x,\alpha\pm\tilde c\,\eta^{-1},\eta)$,
\begin{align}
    \mathcal{L}_\pm \psi(x,\alpha,\eta)= 
    C(\alpha,\eta)\,\psi(x,\alpha\pm \tilde c\,\eta^{-1},\eta)\ ,
\end{align}
where $C(\alpha,\eta)$ is a factor independent of $x$.
Note that $\psi(x,\alpha,\eta)$ and $\psi(x,\alpha\pm\tilde c\,\eta^{-1},\eta)$ have the same form and only their arguments are different, i.e.~$\alpha$ or $\alpha\pm\tilde c\,\eta^{-1}$.
The ladder operator may contain the derivative operator $\partial / \partial x$ and is explicitly obtained depending on the potential $V(x,\alpha,\eta)$, see Sec.~\ref{subsec:Voros_ex1} and Appendices \ref{app:caseII} and \ref{app:caseIII} for concrete examples.
By taking the logarithmic derivative $\partial_x \ln(\mathcal{L}_\pm \psi)$, one can obtain the difference relation for $S^{(+)}$,%
\footnote{
$\psi$ is a solution given as
$    \psi=
    \exp\bigg[
    \int^x S^{(+)}(x') \, \dd x'
    \bigg]
$, formally equivalent to the form in \eqref{eq:WKB_2}.}
\begin{align}
    \mit{\Delta}_\alpha S^{(+)}\equiv S^{(+)}(\alpha+\tilde c \eta^{-1},\eta)-S^{(+)}(\alpha,\eta)=g(\alpha,\eta)\ ,
\end{align}
where $g(\alpha,\eta)$ is given as a concrete function.
This difference relation of $S^{(+)}$ enables us to compute the difference relation of the Voros coefficient,%
\footnote{Remember that the Voros coefficient was defined with the integration of $S^{(+)}$ (and some $S_j$).}
\begin{align}
    \mit{\Delta}_\alpha \mathcal{V}_{\rm voros}(\alpha,\eta)\equiv \mathcal{V}_{\rm voros}(\alpha+\tilde c \eta^{-1},\eta)-\mathcal{V}_{\rm voros}(\alpha,\eta)\ .
\end{align} 
The trick to relate this difference equation to that in \eqref{eq:del_eta,voros} is to use a scaling property of $S_j(x,\alpha)$.%
\footnote{As seen in concrete examples, this property takes the form,
\begin{align}
    S_j(\alpha^p x,\alpha)=\alpha^q S_j(x,\alpha)\ , 
\end{align}
where $p$ and $q$ are rational numbers.}
In all examples we study in the following sections, the Voros coefficient is given as a series in terms of $\alpha\eta$,%
\begin{align}
    \mathcal{V}_{\rm voros}(\alpha,\eta)= \mathcal{V}_{\rm voros}(\alpha\eta)\ .
\end{align}
This property allows to exchange the part of the argument including $\tilde c$ for the Voros coefficient,  
\begin{align}
\label{eq:voros_arg_property}
    \mathcal{V}_{\rm voros}(\alpha+\tilde c\eta^{-1},\eta) =\mathcal{V}_{\rm voros}(\alpha\eta +\tilde c)=\mathcal{V}_{\rm voros}(\alpha,\eta+\tilde c\alpha^{-1})\ .
\end{align}
Consequently, we obtain
\begin{align}
    \mit{\Delta}_\alpha \mathcal{V}_{\rm voros}(\alpha,\eta)= \mit{\Delta}_\eta \mathcal{V}_{\rm voros}(\alpha,\eta)=f(\alpha,\eta)\ ,
\end{align}
which is the desired relation \eqref{eq:del_eta,voros}.
A non-zero Voros coefficient is related to a peculiar type of singularities.
Concretely, \eqref{eq:voros_B} implies 
that there exist infinite number of singularities on the complex $\zeta$-plane,
\begin{align}
    \zeta=2\pi i \, \frac{m}{\tilde c}\ , \quad m\in \mathbb{Z}\ .
\end{align}
This type of singularities are known as the fixed singularities,%
\footnote{A singularity is called as the movable singularity when it depends on the coordinate $x$.}
which do not depend on the coordinate $x$ in contrast to the singularity on the Stokes curves.

\subsection{Degenerate Stokes curves and Voros coefficients}

%%%%
\begin{figure}
\centering
\hspace{1cm}
   \includegraphics[width=0.6\linewidth]{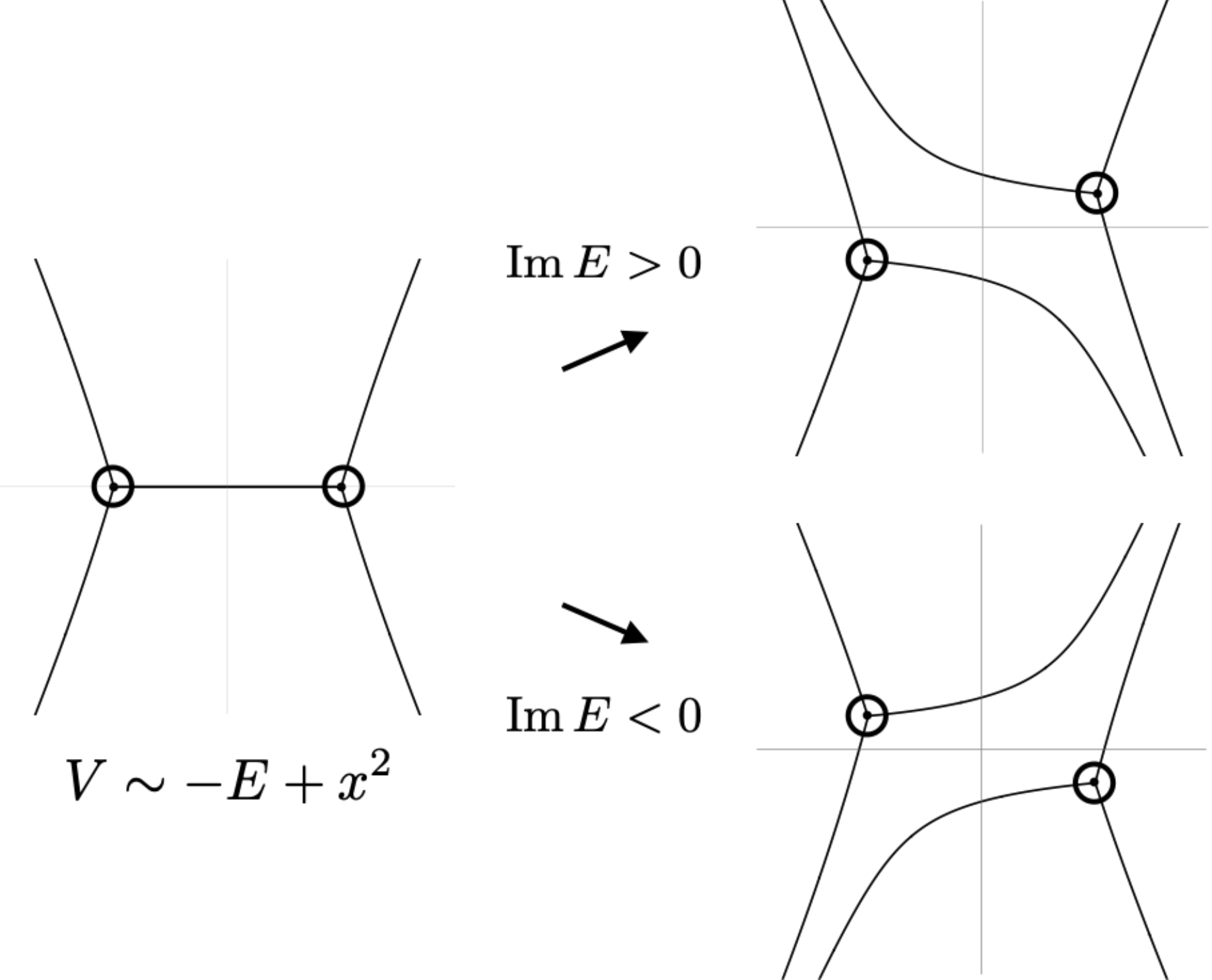}
   \caption{A schematic figure of resolving the degeneracy. The picture uses the potential of the form of $V\sim-E+x^2$ $(E\in \mathbb{R})$, which has the degeneracy as in the left Stokes curves. Introducing a small imaginary number into the potential solves the degeneracy. Depending on the sign of the imaginary parts, we obtain two types of Stokes curves. 
   } 
\label{fig:degenerate_stokes_lines_two_cases}
\end{figure}
%%%% 

A Stokes curve typically emanates out of a turning point and flows into a singular point including infinity. On the other hand, one may occasionally encounter a Stokes curve that terminates at turning points on both ends as in the left panel of Fig.~\ref{fig:degenerate_stokes_lines_two_cases}, forming what is known as a degenerate Stokes graph. In such cases, the series becomes Borel non-summable, and the standard connection formula cannot be applied along the Stokes curve \cite{Aoki:2019}. 

A common approach to resolve this issue is to introduce a small regularization parameter into the potential or the variable, thereby breaking the degeneracy, and taking the limit of the vanishing parameter at the end of the calculation. By slightly perturbing the parameter space, the two turning points are no longer directly connected by a Stokes curve as seen in the right panels of Fig.~\ref{fig:degenerate_stokes_lines_two_cases}. Consequently, one can connect the exact WKB solution in one Stokes region to that in another, within a regularized framework, in a manner similar to the case without degeneracy.  
We eventually identify the obtained results with those of the original model setup by setting this small parameter to zero in the end.
This regularization reminds us the $\epsilon$-prescription in quantum field theory.

There may be multiple choices for the regularization parameter, leading to an apparent ambiguity in the Stokes graph. That is, depending on the value of the small parameter, the resulting Stokes curves may exhibit different structures (see Fig.~\ref{fig:degenerate_stokes_lines_two_cases}). These ambiguities should not affect the final physical results, such as the number density. The physics should remain unchanged, independent of the specific value of the regularization parameter.

The degenerate Stokes curve has a strong connection to the so-called fixed singularities of the Borel transformed Voros coefficient.
These singularities lie on the path of the Laplace integral on the Borel plane when the Stokes curves are degenerate, so that the WKB solution becomes Borel non-summable. Introducing the regularization parameter corresponds to avoiding the integral path hitting the fixed singularities, while the ambiguity of the regularization is the arbitrary choice of how to avoid the singularities.

Put differently, we can understand the relation between Voros coefficients and degenerate Stokes graphs as follows.
Consider a situation where fixed singularities exist on the complex Borel plane. By introducing a regularization parameter, the Stokes graph can be made non-degenerate, and the singularities are no longer on the path of the Laplace integral, making the WKB solution and the Voros coefficient Borel summable. 
Now, changing the regularization parameter continuously, we encounter the degenerate Stokes curve at the point where the parameter crosses zero, which leads to Borel non-summability of the WKB solution and the Voros coefficient. 
Empirically, we hence expect the existence of non-trivial Voros coefficients 
when some of the Stokes curves are degenerate, i.e.~an infinite number of fixed singularities are along the path of Laplace integration,
\begin{align}
\label{eq:voros_c_condition}
    {}^\exists \, \mathcal{V}_{\rm voros}\neq 0 \; \sim \;
    {}^\exists \, \text{degenerate Stokes curves} \; \sim \;
    {}^\exists \, \text{fixed singularties along the integration path} \; .
\end{align}
This offers a possibility to diagnose trivial or non-trivial (i.e.~zero or non-zero) Voros coefficients in a diagramatic way before starting concrete computations of the coefficients. Draw some Stokes curves while changing the parameters of the potential. If you find any degenerate Stokes curves, that signals the existence of non-trivial Voros coefficient. 
If no degenerate Stokes curves exist for any parameter choice, the Voros coefficients may be expected to be zero.

Moreover, to the best of our knowledge, the Voros coefficient is non-trivial when some Stokes curves are degenerate and the singular point defining the Voros coefficient sits in the Borel non-summable region.
For instance with anti-harmomic oscillator potential of $V=-E+x^2/4$, we encounter the degenerate Stokes curves when ${\rm Im} \, (E)=0$ as seen in Sec.~\ref{subsec:Voros_ex1}. The formal exact WKB solutions are Borel non-summable in the entire space. Introducing a small imaginary part to $E$ resolves the degeneracy, which leads to the non-trivial Voros coefficients for the singular points $x\to \infty$ as we show soon below. Another interesting example is a Bessel-type equation, having a loop-type Stokes curve, which emanates from a turning point ends on the same turning point~\cite{Aoki:2019}. Inside the loop including the singular point at $x=0$, the exact WKB solutions are not Borel summable. On the other hand, outside the loop, the solutions are Borel summable.
The Voros coefficient defined with $x=0$ becomes non-trivial.

There are several empirical (non-rigorous) methods to assess whether the Voros coefficients are non-trivial before proceeding with detailed calculation. These methods can be practically useful.
One approach is the diagrammatic method discussed earlier. If multiple Stokes curves descend from a degenerate one, this implies a non-zero Voros coefficient. The clearest scenario arises when a chosen setup (potential with a certain parameter choice) already includes a degenerate Stokes curve.
Generally, however, it may be necessary to identify a degenerate Stokes curve by adjusting potential parameters. 
Another method involves computing the connection matrix without including the Voros coefficient (i.e.~taking a vanishing Voros coefficient) and then verify whether the resulting connection matrix indeed belongs to SU(1,1).%
\footnote{Refer to Sec.~\ref{subsec:part_prod} for the relationship between SU(1,1) and particle production. The connection matrix must be as an element of SU(1,1).}
If not, it implies that the Voros coefficients must be non-trivial.

\subsection{Example: $V=-x$}
\label{subsec:Voros_ex0}

As the first simple example, let us compute the Voros coefficient in the system:
\begin{align}
\left[
    - \frac{\dd^2}{\dd x^2} - \eta^2 V(x)
\right]
\psi(x) = 0, \quad V(x) = -x\ . 
\end{align}
This equation has analytic solutions, Airy functions.
We show that the Voros coefficient is trivial, i.e. $\mathcal{V}_{\rm voros}=0$, in this case.

%%%%%%%%%%%%%%%%%%%%%%
%%%%%%%%%%%%%%%%%%%%%%
\begin{figure}[t]
    \centering
    \includegraphics[width=0.5\linewidth]{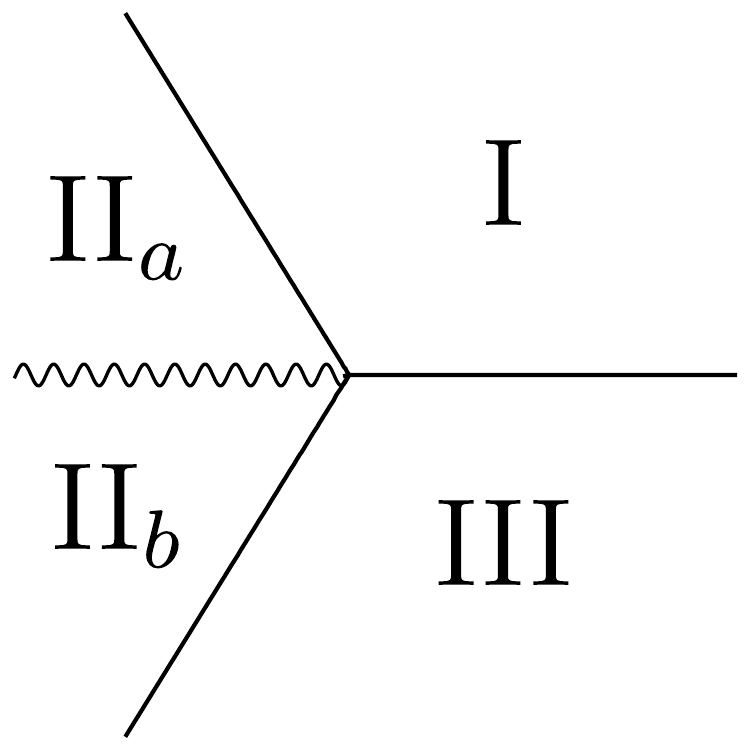}
    \caption{The Stokes graph for $\sqrt{-V}=x$. The solid lines show the Stokes curves (in this case, lines) emanating from the origin $x=0$, and the Roman numbers denote the Stokes regions. Region II is divided into two subregions by the branch cut running between $x=0$ and $x=\infty$, represented by the wavy line. }
    \label{fig:stokesgraph_x}
\end{figure}
%%%%%%%%%%%%%%%%%%%%%%
%%%%%%%%%%%%%%%%%%%%%%

To investigate the analytic structure of this sysmte, we have 
\begin{itemize}
    \item Simple turning point: \(x = 0, \; \dd V / \dd x \, \vert_{x=0} \ne 0 \) 
    \item Irregular singular point: \(x = \infty.\) 
\end{itemize}
We promote the originally real-valued $x$ to a complex variable; we further attempt to compactify the Riemann surface by including the infinity as a point. We assume this procedure throughout this paper, unless otherwise noted. 
The branch cut runs from \(x = 0\) to \(x = \infty\), and the Stokes graph is captured in Fig.~\ref{fig:stokesgraph_x}.
With the potential under consideration, $\sqrt{-V} = \sqrt{x}$ has a structure in which two Riemann spheres, each including the infinity, are glued together by connecting the branch cut on one sphere with that on the other, forming an $S^2$ topology.

Notice from Fig.~\ref{fig:stokesgraph_x} that there is only one set of Stokes curves, associated with one turning point, and therefore there is no room for degenerate Stokes curves in this model. The heuristic argument around \eqref{eq:voros_c_condition} then suggests that there are no fixed singularities on the Borel plane and thus the Voros coefficient is trivial. Let us show this explicitly below.

The formal WKB solution normalized at the turning point is
\begin{align}
\psi_{\pm}(x) = \frac{1}{\sqrt{S_\text{odd}(x)}} \exp \left[ \pm \int_0^x S_\text{odd}(x') \, \dd x' \right],
\end{align}
and 
\begin{align}
    S_\text{odd}(x) \, \dd x = \left( \eta \sqrt{x} + \eta^{-1} \frac{-5}{32 x^{5/2}} 
    + \dots \right) \dd x \; , \quad
    S_\text{even}(x) \, \dd x = \left( \frac{-1}{4x} + \eta^{-2} \frac{-15}{64 x^4} 
    + \dots \right) \dd x \; . 
\end{align}
By changing the variable $y=1/x$, we also obtain
\begin{align}
\label{eq:Sodd_exapmle0}
    S_\text{odd}(x) \, \dd x = \left(-\eta \frac{1}{y^{5/2}} + \eta^{-1} \frac{5y^{1/2}}{32} 
    + \dots \right) \dd y \; ,  \quad
    S_\text{even}(x) \, \dd x = \left(\frac{1}{4y} + \eta^{-2} \frac{15y^2}{64} 
    + \dots \right) \dd y \ .
\end{align}
Thus, the only diverging terms at \(x = \infty\) (or \(y = 0\)) are due to \(S_{-1} = \sqrt{x}\) and \(S_0 = -1/(4x)\). Hence, we define the WKB solution normalized at \(x = \infty\) as:
\begin{align}
\psi^{(\infty)}_{\pm}(x) = \exp\left[ \pm \int_0^x \eta S_{-1}(x') \, \dd x' \right] \frac{1}{\sqrt{S_\text{odd}(x)}} \exp \left[ \pm \int_{\infty}^x \left(S_\text{odd}(x') - \eta S_{-1}(x')\right) \dd x' \right]\ ,
\end{align}
where the first factor can be explicitly computed, $\exp\left[ \pm \int_0^x \eta S_{-1}(x') \, \dd x' \right] = \exp \left(\pm \frac{2}{3} x^{3/2} \eta \right)$.

The relation between the two WKB solutions normalized at different points is
\begin{align}
\psi_{\pm}(x) &= \ee^{\pm \mathcal{V}_{\text{voros}}} \, \psi^{(\infty)}_{\pm}(x)\ , 
\end{align}
where \(\mathcal{V}_{\text{voros}}\) is the Voros coefficient, given by
\begin{align}
\mathcal{V}_{\text{voros}} = \int_0^{\infty} \left[S_\text{odd}(x) - \eta S_{-1}(x)\right] \dd x
= \frac{1}{2} \oint_{\gamma_{0,\infty}} \left[S_\text{odd}(x) - \eta S_{-1}(x)\right] \dd x\ .  
\end{align}
Since the branch cut runs from \(x = 0\) to \(x = \infty\), the integral actually becomes an integral of a closed contour $\gamma_{0,\infty}$ encircling the cut.
This contour integral can be equally regarded as encircling the \textit{outside} region, which contains no singular points or poles.
Therefore, by the residue theorem, this integral evaluates to 0:
\begin{align}
\mathcal{V}_{\pm} = 0\ . 
\end{align}
The Voros coefficient in the case of $V(x)=-x$ thus vanishes, as expected from the heuristic consideration.

\subsection{Example: $V=-E+x^2/4$}
\label{subsec:Voros_ex1}

Next example gives a non-trivial Voros coefficient, which contains a plenty of important basics for the calculation of the Voros coefficient. Consider the Schr\"{o}dinger-type equation with a quadratic potential,
\begin{align}
\label{eq:Sch_eq_example1}
\left[
    - \frac{\dd^2}{\dd x^2} - \eta^2 V(x)
\right]
\psi(x) = 0, \quad V(x) = -E + \frac{x^2}{4}. 
\end{align}
Same as before, we first investigate the analytic structure of this system:
\begin{itemize}
    \item Simple turning points: \(x = \pm 2\sqrt{E} \equiv \tau_{\pm}, \; \dd V / \dd x \, \vert_{x = \tau_\pm} \ne 0 \) 
    \item Irregular singular point: \(x = \infty.\) 
\end{itemize}
%
%%%%
\begin{figure}
\centering
\hspace{1cm}
   \includegraphics[width=0.6\linewidth]{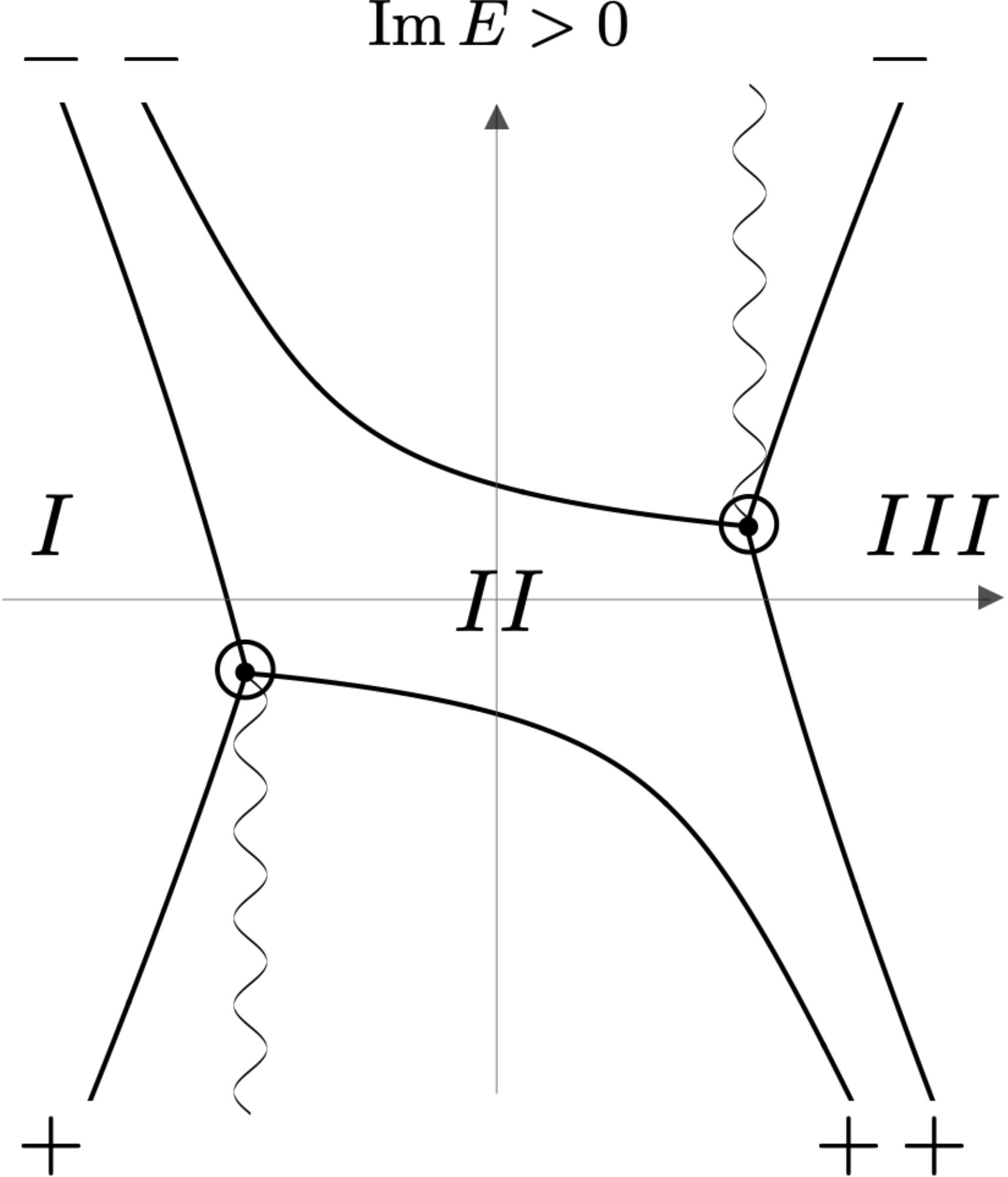}
   \caption{The Stokes curves and regions for $V=-E+x^2/4$ with ${\rm Im} \, (E)>0$. Three (relevant) Stokes regions are labeled by I, II, III. The branch cuts are emanating from the turning points (dot surrounded by a circle). The plus and minus sings denote the sign for the dominance relation.
   } 
\label{fig:stokes_curves_positive_imaginary}
\end{figure}
%%%% 
%
We take the branch cuts running between $\tau_\pm$, going through infinity as in Fig.~\ref{fig:stokes_curves_positive_imaginary}. We again include the infinity as a point on the complex plane. The branch of \(S_{-1}(x)\), corresponding to the value on the first Riemann sheet, is chosen such that
\begin{align}
   \ee^{-i\pi/2} \sqrt{E - \frac{x^2}{4}} > 0, \quad \text{for } E > 0, \; x > 2\sqrt{E} \ ,
\end{align}
when $E$ and $x$ take real values. 
The topology of $\sqrt{-V} = \sqrt{E- x^2/4}$ is the same as the case of $V=x$ in Sec.~\ref{subsec:Voros_ex0}: two Riemann spheres, compactified with the infinity included, are glued together, forming $S^2$.
 
The WKB solution normalized at the turning point is
\begin{align}
\psi_{\pm, \tau_\pm}(x) = \frac{1}{\sqrt{S_\text{odd}(x)}} \exp \left[ \pm \int_{\tau_\pm}^x S_\text{odd}(x') \, \dd x' \right] \ . 
\end{align}
Since the Sch{\"o}dinger-type equation has a reversal symmetry under \(x \to -x\), we impose
\begin{align}
\psi_{+, \tau_+}(-x) = \psi_{-, \tau_-}(x), \quad \psi_{+, \tau_-}(-x) = \psi_{-, \tau_+}(x) \ .
\end{align}
Expanding \(S_\text{odd}\) and \(S_\text{even}\), we obtain
\begin{subequations}
\begin{align}
S_\text{odd}(x) \, \dd x &= \left[ \eta \sqrt{E - \frac{x^2}{4}} + \eta^{-1} \frac{-8E - 3x^2}{4 \left( 4E - x^2 \right)^{5/2}} + \dots \right] \dd x \; ,  \\
S_\text{even}(x) \, \dd x &= \left[ \frac{x}{8E - 2x^2} + \eta^{-2} \frac{3x \left( 6E + x^2 \right)}{\left( 4E - x^2 \right)^4} + \dots \right] \dd x \ ,
\end{align}    
\end{subequations}
and
\begin{subequations}
\begin{align}
S_\text{odd}(x) \, \dd x &= \left[-\frac{\eta}{y^2} \sqrt{E - \frac{1}{4y^2}} + \eta^{-1} \frac{\left( 8Ey^2 + 3 \right) y}{4 \left( 4Ey^2 - 1 \right)^{5/2}} + \dots \right] \dd y,  \\
S_\text{even}(x) \, \dd x &= \left[-\frac{1}{2y \left( 4Ey^2 - 1 \right)} - \eta^{-2} \frac{3y^3 \left( 6Ey^2 + 1 \right)}{\left( 4Ey^2 - 1 \right)^4} + \dots \right] \dd y. 
\end{align}
\end{subequations}
where \(y = 1/x\). We find the only diverging terms at \(x \to \infty\), or \(y \to 0\), are
\[S_{-1} = \sqrt{E - \frac{x^2}{4}}, \quad S_0 = \frac{x}{8E - 2x^2} \ .
\]
Hence, we define the WKB solution normalized at \(x \to \infty\) as
\begin{align}
\psi^{(\pm \infty)}_{\pm}(x) = \exp \left[ \pm \int_{\tau_{\pm}}^x \eta S_{-1}(x') \, \dd x' \right] \frac{1}{\sqrt{S_\text{odd}(x)}} \exp \left[ \pm \int_{\pm \infty}^{x} \left(S_\text{odd}(x') - \eta S_{-1}(x')\right) \dd x' \right] \ . 
\end{align}
The integration paths are assumed not to cross any Stokes curves, and the sign of $-\infty$ (or $+\infty$) in the superscript and the integration bound denotes the path to the infinity that runs in region I (or region III) in Fig.~\ref{fig:stokes_curves_positive_imaginary}.
Owing to the reversal symmetry under $x\to -x$ in $S_{\rm odd}$, we have
\begin{align}
    \psi^{(+\infty)}_+(-x)=\psi_-^{(-\infty)}(x)\ , \quad
    \psi^{(-\infty)}_+(-x)=\psi^{(+\infty)}_-(x)\ .
\end{align}
The relation between the WKB solutions normalized at different points is
\begin{align}
    \psi_{\pm,\tau_+}(x)
    =\ee^{\pm \mathcal{V^{(+\infty)}_{\rm voros}}}\psi_\pm^{(+\infty)}(x)\ , \quad
    \psi_{\pm,\tau_-}(x)
    =\ee^{\pm \mathcal{V^{(-\infty)}_{\rm voros}}}\psi_\pm^{(-\infty)}(x)
\end{align}
where $\mathcal{V}_{\rm voros}^{(\pm\infty)}$ are the Voros coefficients defined as
\begin{align}
    \mathcal{V}^{(\pm\infty)}_{\rm voros}
    & \equiv \frac{1}{2} \int_{\gamma_{\tau_\pm, \infty}} \left[ S_{\rm odd}(x)-\eta S_{-1}(x) \right] \dd x
    \sim \int_{\tau_\pm}^{\pm\infty} \left[ S_{\rm odd}(x)-\eta S_{-1}(x) \right] \dd x\ ,
\end{align}
As discussed before, the integration path of integral $\int_{\gamma_{\tau_\pm, x}} \dd x \sim 2\int_{\tau_\pm}^x \dd x$ starts from the point on the second Riemann sheet corresponding to $x$, runs to $\tau_\pm$, goes
around $\tau_\pm$, and runs to the point $x$ on the first Riemann sheet.
When $x$ extends to $\pm\infty$ (they are actually the same point on the Riemann sphere – the only difference is the path from which the integral
approaches to $\infty$), due to crossing a branch cut, the integral path in $\mathcal{V}^{(\pm\infty)}_{\rm voros}$ is not closed,
unlike the previous case of $V=-x$.
Due to the reversal symmetry, we also have
\begin{align}
\label{eq:Voros_reversal_example1}
    \mathcal{V}^{(+\infty)}_{\rm voros}=- \mathcal{V}^{(-\infty)} _{\rm voros}\equiv \mathcal{V}_{\rm voros}\ ,
\end{align}
which reduces some redundant notations.

Since each term in $S_{\rm even}$ is single-valued around $x=\tau_\pm$ and $\infty$, the integral path in $\mathcal{V}_{\rm voros}$ acts as a closed contour integral for $S_{\rm even}$ encircling $\tau_+$ and $\infty$, or $\tau_-$ and $-\infty$.
Let us call these paths as $\gamma_\pm$, respectively. Noting that the only term in $S_{\rm even}$ who has poles at
$x=\tau_\pm,~\infty$ is $S_0$, and applying the residue theorem, we gather
\begin{align}
    \int_{\tau_+}^\infty S_{\text{even}}(x) \, \dd x 
    = \frac{1}{2} \oint_{\gamma_+} S_0 \, \dd x 
    = \mp \frac{\pi i}{4} \ ,
\end{align}
where the $\mp$ sign corresponds to the counterclockwise or clockwise direction, respectively, of the closed contour, which is determined by the path of analytic continuation.
Since $S_{\rm even}$ has a residue only from the $S_0$ term, we can rewrite the Voros coefficient as
\begin{align}
\mathcal{V}_{\rm voros} 
= \int_{\tau_+}^{+\infty} \left[ S^{(+)}(x) - \eta S_1(x) - S_0(x)\right] \dd x\ ,
\end{align}
which is found a more convenient form in the following computation.

Our first task is to construct the ladder operators. For a function $\psi(x,E,\eta)$ that satisfies \eqref{eq:Sch_eq_example1},
the ladder operators, or the raising and lowering operators $\mathcal{L}_\pm$, respectively, play the role of 
\begin{align}
\label{eq:def_ladders}
    \left[ -\frac{\dd^2}{\dd x^2} + \eta^2 \left( E \pm c - \frac{x^2}{4} \right) \right] \big[ \mathcal{L}_\pm \psi(x, E, \eta) \big] = 0\ ,
\end{align}
where $c$ is a constant yet to be decided later.
To find out such a ladder operator, we may compare both sides of the equation, 
\begin{align}
\mathcal{L}_\pm \left[ -\frac{\dd^2}{\dd x^2}+\eta^2 \left( E-\frac{x^2}{4} \right) \right] = \left[-\frac{\dd^2}{\dd x^2} + \eta^2 \left( E\pm c-\frac{x^2}{4} \right) \right] \mathcal{L}_\pm\ ,
\end{align}
up to some overall constant.
By a straightforward but tedious calculation, we obtain the ladder operators as
\begin{align}
    \mathcal{L}_\pm
    =\frac{\dd}{\dd x}\pm i\eta\frac{x}{2}\ ,
\end{align}
with $c = \pm i/\eta$, where the $\pm$ sign is in a respective order.

Now for a WKB solution 
$\psi(x,E,\eta) = \exp\left[ \int^x S^{(+)}(x') \, \dd x' \right]$,
we see
\begin{align}
    \mathcal{L}_\pm \psi(x, E, \eta) = \left[ S^{(+)}(x, E, \eta) \pm i\eta \, \frac{x}{2} \right] \psi(x, E, \eta) \ .
\end{align}
The same argument works for $S^{(-)}$.
On the other hand, we can write, from the definition of the ladder operators in \eqref{eq:def_ladders},
\begin{align}
    \mathcal{L}_\pm\psi(x,E,\eta)
    =\tilde 
    \psi(x,E\pm i\eta^{-1},\eta),
\end{align}
where $\tilde \psi(x,E\pm i\eta^{-1},\eta)$ denotes a solution to the raised equation.
In principle, $\tilde \psi(x,E\pm i\eta^{-1},\eta)$ can be any linear combination of $\psi_\pm(x,E\pm i\eta^{-1},\eta)$, where the argument $E$ is replaced to $E+i\eta^{-1}$ for $\psi_\pm(x,E,\eta)$; however, the ladder operators $\mathcal{L}_\pm$ do not change the dominant exponent in $\psi_\pm$, and therefore $\tilde \psi_\pm (x,E\pm i\eta^{-1},\eta)$ must be proportional to  $\psi_\pm(x,E\pm i\eta^{-1},\eta)$, up to some constant factor, namely, the above two equations should
be equated through 
\begin{align}
    \mathcal{L}_\pm \psi(x, E, \eta) = \left[ S^{(+)}(x, E, \eta) \pm i\eta \, \frac{x}{2} \right] \psi(x, E, \eta) = C_\pm(E, \eta) \, \psi(x, E \pm i\eta^{-1}, \eta) \ ,
\end{align}
where $C_\pm(E,\eta)$ is a factor independent of $x$. By taking the logarithmic derivative $\partial_x\ln(\mathcal{L}_\pm\psi)$ and slightly rearranging the equation,  
we observe that $S^{(+)}$ satisfies the following difference equation:
\begin{align}
\label{eq:diff_eq_S}
    \Delta_E^{(\pm)} S^{(+)} \equiv S^{(+)}(x, E \pm i\eta^{-1}, \eta) - S^{(+)}(x, E, \eta) = \frac{\dd}{\dd x} \log \left[ S^{(+)}(x, E, \eta) \pm i\eta \, \frac{x}{2} \right] \ .
\end{align}
By defining
\begin{align}
    &I^{x}(x, E, \eta) = \int^x_{\tau_+} S^{(+)}(x', E, \eta) \, \dd x' = \frac{1}{2}\int_{\gamma_x} S^{(+)}(x', E, \eta) \, \dd x'\ ,\\
 &I^x_n(x, E, \eta) = \int_{\tau_+}^x S_n(x', E, \eta) \, \dd x'=\frac{1}{2}
 \int_{\gamma_x}S_n(x',E,\eta) \, \dd x'\ ,
\end{align}
where the integral path $\gamma_x$
goes from $\tilde x$, the point on the 2nd Riemann sheet corresponding to $x$, goes around $\tau_+$, and comes back
to $x$ on the first Riemann sheet.
We can now write the Voros coefficient as
\begin{align}
    \mathcal{V}_{\rm voros}
    =\lim_{x\to\infty}
    \big[
    I^x(x,E,\eta)
    -\eta\, I_{-1}^x(x,E,\eta)
    -I_0^x(x,E,\eta)
    \big]\ ,
\end{align}
amd consider computing $\Delta_E^{(+)} \mathcal{V}_{\rm voros}(E, \eta)$.
Its first term (before taking the $x \to \infty$ limit) can be evaluated by using~\eqref{eq:diff_eq_S}, giving
\begin{align}
    \Delta_E^{(+)} I^{x}(x, E, \eta) = \frac{1}{2} \log \left[ S^{(+)}(x, E, \eta) + i\eta \frac{x}{2} \right] - \frac{1}{2} \log \left[ S^{(+)}(\tilde x, E, \eta) + i\eta \frac{x}{2} \right]
    &=\frac{1}{2}\log\left[
    \frac{S^{(+)}(x,E,\eta)+i\eta\frac{x}{2}}{S^{(-)}(x,E,\eta)+i\eta\frac{x}{2}}
    \right]\ ,
\end{align}
where $S^{(+)}(\tilde x, E, \eta)=S^{(-)}(x,E,\eta)$ is used. In the limit of $x\to\infty$, we obtain
\begin{align}
    \Delta_E^{(+)} I^{x}(x, E, \eta) \simeq \frac{1}{2}\log\left(\frac{2i\eta x^2}{2iE\eta-1}\right),
\end{align}
For the second term in $\Delta_E^{(+)} \mathcal{V}_{\rm voros}(E, \eta)$, 
we first compute $I^x_{-1}$ directly,
\begin{align}
    I_{-1}^{x} 
\simeq i\frac{x^2}{4} - i\frac{E}{2} -i\frac{E}{2}\log \frac{x^2}{E}\ 
\end{align}
in the limit of $x\to \infty$, hence giving
\begin{align}
\Delta_E^{(+)} I_{-1}^{x} \simeq \frac{\eta^{-1}}{2}-i\frac{E}{2}\log\frac{E}{E+i\eta^{-1}}+\frac{\eta^{-1}}{2}\log\frac{x^2}{E+i\eta^{-1}}\ ,
\end{align}
in the same limit.
The last term in $\Delta_E^{(+)} \mathcal{V}_{\rm voros}(E, \eta)$ vanishes on its own: since $S_0(x,E,\eta)$ is a single-valued function of $x$, the path $\gamma_x$ is actually a closed contour, and the residues appearing at $x=\tau_\pm$ and $x=\infty$ are numbers independent of $E$, giving $\Delta^{(+)}_E I^x_0 = 0$.
Combined, we obtain
\begin{align}
    \Delta^{(+)}_E\mathcal{V}_{\rm voros}(E,\eta) 
    & = \lim_{x \to \infty} \left[ \Delta_E^{(+)} I^{x}(x, E, \eta) - \eta \Delta_E^{(+)} I_{-1}^{x}(x, E, \eta) - \Delta_E^{(+)} I_0^{x}(x, E, \eta) \right] \nonumber \\
    & = -\frac{1}{2} + \frac{1+\sigma}{2} \log \left(1+\frac{1}{\sigma} \right) - \frac{1}{2} \log \left(1 + \frac{1}{2\sigma} \right) \ .
\end{align}
where $\sigma\equiv -iE\eta$.

Now notice, there is a scaling relation $S_j(\sqrt{E} \, x, E) = E^{-j-1/2} S_j(x, 1)$, which leads to
\begin{align}
    \mathcal{V}_{\rm voros}(E,\eta) = \int_2^\infty \sum_{j=1}^\infty S_{2j-1}(t,1) \left( E \eta \right)^{-2j+1} \, \dd t\ . 
\end{align}
From this expression, we observe an interesting correspondence,
\begin{align}
    \mathcal{V}_{\rm voros}(E+i\eta^{-1},\eta) = \mathcal{V}_{\rm voros}(E,\eta+iE^{-1})\ .
\end{align}
See also the discussion around \eqref{eq:voros_arg_property}.
Hence,
\begin{align}
    \Delta^{(+)}_\eta \mathcal{V}_{\rm voros}(E,\eta) &\equiv \mathcal{V}_{\rm voros}(E,\eta+iE^{-1}) - \mathcal{V}_{\rm voros}(E,\eta) \nonumber \\
    &= \Delta^{(+)}_E\mathcal{V}_{\rm voros}(E,\eta) \nonumber \\
    &= -\frac{1}{2} + \frac{1+\sigma}{2} \log \left(1+\frac{1}{\sigma} \right) - \frac{1}{2} \log \left(1 + \frac{1}{2\sigma} \right) \ ,
    \label{eq:diff_voros}
\end{align}
where formally the difference operator $\Delta^{(+)}_\eta$
can be written by the differential operator of infinite order
\begin{align}
    \Delta^{(+)}_\eta = \ee^{iE^{-1}\partial_\eta} - 1 = \sum_{k=1}^\infty \frac{(iE^{-1}\partial_\eta)^k}{k!}\ .
\end{align}
The Voros coefficient can be formally written as
\begin{align}
    \mathcal{V}_{\rm voros}(E,\eta) = \sum_{j=1}^\infty \mathcal{V}_{2j-1}(E) \, \eta^{-2j+1}, \quad \mathcal{V}_{2j-1}(E) \equiv \int_{\tau_+}^\infty S_{2j-1}(x, E) \, \dd x\ .
\end{align}
Now consider Borel-transforming~\eqref{eq:diff_voros}. The left-hand side of \eqref{eq:diff_voros} reads
\begin{align}
    \Delta^{(+)}_\eta \mathcal{V}_{\rm voros}(E,\eta) 
    &= \sum_{k=1}^\infty \sum_{j=1}^\infty \frac{\mathcal{V}_{2j-1}(E)}{k!} \left( -iE^{-1} \right)^k \frac{\left( 2j+k-2 \right)!}{\left( 2j-2 \right)!} \eta^{-2j-k+1}\ ,
\end{align}
and one can directly compute its Borel transformation, which in the end amounts to
\begin{align}
     \mathcal{B}\left[\Delta_\eta^{(+)}\mathcal{V}_{\rm voros}(E, \eta)\right](y) 
    = \left[ \ee^{y/(iE)}-1 \right] \mathcal{V}_{{\rm voros},B}(E,y)\ ,
\end{align}
where $\mathcal{B}$ indicates the Borel transformation.
Now let us consider the right-hand side of~\eqref{eq:diff_voros}.
Using
\begin{align}
   & \mathcal{B}\left[
    \log\left(1+\frac{1}{\lambda\eta}\right)
    \right]
    =-\frac{1}{y} \left( \ee^{-y/\lambda}-1 \right)\ ,\\
    &\mathcal{B}\left[
    -1+\lambda\eta\log\left(1+\frac{1}{\lambda\eta}\right)
    \right]
    =\frac{\lambda}{y^2} \left( \ee^{-y/\lambda}-1 \right) + \frac{\ee^{-y/\lambda}}{y}\ ,
\end{align}
we obtain
\begin{align}
    \mathcal{B} \left[ \text{rhs of~\eqref{eq:diff_voros}} \right]
    =\frac{-iE}{2y^2} \left[ \ee^{y/(iE)} - 1 \right] + \frac{\ee^{y/(2iE)} }{2y} \ .
\end{align}
Supplying the above expressions into the Borel-transformed version of \eqref{eq:diff_voros} and solving for $\mathcal{V}_{{\rm voros},B}$, we find
\begin{align}
\label{eq:Voros_B_example1}
    \mathcal{V}_{{\rm voros},B}
    =\frac{-iE}{2y^2}+\frac{1}{4y}
    \left[\frac{1}{\ee^{y/(2iE)}-1}+\frac{1}{\ee^{y/(2iE)}+1}\right]\ ,
\end{align}
which is exactly the same expression as the one in~\cite{Takei:2008s}.
The final step is to Laplace-transform \eqref{eq:Voros_B_example1}. This integral is convergent if and only if ${\rm Im} \, (E \eta) < 0$; for the parameter region ${\rm Im} \, (E\eta) > 0$, we utilize the reversal symmetry in \eqref{eq:Voros_reversal_example1}. Using the known formula,
\begin{align}
    \int_0^\infty \ee^{-z t} \left( \frac{1}{\ee^t - 1} - \frac{1}{t} + \frac{1}{2} \right) \frac{\dd t}{t}
    = \log \frac{\Gamma(z)}{\sqrt{2 \pi}} - \left( z - \frac{1}{2} \right) \log z + z \ , \qquad
    {\rm Re} \, (z) > 0 \ ,
\end{align}
and the reversal relation, 
the Borel summed Voros coefficient, denoted by $ \mathscr{V}$, is computed as,
\begin{align}
\label{eq:voros_coefficient_result}
    \mathscr{V}(E\eta)=
    \begin{cases}
    \displaystyle
    \frac{1}{2}\left[\log\frac{\Gamma(iE\eta+1/2)}{\sqrt{2\pi}}-iE\eta\log(iE\eta)+iE\eta\right]\ , & {\rm Im}(E\eta)<0 \ ,
    \vspace{2mm} \\
    \displaystyle
    -\frac{1}{2}\left[\log\frac{\Gamma(-iE\eta+1/2)}{\sqrt{2\pi}}+iE\eta\log(-iE\eta) -iE\eta
    \right]\ , & {\rm Im}(E\eta)>0\ ,
    \end{cases}
\end{align}
with the identical result given in \cite{Takei:2008s}.
This concludes the calculation of the Voros coefficients in the case of $V=-E+x^2/4$.
See the appendices for other examples of non-trivial Voros coefficients.

\section{Analyzing particle production with exact WKB analysis}
\label{sec:particle_production_exact_WKB}

In previous sections, we made preparations to describe particle production, utilizing the exact WKB analysis. 
In this section, we first overview a generic procedure of the calculation of particle number density with the exact WKB method. We then demonstrate the validity of our method with a concrete potential. 
%See the appendices for other examples.

\subsection{A generic flow of calculation}

\noindent
{\bf 1. SET UP 1-D SCHR\"ODINGER-LIKE EQUATION}

\vspace{3mm}
Consider a model of a Schr\"{o}dinger-type equation with the potential $V(x,\bm{\vartheta})$
\begin{align}
\label{eq:Scheq_sec_ptclprod}
    \left[ \frac{\dd^2}{\dd x^2} + 
    V(x,\bm\vartheta) \right] 
    \psi(x,\bm\vartheta) = 0 \ ,
\end{align}
where $\bm\vartheta$ denotes a collection of model parameters, which may include $\hbar$ for a quantum-mechanical problem.
In order to construct WKB series solutions, we introduce a ``formally large'' parameter $\eta$.
It should be introduced in such a way that the asymptotic behaviors of solutions to \eqref{eq:Scheq_sec_ptclprod} are reproduced by the leading order of the WKB approximation with respect to $\eta \gg 1$.
If quantum effects are assumed to be small, $\eta$ may be identified with $1/\hbar$, but this is not necessary in general, as
$\eta$ is just a formal expansion parameter to utilize the exact WKB analysis.
If $\bm\vartheta$ contains some coupling constants with different strengths, it may be found useful to introduce $\eta$ to capture the hierarchy.
We thus deform the potential such that it recovers $V(x,\bm\vartheta)$ when $\eta=1$ is taken.
With some abuse of notation, our equation to be studied is then
\begin{align}
    \left[ \frac{\dd^2}{\dd x^2} + \eta^2 V(x,\eta, \bm\vartheta) \right] \psi(x,\eta,\bm\vartheta)=0 \ .
\end{align}
We analyze particle production using 
this deformed potential via exact WKB analysis, and then set $\eta=1$ at the end of the computation to obtain the result for the original model with $V(x,\bm\vartheta)$.

We determine how to introduce $\eta$ in order to match approximate solutions via the WKB approximation (i.e.~solutions up to the 
leading order) to the asymptotic solutions.
The asymptotic solutions are obtained by solving the Shr\"odinger-like equation with the approximate potentials in the asymptotic regions.  
We choose a way to introduce $\eta$ so that the WKB approximated solutions reproduce the asymptotic solutions. This choice becomes relevant in finding mode functions as become clear later in this section.
To be more concrete, let us consider an example potential $V=\frac{1}{4}-\frac{\nu^2-1/4}{x^2}$, as considered in Appendix~\ref{app:caseII}.
For the two asymptotic states in the limit of $x\to 0$ and $x\to \infty$, the approximated potentials are given as $V=-\frac{\nu^2-1/4}{x^2}$ and $V=1/4$ before introducing $\eta$,
respectively. The asymptotic solutions via those potentials are computed as 
\begin{align}
\begin{array}{rll}
x \to 0 \quad & \displaystyle \implies \quad V \simeq -\frac{\nu^2-1/4}{x^2} \quad & \displaystyle \implies \quad \psi \sim x^{\frac{1}{2}\pm\nu} \ , \vspace{1mm}\\
x \to \infty \quad & \displaystyle \implies \quad V \simeq -\frac{1}{4} \quad & \displaystyle \implies \quad \psi \sim \ee^{\pm i x/2}\ .
\end{array}
\end{align}
In order that the WKB approximation reproduces those asymptotic solutions, one needs to deform $V$ by introducing $\eta$ as
\begin{align}
    V \to \frac{1}{4}-\frac{\nu^2-\eta^{-2}/4}{x^2} \ .
\end{align}
This choice of introduction of $\eta$ is consistent with the assumption made to ensure the Borel summability of the WKB solutions. See e.g.~\cite{Iwaki:2014vad} for more detailed discussion.%
\footnote{The potential $Q$ in~\cite{Iwaki:2014vad} is related to our $V$ by $Q=-V$. They assume $Q_2$ has a double pole at $x=p$ and satisfies $Q_2=-\frac{1}{4x^2}$ in the limit $z\to 0$ where $z$ is a local coordinate near $p$ with $z(p)=0$. To compare with our example potential, we have a double pole at $x=0$ and $z=x$.}

\vspace{5mm}
\noindent
{\bf 2. STOKES CURVES, STOKES REGIONS, AND DOMINANCE RELATION}

\vspace{3mm}
Once a way to introduce $\eta$ is determined, one can obtain the turning points,
draw the Stokes curves, identify the corresponding Stokes regions, and compute the dominance relation on the curves,
\begin{itemize}
    \item Turning points $\tau_i$: $V_0(\tau_i)=0$ ($V_0$: the leading term of the potential $V$ in terms of ``large'' $\eta$) 
    \item Stokes curves: $\displaystyle {\rm Im}\int_{\tau_i}^x\sqrt{-V_0(\chi)} \, \dd \chi=0$
    \item  Dominance relation:
    $\displaystyle \mathop{\rm sign} \left[ {\rm Re}\int_{\tau_i}^x\sqrt{-V_0(\chi)} \, \dd\chi \right]$
    on the Stokes curves
\end{itemize}
Those quantities can depend  on a way to introduce $\eta$.
One needs to check if each turning point is simple, i.e.,
\begin{itemize}
  \item Simple turning points: $\displaystyle \frac{\dd V_0}{\dd x} \bigg\vert_{x=\tau_i} \neq 0$
\end{itemize}
Otherwise, the connection formulae would not be valid.%
\footnote{A simple pole behaves as a turning point with modified connection formulae, as previously mentioned.}

\vspace{5mm}
\noindent
{\bf 3. DEGENERATE STOKES CURVES}

\vspace{3mm}
A Stokes curve in general emanates from a turning point and flows into a singular point of the potential. Sometimes, however, one may encounter a curve that connects two turning points for some particular choice of model parameters. This is called a degenerate Stokes curve, and it is formed by overlapping would-be two curves for other parameter values.
Degenerate Stokes curves give pathological graphs, which lead to Borel non-summability of the WKB series solutions due to the existence of fixed singularities on the Borel plane along the path of the Laplace integral. Introducing a small parameter to resolve the degeneracy is one prescription. Such a regularization parameter is taken to be zero at the end of the computation. Often, there are some ambiguities in how to introduce such a parameter, such as its sign.
However, focusing on physical quantities, these ambiguities should disappear in the end. Indeed, all our examples give the same number density without depending on the sign of regularization parameters.

A degenerate Stokes graph, as in the left panel of Fig.~\ref{fig:degenerate_stokes_lines_two_cases}, can be made non-degenerate by a small regularization parameter, e.g.~in the right panel of Fig.~\ref{fig:degenerate_stokes_lines_two_cases}.
One has to be careful about branches and Riemann surfaces for precise calculations because the signs of various quantities can be easily flipped if a wrong branch/Riemann surface is considered.
Also, note that when $S_{-1}$ is a two-valued function, the Borel summed WKB solutions are four-valued functions due to the presence of the factor $1/\sqrt{S_{\rm odd}}$ in the formal solution.

\vspace{5mm}
\noindent
{\bf 4. CONNECTION FORMULA AND REPLACING TURNING POINTS}

\vspace{3mm}
Once steps 2 and 3 above are completed, the exact WKB solutions normalized at a turning point in a Stokes region can be related to those in a different region using the connection formulae given in \eqref{eq:connection_formula_1} and \eqref{eq:connection_formula_2}. 

When crossing the Stokes curves emanating from distinct turning points, it becomes necessary to shift the turning points used to normalize the exact WKB solutions in order to apply the connection formulae. For instance in the case of Fig.~\ref{fig:stokes_curves_positive_imaginary}, 
the shift from the left turning point $\tau_-$ to the right one $\tau_+$ may take place for the solutions constructed in region II, which can be accounted for by changing the normalization of $\psi_\pm^{\rm II}$ as
\begin{align}
\left(
\begin{array}{c}
\displaystyle
\psi^{\rm II}_{+,\tau_-} \vspace{1mm} \\
\displaystyle
\psi^{\rm II}_{-,\tau_-}
\end{array}
\right)&=
\left(
\begin{array}{cc}
\displaystyle
\ee^{+\int_{\tau_-}^{\tau_+}S_{\rm odd} \, \dd x} & 0 \vspace{1mm}\\
0 & 
\displaystyle
\ee^{-\int_{\tau_-}^{\tau_+}S_{\rm odd} \, \dd x} 
\end{array}
\right)
\left(
\begin{array}{c}
\displaystyle
\psi^{\rm II}_{+,\tau_+} \vspace{1mm}\\
\displaystyle
\psi^{\rm II}_{-,\tau_+}
\end{array}
\right)\ .
\end{align}
Combining this effect with the connection formulae, the overall connection matrix for analytical continuation along the path from the left (region I) to right (region III) on the real axis is expressed as follows,
\begin{align}
\label{eq:connection_eg}
\left(
\begin{array}{c}
\displaystyle
\psi^{\rm I}_{+,\tau_-} \vspace{1mm}\\
\displaystyle
\psi^{\rm I}_{-,\tau_-}
\end{array}
\right)&=\left(
\begin{array}{cc}
1 & 0\\
-i & 1
\end{array}
\right)
\left(
\begin{array}{cc}
\displaystyle
\ee^{+\int_{\tau_-}^{\tau_+} S_{\rm odd} \, \dd x} & 0 \vspace{1mm}\\
0 & 
\displaystyle
\ee^{-\int_{\tau_-}^{\tau_+} S_{\rm odd} \, \dd x} 
\end{array}
\right)
\left(
\begin{array}{cc}
1 & i\\
0 & 1
\end{array}
\right)
\left(
\begin{array}{c}
\displaystyle
\psi^{\rm III}_{+,\tau_+} \vspace{1mm}\\
\displaystyle
\psi^{\rm III}_{-,\tau_+}
\end{array}
\right)
\end{align}
where the non-diagonal matrices arise from the connection formulae, and the diagonal matrix in the middle 
is due to the shift in the turning points used for normalization.
The computation of the integral $\int_{\tau_-}^{\tau_+}S_{\rm odd} \, \dd x$ is
feasible in the specific examples presented later.
However, in general, a precise calculation of the integral poses a significant challenge.

\vspace{5mm}
\noindent
{\bf 5. VOROS COEFFICIENTS AND MODE FUNCTIONS}

\vspace{3mm}
Let us now define the mode functions, a central ingredient to define vacua. The first step here is to find the Voros coefficients and to construct the WKB solutions normalized in the asymptotic limits, e.g.,
\begin{align}
\label{eq:decomp_psi_voros}
    \psi_{\pm,\tau_1}
    =\exp \left( \pm \mathcal{V}_{\rm voros}^{(\sigma_1)} \right) \, \psi^{(\sigma_1)}_{\pm}\ , \quad
     \psi_{\pm,\tau_2}
     = \exp \left( \pm \mathcal{V}_{\rm voros}^{(\sigma_2)} \right) \,\psi^{(\sigma_2)}_{\pm}\ . 
\end{align}
Here, we assume two singular points $x\to \pm \sigma_{1,2}$,%
\footnote{In fact, a single singular point may be denoted by $\sigma_{1,2}$, where the subscript only indicates the direction from which the integration path approaches to this singular point. An exapmle of this type is the case with $V=-E+x^2/4$ in Sec.~\ref{subsec:Voros_ex1}.}
$\mathcal{V}_{\rm voros}^{(\sigma_{1,2})}$ denote the Voros coefficients connecting these singular points to the turning points $\tau_{1,2}$, and $\psi^{(\sigma_{1,2})}_{\pm}$ are the formal WKB solutions normalized at $\sigma_{1,2}$, respecitvely,
which are defined as
\begin{subequations}
\label{eq:def_psi_voros}
\begin{align}
    \psi^{(\sigma_i)}_{\pm}& \equiv \frac{1}{\sqrt{S_{\rm odd}}} \exp \left( \pm\eta \int_{\tau_i}^x S_{-1} \, \dd x \right) \,
    \exp \left[ \pm \int_{\sigma_i}^x (S_{\rm odd}-\eta S_{-1}) \, \dd x \right]\ ,\\
    \mathcal{V}_{\rm voros}^{(\sigma_i)} & \equiv
    \int_{\tau_i}^{\sigma_i} \left( S_{\rm odd}-\eta S_{-1} \right) \dd x \ .
\end{align}
\end{subequations}
The calculation of the Voros coefficients have been discussed in details in Sec.~\ref{sec:voros_coefficient}.
The decomposition in \eqref{eq:decomp_psi_voros} with \eqref{eq:def_psi_voros} is done at the level of formal (divergent) WKB series. However, the same decomposition holds after the Borel summation, as long as the Borel summability is respected. This is due to the nice property of the Borel sum for functional products discussed in Sec.~\ref{sec:review_exact_WKB}.

The mode functions are favorably defined in terms of the Borel summed $\psi^{(\sigma_i)}_{\pm}$, whose behavior is well approximated by the WKB approximated solution in the asymptotic limit,%
\footnote{Conversely, if one should $\psi_{\pm,\tau_i}$ for the mode functions, their amplitudes would differ by the values of the Voros coefficients compared to the WKB approximation. This would result in a wrong commutation relation differing from \eqref{eq:commutation_condition_eWKB}, failing quantization.}
\begin{align}
\psi^{(\sigma_i)}_{\pm} & \sim \frac{1}{\sqrt{\eta S_{-1}}} \exp\left(\pm\eta \int_{\tau_i}^x S_{-1} \, \dd x \right) \, , \quad x\to \sigma_i \ .
\end{align}
Owing to this property, we may ``define'' the conjugate functions of these solutions without ambiguity by matching their asymptotic expansions.%
\footnote{Complex conjugation generically breaks the analyticity of functions under consideration. Thus we only identify their conjugates with those exact WKB solutions, which are analytic functions, that provide appropriate asymptotic behaviors and the correct Wronskian.}
Overall, we decompose the mode functions in terms of a linear combination of $\psi^{(\sigma_i)}_\pm$,
\begin{align}
\label{eq:upm_decompose_eWKB}
    u^{(\sigma_i)}_{+} & \equiv \frac{1}{\sqrt{2}} \left[ \alpha^{(\sigma_i)} \, \psi^{(\sigma_i)}_{-} + \beta^{(\sigma_i)} \, \psi^{(\sigma_i)}_{+} \right] \ , %\\
    \qquad
    u^{(\sigma_i)}_{-}
    \equiv \frac{1}{\sqrt{2}} \left[
    \beta^{(\sigma_i) \, *} \, \psi^{(\sigma_i)}_{-} + \alpha^{(\sigma_i) \, *} \, \psi^{(\sigma_i)}_{+} \right]
    \equiv \overline{u^{(\sigma_i)}_{+}} \ ,
\end{align}
where $\psi^{(\sigma_i)}_{\pm}$ should be interpreted as Borel-resummated ones.
The coefficients $\alpha^{(\sigma_i)}$ and $\beta^{(\sigma_i)}$ are constant complex numbers, which are determined from the inner product conditions as in \eqref{eq:modes},
and the boundary conditions in the asymptotic limits. We adopt the  boundary conditions by making $\psi^{(\sigma_i)}_-$ 
the positive 
frequency mode, 
which implies 
$\vert \alpha^{(\sigma_i)} \vert = 1$ and $\beta^{(\sigma_i)} = 0$ to satisfy the boundary condition in each asymptotic region.
Note that these mode functions are defined in the limit that the potential is real function, while we often introduce small imaginary part to the potential when the Stokes curves are degenerate and take it to zero at the end of calculation. This procedure may cause an apparent ambiguity depending on how this small parameter deforms the Stokes graph. The determination of the mode functions outlined above is a physical requirement, together with which the apparent ambiguity ceases, at least in the models we consider in this paper.

\vspace{5mm}
\noindent
{\bf 6. NUMBER DENSITY}

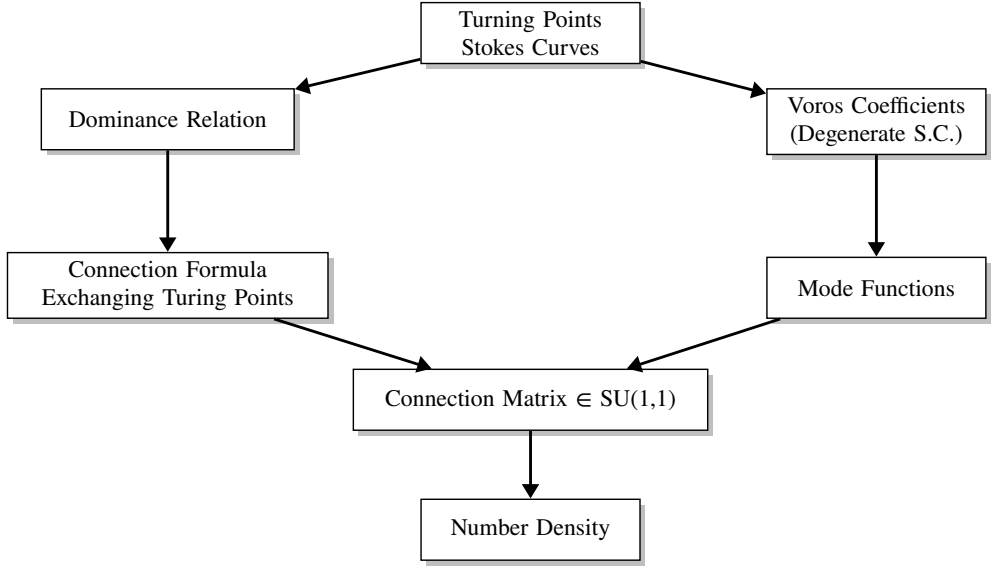
\begin{figure}[t]
\centering
\scalebox{0.9}{
\begin{tikzpicture}[
align=center, node distance = 5mm,
    arr/.style = {-Triangle, very thick},
    box/.style = {rectangle, draw, semithick,
                  minimum height=9mm, minimum width=30mm,
                  text width=30mm, align=center, fill=white, drop shadow},
]

% Nodes
\node (B) [box] {Turning Points\\
Stokes Curves};
\node (D) [box, below left=of B, xshift=-15mm, text width=35mm] {Dominance Relation};
\node (E) [box, below right=of B, xshift=15mm] {Voros Coefficients \\ (Degenerate S.C.)};
\node (F) [box, below=of D, yshift=-10mm, text width=45mm] {Connection Formula\\ Exchanging Turing Points};
\node (G) [box, below=of E, yshift=-10mm] {Mode Functions};
\node (I) [box, below=of B, yshift=-40mm, text width=50mm] {Connection Matrix $\in$ SU(1,1)};
\node (J) [box, below=of I, yshift=-5mm] {Number Density};

% Arrows
\draw[arr] (B) -- (D);
\draw[arr] (B) -- (E);
\draw[arr] (D) -- (F);
\draw[arr] (E) -- (G);
\draw[arr] (F) -- (I);
\draw[arr] (G) -- (I);
\draw[arr] (I) -- (J);
\end{tikzpicture}
}
\caption{The summary of the calculation flow for the overall connection matrix using the exact WKB analysis. Note that Stokes curves are abbreviated as S.C..}
  \label{fig:wkb-flow}
\end{figure}

\vspace{3mm}
A leading indicator of particle production is the number density, which is evaluated as an expectation value of the number operator $\hat{N} = \hat a^\dagger \hat a$. The operator $\hat a$ is defined as the one that annihilates the vacuum at one asymptotic limit, say $\sigma_1$, while the expectation is taken with respect to the vacuum at another asymptotic limit $\sigma_2$, see Sec.~\ref{subsec:part_prod} for details. 
Given that two pairs of mode functions constructed with respect to two different asymptotic limits are related through \eqref{eq:decomposition}, the number density, or rather the occupation number $n$, is given by
\begin{align}
    n \equiv \langle \hat{N} \rangle = \vert \beta \vert^2 \ ,
\end{align}
as derived in \eqref{eq:number_operator}. In the language of this section, the Bogoliubov coefficients $\alpha$ and $\beta$ relate the mode functions $u_\pm^{(\sigma_i)}$ in \eqref{eq:upm_decompose_eWKB} that are properly normalized at two different asymptotic points.%
\footnote{We are using somewhat confusing notations of $\alpha$ and $\beta$. Those appearing in \eqref{eq:upm_decompose_eWKB} are Bogoliubov-type coefficients relating between $u_\pm^{(\sigma_i)}$ and $\psi_\pm^{(\sigma_i)}$ for a single asymptotic point $\sigma_i$, while $\alpha$ and $\beta$ without superscripts are being used as the Bogoliubov coefficients between $u^{(\sigma_{1,2})}_\pm$ for different asymptotic points, say $\sigma_{1,2}$.}
Therefore, computing $n$ is equivalent to finding the connection formulae similar to \eqref{eq:connection_eg}, but those for $u_\pm^{(\sigma_i)}$ from one asymptotic to another.
Let us outline procedure below, combining the previous steps, summarized in Fig.~\ref{fig:wkb-flow}.

Suppose we find the evolution matrix $\mathcal{U}_{AB}$ that connects the exact WKB solutions in region $A$ to those in region $B$, normalized at the turning points $\tau_{1,2}$, respectively, that is,
\begin{align}
\left(
\begin{array}{c}
\displaystyle
\psi^A_{+,\tau_1} \vspace{1mm}\\
\displaystyle
\psi^A_{-,\tau_1}
\end{array}
\right)&=
\mathcal{U}_{AB}
\left(
\begin{array}{c}
\displaystyle
\psi^{B}_{+,\tau_2} \vspace{1mm}\\
\displaystyle
\psi^{B}_{-,\tau_2}
\end{array}
\right)\ ,
\end{align}
where $\mathcal{U}_{AB}$ may contain the contributions from the connection formulae when crossing the Stokes curves and the exchange of the turning points used for normalization (and possibly the crossing of branch cuts, which we do not encounter in this paper).
In the example case of Fig.~\ref{fig:stokes_curves_positive_imaginary}, this equation corresponds to \eqref{eq:connection_eg}, and $\mathcal{U}_{AB}$ to the product of three matrices in \eqref{eq:connection_eg}.

The next step is to compute the Voros coefficients $\mathcal{V}_{\rm voros}$, and relate $\psi^{A,B}$ to the corresponding asymptotic solutions $\psi^{(\sigma_i)}$,
\begin{align}
\left(
\begin{array}{c}
\displaystyle
\psi^A_{+,\tau_i} \vspace{1mm} \\
\displaystyle
\psi^A_{-,\tau_i}
\end{array}
\right)
=
\mathscr{V}^{(\sigma_i)}
\left(
\begin{array}{c}
\displaystyle
\psi^{(\sigma_i)}_{+} \vspace{1mm} \\
\displaystyle
\psi^{(\sigma_i)}_{-}
\end{array}
\right)\ , \qquad
\mathscr{V}^{(\sigma_i)}
\equiv 
\left(
\begin{array}{cc}
\displaystyle
\exp \left( \mathcal{V}^{(\sigma_i)}_{\rm voros} \right) & 0\\
0 & 
\displaystyle
\exp \left( -\mathcal{V}^{(\sigma_i)}_{\rm voros} \right)
\end{array}
\right) \ ,
\end{align}
where the superscript $(\sigma_i)$ of the wave functions denote the WKB solutions normalized at the appropriate asymptotic points $\sigma_i$.

We then determine the mode functions $u_{\pm}$ given as a linear combination of $\psi_\pm$ normalized with respect to $\sigma_i$,
\begin{align}
 \left(
\begin{array}{c}
\displaystyle
u^{(\sigma_i)}_{+} \vspace{1mm}\\
\displaystyle
u^{(\sigma_i)}_{-}
\end{array}
\right)
=\mathcal{U}^{(\sigma_i)}
\left(
\begin{array}{c}
\displaystyle
\psi^{(\sigma_i)}_{+} \vspace{1mm}\\
\displaystyle
\psi^{(\sigma_i)}_{-}
\end{array}
\right)
\end{align}
where the components of $\mathcal{U}^{(\sigma_i)}$ consist of $\alpha^{(\sigma_i)}$ and $\beta^{(\sigma_i)}$ in \eqref{eq:upm_decompose_eWKB}.
Computing the inner product require the complex conjugation of the wave functions. Since complex conjugation does not preserve analyticity, we make an identification of the conjugates with appropriate exact WKB solutions, which can be done by matching their asymptotic expansions along the anti-Stokes curves.

Putting all the above together, we obtain the evolution matrix that connects the mode functions for two asymptotic regions $\sigma_{1,2}$ as,
\begin{align}
\label{eq:connection_final_general}
\left(
\begin{array}{c}
\displaystyle
u^{(\sigma_1)}_{+} \vspace{1mm}\\
\displaystyle
u^{(\sigma_1)}_{-}
\end{array}
\right)
=&\mathcal{U}^{(\sigma_1)}
\left( \mathscr{V}^{(\sigma_1)} \right)^{-1}
\mathcal{U}_{AB} \,
\mathscr{V}^{(\sigma_2)}
\left( \mathcal{U}^{(\sigma_2)} \right)^{-1}
 \left(
\begin{array}{c}
\displaystyle
u^{(\sigma_2)}_{+} \vspace{1mm}\\
\displaystyle
u^{(\sigma_2)}_{-}
\end{array}
\right)\ .
\end{align}
The Bogoliubov coefficients $\alpha$ and $\beta$ relating the two pairs of mode functions can be written as the inner products \eqref{eq:innerproduct} between $u_\pm^{(\sigma_{1,2})}$, i.e.,
\begin{align}
    \alpha = \left( u_+^{(\sigma_2)}, \, u_+^{(\sigma_1)} \right) \ , \quad
    \beta = \left( u_+^{(\sigma_2)}, \, u_-^{(\sigma_1)} \right) \ ,
\end{align}
while respecting $|\alpha|^2- |\beta|^2=1$. Then \eqref{eq:connection_final_general} can be rewritten in terms of $\alpha$ and $\beta$ as,
\begin{align}
 \left(
\begin{array}{c}
u^{(\sigma_2)}_{+} \vspace{1mm}\\
u^{(\sigma_2)}_{-}
\end{array}
\right)
=
    \begin{pmatrix}
\alpha & \beta \vspace{1mm} \\
\bar\beta & \bar\alpha
\end{pmatrix}
   \left(
\begin{array}{c}
u^{(\sigma_1)}_{+} \vspace{1mm}\\
u^{(\sigma_1)}_{-}
\end{array}
\right)\ , 
\end{align}
where we have switched the order of $\sigma_{1,2}$ just for convention.
As noted in Sec.~\ref{subsec:part_prod}, the connection matrix belongs to SU(1,1), i.e.,
\begin{align}
\begin{pmatrix}
\alpha & \beta \\
\bar\beta & \bar\alpha
\end{pmatrix} 
\in {\rm SU}(1,1)\ .
\end{align}
This property provides us a way to crosscheck whether the Voros coefficients are non-trivial or not. 
If we compute the connection matrix with all the Voros coefficients taken to be zero and then the connection matrix is not in SU(1,1), the Voros coefficients should be non-trivial. Note that the Voros coefficients contain additional information, such as phase, and cannot be fully determined by requiring this SU(1,1) condition alone.

\subsection{Example: $V=-E+x^2/4$}
\label{subsec:example_x2}

In this subsection, we apply our generic prescription for particle production described above to the case of $V=-E+x^2/4$ and demonstrate its validity.
This potential has been well studied before as the approximated potential to investigate the analytic structure of an oscillatory potential~\cite{Kofman:1997yn,Salehian:2020dsf}. 
The Voros coefficients associated with this potential has been thoroughly computed in Sec.~\ref{subsec:Voros_ex1}, and readers may refer to the derivation there for definitions and notations.

We have Stokes curves emanating from the two turning points shown as in Fig.~\ref{fig:degenerate_stokes_lines_two_cases}.
Our interested parameter regime is where $E$ is a real constant, but in this case, the two turning points are connected by a Stokes curve. This is the degenerate Stokes curve, and the WKB series is not Borel summable.
To overcome this problem, we introduce a small parameter $i\epsilon$ where $\epsilon\in \mathbb{R}$, i.e.~we replace $E$ by $E\to E+i\epsilon$.
We have two choices of the sign of ${\rm Im}\,E$. Either should give the same physics after taking $\epsilon\to 0$ in the end of the calculation. Thus without loss of generality, we take $\epsilon>0$ in the following computation of this section, while the case with $\epsilon < 0$ is verified in Appendix~\ref{app:negative_epsilon}.
The turning points are now shifted to the imaginary direction,
\begin{align}
    \tau_{\pm}\approx
    \pm 2\sqrt{E}  \pm i\epsilon\ ,
\end{align}
which is expanded by a small $\epsilon$. The Stokes curves are given in Fig.~\ref{fig:stokes_curves_positive_imaginary}, and turning points are no longer connected by any Stokes curves. 
All Stokes curves emanating from the turning points flow into the singular point $x=\infty$, and the dominance relation, given by the sign of ${\rm Re} \int_{\tau_\pm}^x \sqrt{-V(x')}\,\dd x'$, is indicated in Fig.~\ref{fig:stokes_curves_positive_imaginary}. 
We also call the Stokes regions from left to right by I, II, and III, and place the branch cuts as in Fig.~\ref{fig:stokes_curves_positive_imaginary}.

%%%%
\begin{figure}[t]
\centering
\hspace{1cm}
   \includegraphics[width=0.4\linewidth]{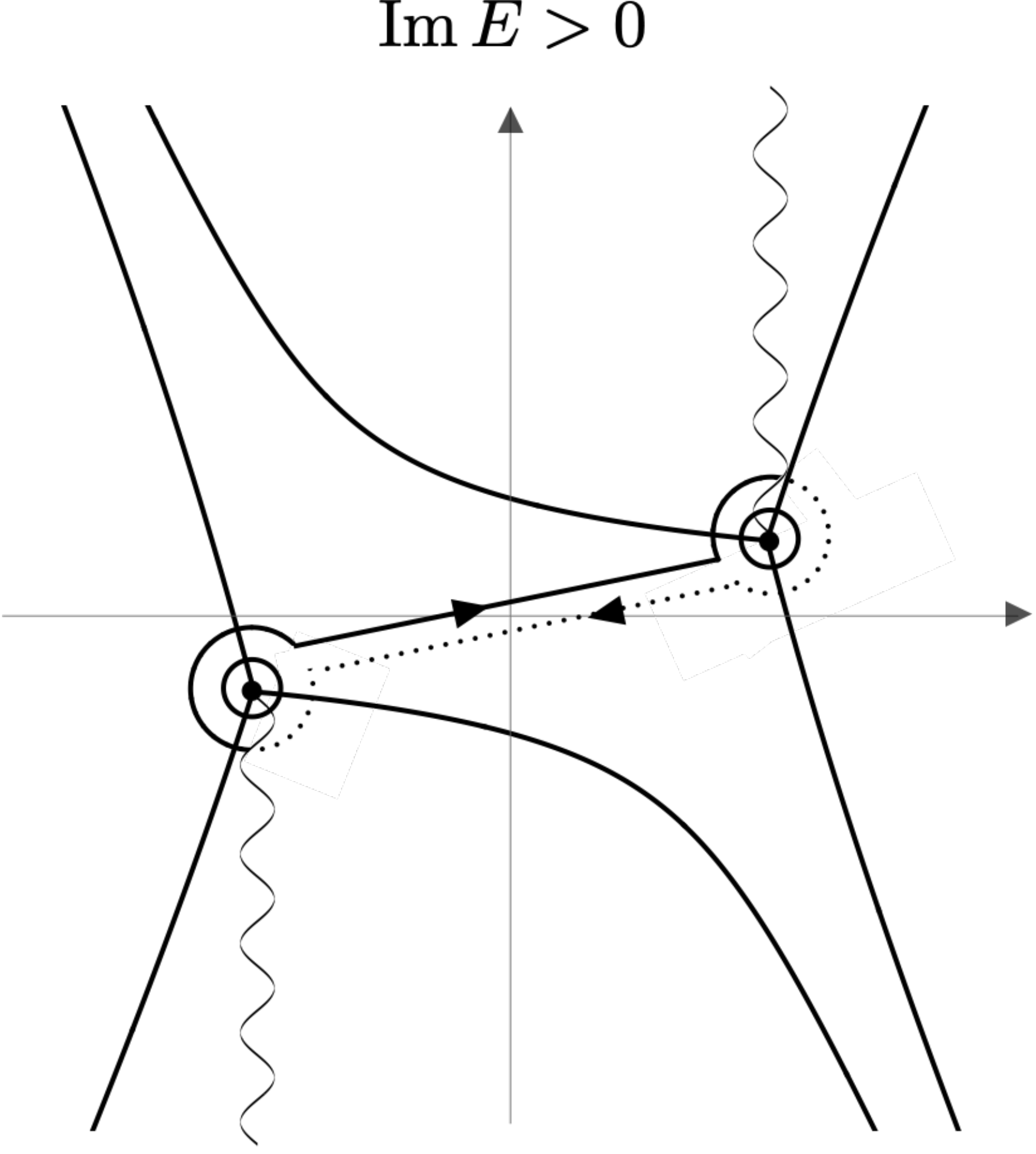}
   \caption{The integration path between the two turning points. The solid line denotes the first Riemann sheet and the dashed line denotes the second Riemann sheet. 
   %We take the clockwise path as mentioned before.
   } 
\label{fig:integration_between_turning_points.pdf}
\end{figure}
%%%% 

Our goal is to connect the mode functions constructed to define the vacuum at $x=\infty$, approached from region I, to those approached from region III.
The evolution matrix from region I to region III for the exact WKB solutions normalized at $\tau_\pm$ is given in \eqref{eq:connection_eg}.
Our task now is to take the integral of $S_{\rm odd}$, which is defined as the closed contour integration that surrounds the two turning points (see Fig.~\ref{fig:integration_between_turning_points.pdf}),
\begin{align}
    \int_{\tau_-}^{\tau_+}S_{\rm odd} \, \dd x
    = \frac{1}{2} \oint_{\gamma_{\tau_\pm}} S_{\rm odd} \, \dd x
    = \pi E\eta \ .
\end{align}
In more detail, the integration contour can be deformed to encircle the branch cut counterclockwise on the first Riemann sheet.
This is equivalent to encircling infinity in a clockwise manner on the same sheet, allowing us to evaluate the integral using the residue theorem.%
\footnote{Although there appear to be two distinct ways to draw the integration contour between the turning points - by exchanging the solid and dotted lines and reversing the direction of the arrows - both contours yield the same result, $\frac{1}{2}\oint_{\gamma_{\tau_\pm}} S_{\rm odd} \,\dd x = \pi E\eta$. Therefore, despite this apparent ambiguity, the result is unambiguous.}

To define the mode functions, we consider the WKB solutions normalized at the asymptotic limit $x=\infty$ approached from region I and region III, respectively $\psi_\pm^{(\pm\infty)}$. 
Following the protocol in the previous subsection, we define the mode functions as
\begin{align}
   & u_+^{(\pm\infty)}=\frac{1}{\sqrt{2}} \, \psi^{(\pm\infty)}_- \ , \quad
    u_-^{(\pm\infty)}=\frac{i}{\sqrt{2}} \, \psi^{(\pm\infty)}_+ \ .
\end{align}
Combining the above results, we have
\begin{align}
\left(
\begin{array}{c}
u^{(-\infty)}_{+} \vspace{1mm}\\
u^{(-\infty)}_{-}
\end{array}
\right)
& =
\begin{pmatrix}
    0 & 1\\
    i & 0
\end{pmatrix}
\left(
\begin{array}{cc}
\exp \left( -\mathcal{V}^{(-\infty)}_{\rm voros} \right) & 0 \vspace{1mm}\\
0 & \exp \left( \mathcal{V}^{(-\infty)}_{\rm voros} \right)
\end{array}
\right)
\left(
\begin{array}{cc}
1 & 0\\
-i & 1
\end{array}
\right)
\left(
\begin{array}{cc}
\ee^{\pi E\eta} & 0 \vspace{1mm}\\
0 & \ee^{-\pi E\eta} 
\end{array}
\right)
\left(
\begin{array}{cc}
1 & i\\
0 & 1
\end{array}
\right)
\nonumber\\
& \quad \times
\left(
\begin{array}{cc}
\exp \left( \mathcal{V}^{(\infty)}_{\rm voros} \right) & 0 \vspace{1mm}\\
0 & \exp \left( -\mathcal{V}^{(\infty)}_{\rm voros} \right)
\end{array}
\right)
\begin{pmatrix}
    0 & -i\\
    1 & 0
\end{pmatrix}
 \left(
\begin{array}{c}
u^{(\infty)}_{+} \vspace{1mm}\\
u^{(\infty)}_{-}
\end{array}
\right) \\
& =
     \begin{pmatrix}
      \sqrt{1+\ee^{2\pi E\eta}} \, \ee^{i\theta}& -\ee^{\pi E\eta}\\
        -\ee^{\pi E\eta} &  \sqrt{1+\ee^{2\pi E\eta}} \,\ee^{-i\theta}
    \end{pmatrix}
 \left(
\begin{array}{c}
u^{(\infty)}_{+} \vspace{1mm}\\
u^{(\infty)}_{-}
\end{array}
\right)\ ,
\end{align}
where the explicit formula \eqref{eq:voros_coefficient_result} of $\mathcal{V}_{\rm voros}$ is plugged in in the last equality and $\theta$ is defined as,%
\footnote{$\log(z)\equiv \log(|z|)+i\arg(z)$ and $-\pi< \arg(z)\leq \pi$. 
}
\begin{align}
    \theta\equiv 
    \arg \Gamma\left(\frac{1}{2}-i E\eta\right)
    -E\eta +E\eta\log(E\eta)\ .
\end{align}
To obtain the above connection matrix, we have used the following relation,%
\footnote{
Our notation and the notation in~\cite{Shen:2006o} are related by
$\ee^{2\mathcal{V}_{\rm voros}^{(\infty)}}~({\rm Im} \, E>0)= N^2~(\arg\hbar<0)$. Also, $-E/\hbar$ in~\cite{Shen:2006o} equals to our $E\eta$.}
\begin{align}
    \big|\ee^{2\mathcal{V}_{\rm voros}^{(\infty)}}\big |=\sqrt{1+\ee^{-2\pi E\eta}}\ .
\end{align}
Including the phase redundancy of an overall global ${\rm U}(1)$ as in \eqref{eq:redundancy}, the connection matrix is expressed as 
\begin{align}
     \begin{pmatrix}
     \ee^{i(\varphi_1+\varphi_2)} \sqrt{1+\ee^{2\pi E\eta}} \, \ee^{i\theta}& -\ee^{\pi E\eta} \, \ee^{i(\varphi_1-\varphi_2)} \vspace{1mm}\\
        -\ee^{\pi E\eta} \, \ee^{-i(\varphi_1-\varphi_2)} &  \sqrt{1+\ee^{2\pi E\eta}} \, \ee^{-i\theta} \, \ee^{-i(\varphi_1+\varphi_2)}
    \end{pmatrix}\ .
\end{align}
where $\varphi_{1,2}$ are arbitrary phases which do not affect the number density.
The connection matrix indeed belongs to SU(1,1), and the result is exactly the same as the previously known results by~\cite{Kofman:1997yn,Salehian:2020dsf}.
Then the Bogoliubov coefficients are given by
\begin{align}
    \alpha = \left( u_+^{(\infty)} , \, u_+^{(-\infty)} \right)
    = \sqrt{1 + \ee^{2 \pi E \eta}} \, \ee^{-i\theta} \, \ee^{-i \left( \varphi_1 + \varphi_2 \right)} \ , \quad
    \beta = \left( u_+^{(\infty)} , \, u_-^{(-\infty)} \right)
    = \ee^{\pi E \eta} \, \ee^{i \left( \varphi_1 - \varphi_2 \right)} \ ,
\end{align}
and thus the occupation number is $n = \vert \beta \vert^2 = \exp\left( 2 \pi E \eta \right)$.

Our formulation provides a systematic method to compute the Bogoliubov coefficients that relate one vacuum to another, which capture the whole effect of particle production. It enables the direct computation without resorting to some known but limited special functions and without approximation. The formalism can be straightforwardly extended to other more complicated systems. The explicit calculation of the Voros coefficients as well as the shift of turning points may become increasingly harder as the analytic structure of the potential becomes more nontrivial. Nonetheless, we can formally obtain exact expressions for robust results using this method; we do not need to rely on approximations until the end of calculation, where one can then employ some approximations to compute those involved quantities.

\section{Conclusion}
\label{sec:conclusion}

In this paper, we have made a step forward to rigorous formulation of particle production using the exact WKB analysis. The conventional methods based on approximate WKB solutions suffer several conceptual and computational issues, such as breakdown of the approximations against exact solutions, ambiguity in defining vacuum, the WKB expansion as a divergent series, and ignorance of the global behavior of solutions.
By leveraging the Borel resummation of the WKB series and employing the exact WKB solutions, we overcome these limitations and provide a more robust framework for describing particle production processes.

One of the key facts in quantizing a system is that vacua are defined at the asymptotic points of the given Schr\"{o}dinger-type differential equation. A time-dependent background typically admits no unique vacuum, but each asymptotic point can locally host a single vacuum, provided that the Hamiltonian can be diagonalized while keeping the commutation relation. For bosonic particles, different vacua at different asymptotic points are related by a special pseudo-unitary transformation, and this operation forms an ${\rm SU}(1,1)$ group, up to an arbitrary diagonal phases in ${\rm U}(1)$.%
\footnote{In the case of fermions, the group corresponds to ${\rm SU}(2)$ instead of ${\rm SU}(1,1)$.}
For a system of $N$ boson species, we expect this to be upgraded to ${\rm U}(N,N) \, \cap \, {\rm Sp}(2N, \mathbb{C})$.

Due to this fact, particle production, which can be regarded as a change of vacua, is essentially how to connect two sets of solutions for a Schr\"{o}dinger-type equation that have different asymptotic behaviors for different asymptotic points. This is where the Voros coefficients take a crucial role. While the exact WKB analysis provides the understanding of the global behavior of exact solutions, the corresponding solutions need to be normalized at a desired asymptotic point. The connection formulae typically known in the literature, given in \eqref{eq:connection_formula_1} and \eqref{eq:connection_formula_2}, are valid for the solutions normalized at turning points, which are not where quantization is conducted. These solutions do not expand to a correct asymptotic behavior in general, unless the Voros coefficients vanish, which we never observe with the potential other than $V=-x$. Therefore, the Voros coefficients are one of the essential ingredients in constructing the connection formulae between different asymptotic points, i.e.~vacuum states, by relating the exact WKB solutions normalized at a turning point to the asymptotic mode functions. Moreover, the ${\rm SU}(1,1)$ symmetry structure discussed above cannot be retrieved unless the Voros coefficients are properly taken into account.

We have demonstrated how to quantize a given system in detail with concrete calculations. Several components are found to contribute to the mechanism of particle production: the connection formulae given in \eqref{eq:connection_formula_1} and \eqref{eq:connection_formula_2}, replacing the turning points for integration bounds as in \eqref{eq:connection_eg}, and the Voros coefficients \eqref{eq:Voros_1}. While calculating the last two ingredients for arbitrary potentials remains a nontrivial task, we have outlined strategies to address this challenge and demonstrated computations for some specific examples, including the potential of an anti-harmonic oscillator. For more complicated potentials, we may need to consider effects due to crossing branch cuts, see e.g.~\cite{Miyachi:2025ptm}.

Our work differs from previous studies on cosmological particle production that employed exact WKB analysis or the Stokes phenomenon~\cite{Kim:2013jca,Li:2019ves,Enomoto:2020xlf,Hashiba:2020rsi,Hashiba:2021npn,Enomoto:2021hfv,Yamada:2021kqw,Hashiba:2022bzi,Enomoto:2022nuj} in several key aspects. One crucial distinction is that we have rigorously formulated the quantization procedure using exact WKB solutions so that the particle production is described in a self-contained manner.
In this approach, the Voros coefficients play a central role, as they provide a precise connection between WKB solutions normalized at different points. To the best of our knowledge, these aspects have not been systematically established in previous literatures.
A primary goal of our study is to derive an exact analytical description of particle production. Achieving this requires obtaining explicit expressions for the Voros coefficients, which are essential for accurately characterizing the system’s evolution.%
\footnote{
The Voros coefficients cannot be determined only by the consistency condition of the Bogoliubov coefficients, i.e.~$|\alpha|^2-|\beta|^2=1$ in general. Besides, some undetermined phase factors can lead to larger uncertainties in repeating process when e.g.~applying to the oscillatory potential. Also, to study particle production in narrow resonance region heavily depends on the phase factors.
Moreover, when the potential does not have the parity $x\to -x$, the Voros coefficients will differ depending on asymptotic regions. In such a case, we expect that even the absolute value of the Voros coefficients cannot be determined only by the consistency condition.
}

Our formulation potentially enables a precise description of particle production without requiring full analytical solutions.
A broad spectrum of future applications can be considered. We have considered a simple system with a single degree of freedom in this paper; its multi-particle extension is an obvious direction to investigate, engaging our speculation of the corresponding symmetry structure. In considering such a multi-particle system, it can sometimes be considered equivalent to a system of a higher-order differential equation.%
\footnote{See e.g.~\cite{Aoki:higherorder1994,Aoki:thirdorder1998,Howls:book2000,Aoki_2005,MOTEKI2017327,Ito:2021boh,Enomoto:2022nuj} for a partial list for the attempt of application to higher-order differential equations.}
While we have considered bosonic systems in this paper, fermionic systems can also be studied, which falls in an ${\rm SU}(2)$ symmetry structure.
In the context of cosmological perturbations for example, non-adiabatic and tachyonic behaviors often arise as asymptotic states. In the conventional approach, it is impractical to quantize a system at such asymptotic regions. Also, the regularization of UV divergences in cosmology, such as the method of adiabatic subtraction, is rather \textit{ad hoc}, and a systematic approach is lacking. The exact WKB formulation may shed light on these types of problems.

%%%%%%%%%%%%%%%%%%%%%%%%%%%%%%%%%%%%%%%%%%%%%%%%%%
%%%%%%%%%%%%%%%%%%%%%%%%%%%%%%%%%%%%%%%%%%%%%%%%%%
\section*{Acknowledgments}
We are deeply grateful to Takashi Aoki for insightful discussions and for sharing the Mathematica code used to visualize the Stokes curves, as well as for providing invaluable guidance on aspects of the exact WKB analysis. We also thank Shofu Uchida and Takao Suzuki for patiently addressing our numerous questions and for engaging in productive discussions.
 R.S.~thanks Seishi Enomoto, Yasuyuki Hatsuda, Masazumi Honda and Tatsuhiro Misumi for multiple useful discussions. M.S.~thanks to Lingfeng Li for discussions. We also thank to Hidetoshi Taya for discussions.
 M.S. is supported by the MUR projects 2017L5W2PT.
 M.S. also acknowledges the European Union - NextGenerationEU, in the framework of the PRIN Project “Charting unexplored avenues in Dark Matter” (20224JR28W).
 Part of this work was carried out during the 2025 “The Dawn of Gravitational Wave Cosmology” workshop, supported by the Fundacion Ramon Areces and hosted by the “Centro de Ciencias de Benasque Pedro Pascual”. We thank both the CCBPP and the Fundación Areces for providing a stimulating and productive research environment.

\appendix

\section{The connection matrix of $N$ bosonic fields system belongs to $U(N,N)\cap Sp(2N,\mathbb{C})$ }
\label{app:n_system}
A coupled system with $N$ bosonic fields in a time-evolving background is described by the connection matrix,
\begin{align}
\label{eq:connection_matrix_n}
\begin{pmatrix}
    \hat {\tilde a}\\
    \hat {\tilde a}^\dagger
\end{pmatrix}
=
    \begin{pmatrix}
    \alpha & \bar\beta\\
    \beta & \bar\alpha
    \end{pmatrix}
    \begin{pmatrix}
       \hat a\\
       \hat a^\dagger
    \end{pmatrix}\ ,
\end{align}
where $\hat a,~\hat a^\dagger$ and $\hat {\tilde a},~\hat{\tilde a}^\dagger$ denote annihilation/creation operators in different basis, decribed with $(N,1)$ matrices.
$\alpha$ and $\beta$ denote the Bogoliubov coefficients expressed as $N\times N$ complex matrices. The inverse relation is given as
\begin{align}
 \begin{pmatrix}
       \hat a\\
       \hat a^\dagger
    \end{pmatrix}
=
    \begin{pmatrix}
    \alpha^\dagger & -\beta^\dagger\\
    -\beta^T & \alpha^T
    \end{pmatrix}
    \begin{pmatrix}
    \hat {\tilde a}\\
    \hat {\tilde a}^\dagger
\end{pmatrix}\ .
\end{align}
From the quantization condition, the Bogoliubov coefficients satisfy~\cite{Nilles:2001fg},
\begin{align}
\label{eq:quantization_n}
    &\alpha\alpha^\dagger-\bar\beta\beta^T=1\ ,~\alpha\beta^\dagger-\bar\beta\alpha^T=0\ ,
    \end{align}
and    
    \begin{align}
    \label{eq:quantizationn_n_2}
    &\alpha^\dagger\alpha-\beta^\dagger \beta=1\ ,~\alpha^\dagger\bar\beta-\beta^\dagger\bar\alpha=0\ .
\end{align}
We show that the connection matrix in \eqref{eq:connection_matrix_n}, subject to these quantization conditions, 
corresponds to an element of $U(N,N)\cap Sp(2N,\mathbb{C})$, and vice versa, i.e.
\begin{align}
    \hat A= \begin{pmatrix}
    \alpha & \bar\beta\\
    \beta & \bar\alpha
    \end{pmatrix}\ \text{with constraints from quantization}
    \Leftrightarrow
    \hat  A\in U(N,N)\cap Sp(2N,\mathbb{C})\ .
\end{align}

{\bf Proof of $\Rightarrow$:}
First, we verify that $ \hat  A$ belongs to $U(N,N)$. From the quantization condition \eqref{eq:quantizationn_n_2}, it follows that
\begin{align}
\label{eq:unn}
    \hat A^\dagger \eta \hat A=\eta\ ,
\end{align}
where $\eta$ is a diagonal matrix with $N$ ones and minus ones,
\begin{align}
    \eta=\text{diag}(1,..,1,-1,...,-1)\ .
\end{align} 
Similarly, we check that the connection matrix belongs to $Sp(2N,\mathbb{C})$. That is,
\begin{align}
   \hat   A^T J \hat  A=J\ ,
\end{align}
where 
\begin{align}
    J=\begin{pmatrix}
        {\bf 0} & {\bf 1}\\
        -{\bf 1} & {\bf 0}
    \end{pmatrix}\ .
\end{align}
From this condition, we obtain
\begin{align}
    \beta^T \alpha-\alpha^T\beta={\bf 0}\ ,~\alpha^\dagger\alpha-\beta^\dagger\beta={\bf 1}\ ,
\end{align}
which are consistent with the quantization conditions. Thus, we have shown that 
\begin{align}
     \begin{pmatrix}
    \alpha & \bar\beta\\
    \beta & \bar\alpha
    \end{pmatrix}
\end{align}
with quantization condition belongs to $U(N,N)\cap Sp(2N,\mathbb{C})$.

{\bf Proof of $\Leftarrow$:}
Now, we show the converse. 
Consider a matrix $\hat A$ of the form $ \hat  A=\exp(\epsilon\, T)$. In the limit of $\epsilon\to 0$, the condition in~\eqref{eq:unn} gives
\begin{align}
    T^\dagger \eta+\eta\, T=0\ .
\end{align}
Decomposing $T$ as
\begin{align}
    T=
 \begin{pmatrix}
     A & B\\
     C & D
 \end{pmatrix}\ ,
\end{align}
we obtain the conditions
\begin{align}
    A+A^\dagger=0,~D+D^\dagger=0,~C^\dagger=B\ .
\end{align}
Meanwhile, $Sp(2N,\mathbb{C})$ requires
\begin{align}
    C^T=C\ ,~B^T=B\ ,~A^T+D=0\ .
\end{align}
Together, these yield
\begin{align}
    D=\bar A,~C=B^\dagger,~A=-A^\dagger, B^T=B\ .
\end{align}
We can express these with
\begin{align}
    A=i H\ ,~ B=K+K^T\ ,
\end{align}
where $H$ is a hermitian matrix and $K$ is a $N\times N$ complex matrix.
Now, we have
\begin{align}
    \exp(\epsilon\, T)\ ,
    ~T=
    \begin{pmatrix}
        A & B\\
        \bar B & \bar A
    \end{pmatrix}\ .
\end{align}
The general form of an element in $U(N,N)\cap Sp(2N,\mathbb{C})$ is then
\begin{align}
    \exp(T)=    
    \begin{pmatrix}
    \alpha & \bar\beta\\
    \beta & \bar\alpha
    \end{pmatrix}\ .
\end{align}
Here, we used the fact 
\begin{align}
 \begin{pmatrix}
        A & B\\
        \bar B & \bar A
    \end{pmatrix}   
    \begin{pmatrix}
        E & F\\
        \bar F & \bar E
    \end{pmatrix}=
     \begin{pmatrix}
        G & H\\
        \bar H & \bar G
    \end{pmatrix}\ ,
\end{align}
where $E,~F,~G,~H$ are $N\times N$ complex matrices.
Furthermore, we also have
\begin{align}
    \det(\exp(T))=\exp(\text{tr}(T))=1\ ,
\end{align}
using $\text{tr}(T)=0$.

Consequently, we have shown that
\begin{align}
  \hat  A= \begin{pmatrix}
    \alpha & \bar\beta\\
    \beta & \bar\alpha
    \end{pmatrix}\ \text{with constraints from quantization}
    \Leftrightarrow
 \hat   A\in U(N,N)\cap Sp(2N,\mathbb{C})\ .
\end{align}

\section{More on basics of Borel sum and exact WKB analysis}
\label{app:basics_eWKB}

In this appendix, we provide additional details on the basic aspects of exact WKB analysis. For completeness, we repeat certain points already discussed in Sec.~\ref{sec:review_exact_WKB}. We also present a simple illustrative example with the potential $V(x) = -x$, which serves to demonstrate how the exact WKB method can be applied in practice.

We consider the one-dimensional Schr\"odinger equation
\begin{align}
\label{appeq:Sheq}
    \left(-\frac{\dd^2}{\dd x^2}-\eta^2 V(x,\eta)\right)\psi(x,\eta)=0\ ,
\end{align}
where $\eta$ is the expansion parameter in the WKB analysis, and $V(x,\eta)$ denotes the potential, which can be expanded as
\begin{align}
    V(x,\eta)
    =\sum_{n=0}^N \eta^{-i} V_i(x)
    = V_0(x) + \eta^{-1} V_1(x) + \eta^{-2} V_2(x) + \dots\ ,
\end{align}
with $N$ being a non-negative integer. 
In this work, we focus on the case where each $V_i(x)$ is a rational function, i.e., $V_i(x) = F_i(x)/G_i(x)$, where $F_i(x)$ and $G_i(x)$ are polynomials in $x$. We also assume that $G_0(x) V_j(x)$ is a polynomial for $j = 1, 2, \dots, N$, with $G_0(x)$ and $F_0(x)$ being coprime polynomials. Outside of this assumption, the order of the poles of some $V_i(x)$ exceeds that of $V_0(x)$, which can lead to a breakdown of the subsequent exact WKB analysis.%
\footnote{In such cases, turning points or Stokes lines defined solely by $V_0(x)$ may not be sufficient to capture the Borel summability.}

Let us assume a solution of the form $\psi = \exp R(x,\eta)$ and introduce 
\begin{align}
    S(x,\eta) \equiv \frac{\partial R}{\partial x}\,.
\end{align}
Then, the Schr\"odinger equation \eqref{appeq:Sheq} can be rewritten as
\begin{align}
    -\left(S^2 + \frac{\partial S}{\partial x}\right) - \eta^2 V = 0\ .
\end{align}
In terms of $S$, the wave function can be expressed as
\begin{align}
\label{appeq:WKB_1}
    \psi = \exp\left(\int_{x_0}^x S(x,\eta)\, \dd x\right)\ ,
\end{align}
where $x_0$ is a constant that determines the overall normalization of $\psi$.

We solve this equation by expanding $S(x,\eta)$ in powers of $\eta$,%
\footnote{Terms with higher powers of $\eta$ vanish, as implied by \eqref{appeq:Sheq}.}
\begin{align}
    S = S_{-1}(x)\,\eta + S_0(x) + S_1(x)\,\eta^{-1} + S_2(x)\,\eta^{-2} + \dots\ ,
\end{align}
and by comparing terms of like powers of $\eta$, which yields
\begin{align}
\label{appeq:Sm1}
    &S_{-1}^2 = -V_0\,,\\
\label{appeq:Sj}
    &S_{j+1} = -\frac{1}{2 S_{-1}} 
    \left(
        \frac{\dd S_j}{\dd x} + \sum_{k=0}^j S_{j-k} S_k + V_{j+2}
    \right),\quad (j=-1,0,1,2,\dots)\,,
\end{align}
with $V_j = 0$ for $j>N$.
For later convenience, we separate $S$ into terms with odd and even powers of $\eta$,
\begin{align}
\label{appeq:odd_even}
    &S = S_{\rm odd} + S_{\rm even}\,,\\
    &S_{\rm odd} \equiv \sum_{j\ge 0} S_{2j-1}\,\eta^{1-2j}, \quad
    S_{\rm even} \equiv \sum_{j\ge 0} S_{2j}\,\eta^{-2j}\,.
\end{align}
Note that this is a {\it formal} expansion; the reordering of terms does not need to be considered at this stage, although it may affect the sum once numerical values are substituted.
From \eqref{appeq:Sheq}, one finds the relation between $S_{\rm odd}$ and $S_{\rm even}$:
\begin{align}
\label{appeq:S_even}
    S_{\rm even} = -\frac{1}{2}\frac{\partial}{\partial x} \ln S_{\rm odd}\,.
\end{align}
Equation~\eqref{appeq:WKB_1} then becomes
\begin{align}
\label{appeq:WKB_2}
    \psi_\pm = \frac{1}{\sqrt{S_{\rm odd}}} \exp\left(\pm \int_{x_0}^x S_{\rm odd}\, \dd x\right)\,.
\end{align}
The two solutions, denoted by $\pm$, arise from \eqref{appeq:Sm1}, i.e., we have two choices for $S_{-1} = \pm \sqrt{-V_0}$. In what follows, we take $S_{-1} = +\sqrt{-V_0}$.  
We also note that $S_{\rm even}$ does not depend on this choice of sign, as seen in \eqref{appeq:S_even}.

In general, the WKB series does not converge. Non-convergence arises even in the simple case of $V=-x$ (see, e.g., \cite{Kawai:1998book}). However, Borel summation can provide an analytic function corresponding to the divergent WKB series~\cite{Voros:1983t}. 
Consider an infinite series in $\eta$ of the form
\begin{align}
\label{appeq:asym_series}
    \psi(\eta) = \ee^{\zeta_0 \eta} \sum_{n=0}^\infty \eta^{-n-\alpha} f_n\ ,
\end{align}
where $\zeta_0$ and $f_n$ are complex numbers, $\alpha$ is a real number with $\alpha \notin \{0,-1,-2,\dots\}$.  
The Borel transform of the series, denoted by $\psi_B(\zeta)$, is defined as
\begin{align}
\label{appeq:Boreltransdef}
    \psi_B(\zeta) = \sum_{n=0}^\infty \frac{f_n}{\Gamma(\alpha+n)} (\zeta + \zeta_0)^{\alpha+n-1}\ ,
\end{align}
where $\Gamma(x)$ is the gamma function.%
\footnote{When $f_0 = 0$, $\psi_B(\zeta)$ can also be defined for $\alpha = 0$.}  
The Borel sum of $\psi(\eta)$ is then given by
\begin{align}
\label{appeq:Borelsumdef}
    \Psi(\eta) = \int_{-\zeta_0}^\infty \ee^{-\eta \zeta} \psi_B(\zeta)\, d\zeta\ ,
\end{align}
where the integration path is taken parallel to the real axis unless otherwise specified.  
The series $\psi(\eta)$ is said to be Borel summable if the following conditions are satisfied:
\begin{enumerate}[label=(\roman*)]
    \item $(\zeta+\zeta_0)^{1-\alpha} \psi_B(\zeta) = \sum_{n=0}^\infty \frac{f_n}{\Gamma(\alpha+n)} (\zeta+\zeta_0)^n$ converges at $\zeta=-\zeta_0$;
    \item $\psi_B(\zeta)$ can be analytically continued to a domain containing $\{\zeta \in \mathbb{C} \mid {\rm Im}(\zeta+\zeta_0)=0~\text{and}~{\rm Re}(\zeta+\zeta_0)>0\}$ in the $\zeta$-plane;
    \item For sufficiently large $\eta$, the integral $\int_{-\zeta_0}^\infty \ee^{-\eta \zeta} \psi_B(\zeta)\, d\zeta$ converges to a finite value.
\end{enumerate}

The Borel summability of a WKB solution generally depends on $x$, i.e., $\zeta_0$ and $f_n$ in \eqref{appeq:asym_series} become functions of $x$.  
To systematically discuss Borel summability and the structure of the exact WKB solution, it is convenient to introduce certain distinguished points and curves in the complex $x$-plane associated with the potential, namely turning points, Stokes curves, and Stokes regions.
A turning point, denoted by $x = \tau_0$, is defined as a zero of $V_0(x)$,%
\footnote{A simple pole of $V_0(x)$ (of order one) is sometimes also called a turning point. In our subsequent analysis, such simple-pole turning points will not play a role.}  
\begin{align}
    V_0(\tau_0) = 0\ .
\end{align}
A simple turning point is further characterized by%
\footnote{This condition does not apply to turning points of the simple-pole type}
\begin{align}
    \left. \frac{\dd V_0}{\dd x}\right|_{x=\tau_0} \neq 0\ .
\end{align}
For example, in the case $V_0 = -x$, $x = 0$ is a simple turning point.  
Depending on the form of $V_0$, multiple turning points may exist.
Associated with each turning point $\tau_0$, one can define Stokes lines (or Stokes curves) by
\begin{align}
    {\rm Im} \int_{\tau_0}^x \sqrt{-V_0}\, \dd x = 0\ .
\end{align}
Along these curves, one can also evaluate the sign of the real part of the integral,
\begin{align}
\label{appeq:dominance}
    \text{Sign}\left({\rm Re} \int_{\tau_0}^x \sqrt{-V_0}\, \dd x\right)\ .
\end{align}
The sign of this quantity on the Stokes lines determines the so-called dominance relation, which plays a crucial role in analyzing the Stokes phenomenon.  
By definition, at least one endpoint of a Stokes line is a turning point. The Stokes lines divide the complex $x$-plane into distinct regions, called Stokes regions.

It is convenient to normalize the WKB solution at a turning point. For a turning point $\tau_i$ ($i=1,2,3,\dots$), we write
\begin{align}
\label{psi_turning}
    \psi_{\pm,\tau_i} = \frac{1}{\sqrt{S_{\rm odd}}} \exp\left(\pm \int_{\tau_i}^x S_{\rm odd}\, \dd x\right)\ .
\end{align}
This choice of normalization ensures that the behavior of the WKB solution is well-defined in the vicinity of the turning point and provides a natural reference for analyzing Stokes transitions.

\subsection{Example: Exact WKB analysis for $V=-x$}

\begin{figure}
\centering
\hspace{1cm}
   \includegraphics[width=0.4\linewidth]{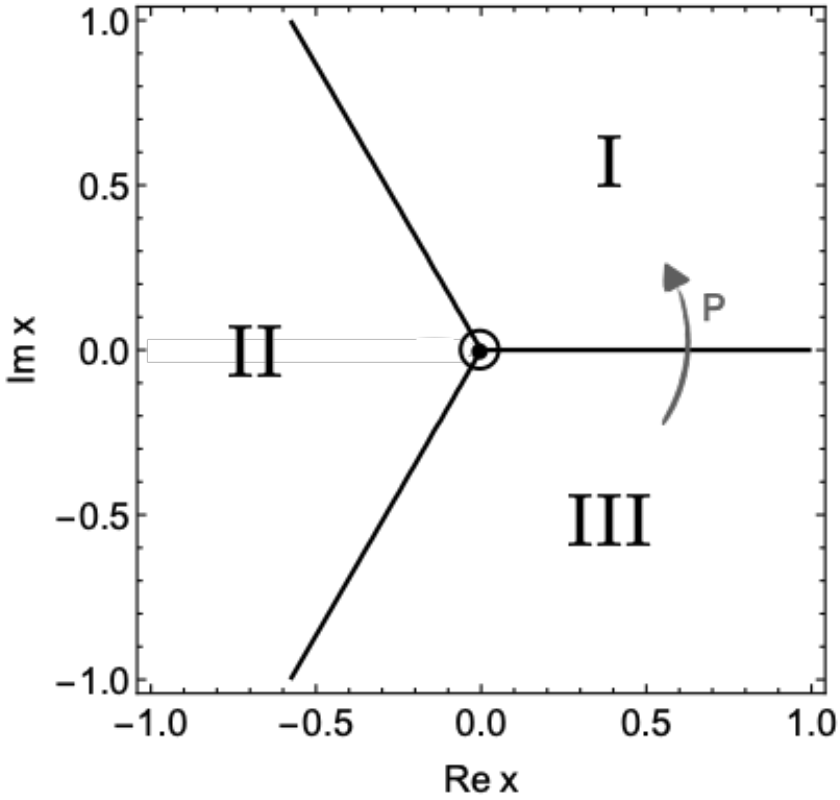}
   \vspace{0.2cm}
   \caption{The Stokes lines and Stokes regions for $V=-x$. The turning point, $x=0$, is indicated by $\odot$.
   Three Stokes lines in direction of the angle $\theta=0,2\pi/3,4\pi/3$ are denoted by solid black lines. Three Stokes regions are labeled by I,~II,~III.
   The gray arrow denoted by $P$ is used when we discuss analytical continuation of the Borel sum of the WKB solution from the Stokes region III to the Stokes region I (see the following sections for more details).}
\label{fig:x_stokes_lines_x}
\end{figure}

Let us now illustrate the exact WKB analysis using the simple potential $V=-x$.  
The simple turning point is located at $\tau_0 = 0$ as mentioned before, determined by
\begin{align}
    V(0) = 0\ , \qquad \left. \frac{\dd V}{\dd x}\right|_{x=0} \neq 0\ .
\end{align}
The Stokes lines are obtained by
\begin{align}
\label{appeq:Stokesline_x}
    0 = {\rm Im} \int_0^x \sqrt{-V}\, \dd x = \int_0^x \sqrt{x}\, \dd x = \frac{2}{3} r^{3/2} \ee^{3 i \theta/2}, \quad (x = r \ee^{i \theta})\ .
\end{align}
From $x=0$, three Stokes lines emanate in the directions $\theta = 0,~2\pi/3,~4\pi/3$, as illustrated in Fig.~\ref{fig:x_stokes_lines_x}.  
%%%
%%%% 
The Stokes lines divide the complex $x$-plane into three Stokes regions, which are labeled I, II, and III in the figure.

The WKB solution normalized at the turning point is
\begin{align}
    \psi_{\pm,0} = \frac{1}{\sqrt{S_{\rm odd}}} \exp\left(\pm \int_0^x S_{\rm odd}\, \dd x\right)\ .
\end{align}
The WKB solution is Borel summable as long as the integration path from $x=0$ to $x$ does not cross any Stokes lines.

Let us now consider the explicit form of the WKB solution and examine its Borel transform and Borel sum.  
From \eqref{appeq:Sm1} and \eqref{appeq:Sj}, we obtain the formal series for $S(x,\eta)$:  
\begin{align}
    S(x,\eta) = \eta \sqrt{x} - \frac{1}{4x} - \eta^{-1} \frac{5}{32 x^{5/2}} + \dots\ ,
\end{align}
where the coefficients $S_n(x)$ take the form $S_n(x) = c_n x^{-1-3n/2}$ with constants $c_n$.  
The corresponding WKB solution can then be expressed as the series
\begin{align}
\label{appeq:psi_dn}
    \psi_{\pm,0} = \ee^{\pm \frac{2}{3} x^{3/2} \eta} \sum_{n=0}^\infty d_{\pm,n} \, x^{- \frac{3}{2} n - \frac{1}{4}} \, \eta^{-n - \frac{1}{2}}\ .
\end{align}
The constants $d_{\pm,n}$ are determined by substituting \eqref{appeq:psi_dn} into \eqref{appeq:Sheq}, deriving a recurrence relation, and solving it with the initial condition $d_{\pm,0} = 1$.  
Explicitly, one finds
\begin{align}
    d_{\pm,n} = \frac{1}{2\pi} \left(\pm \frac{3}{4}\right)^n \frac{\Gamma(n+1/6)\, \Gamma(n+5/6)}{n!}\ .
\end{align}
It is clear that $\psi_{\pm,0}$ is a divergent series, since $d_{\pm,n}$ grows with $n$.  
To obtain a {\it well-defined} function of $x$, we apply the Borel transformation and Borel summation, which will be discussed in the remainder of this subsection.

The Borel transforms of $\psi_{\pm,0}$ are given by
\begin{align}
    &\psi_{+,B}(x,\zeta) = \left. \sqrt{\frac{3}{4\pi}} \frac{1}{x} s^{-1/2} F\Big(\frac{1}{6}, \frac{5}{6}, \frac{1}{2}; s\Big) \right|_{s = \frac{3\zeta}{4x^{3/2}} + \frac{1}{2}},\\
    &\psi_{-,B}(x,\zeta) = \left. -i \sqrt{\frac{3}{4\pi}} \frac{1}{x} (1-s)^{-1/2} F\Big(\frac{1}{6}, \frac{5}{6}, \frac{1}{2}; 1-s\Big) \right|_{s = \frac{3\zeta}{4x^{3/2}} + \frac{1}{2}},
\end{align}
where $F(\alpha,\beta,\gamma;z)$ denotes the hypergeometric function (see~\cite{Kawai:1998book} for our notation).  
Let us focus on $\psi_{+,B}(x,\zeta)$. Its Borel sum is given by
\begin{align}
    \Psi_+(x,\eta) =
    \sqrt{\frac{3}{4\pi}} \frac{1}{x} 
    \int_{-\frac{2}{3}x^{3/2}}^\infty 
    \ee^{-\eta \zeta} 
    \Bigg(\frac{3\zeta}{4x^{3/2}} + \frac{1}{2}\Bigg)^{-1/2} 
    F\Big(\frac{1}{6}, \frac{5}{6}, \frac{1}{2}; \frac{3\zeta}{4x^{3/2}} + \frac{1}{2}\Big) \, d\zeta,
\end{align}
where the integration path is taken parallel to the real axis.  
From the properties of the hypergeometric function, $\psi_{+,B}$ has singularities at
\begin{align}
\label{appeq:sing_x}
    \zeta = \pm \frac{2}{3} x^{3/2} = \pm \int_0^x \sqrt{-V}\, dx'\ .
\end{align}
A singularity lies on the integration path if
\begin{align}
    0 = {\rm Im}\, x^{3/2} \propto {\rm Im} \int_0^x \sqrt{-V}\, dx'\ .
\end{align}
This condition coincides with \eqref{appeq:Stokesline_x}, which defines the Stokes lines.

%%%%
\begin{figure}
\centering
\hspace{1cm}
   \includegraphics[width=0.4\linewidth]{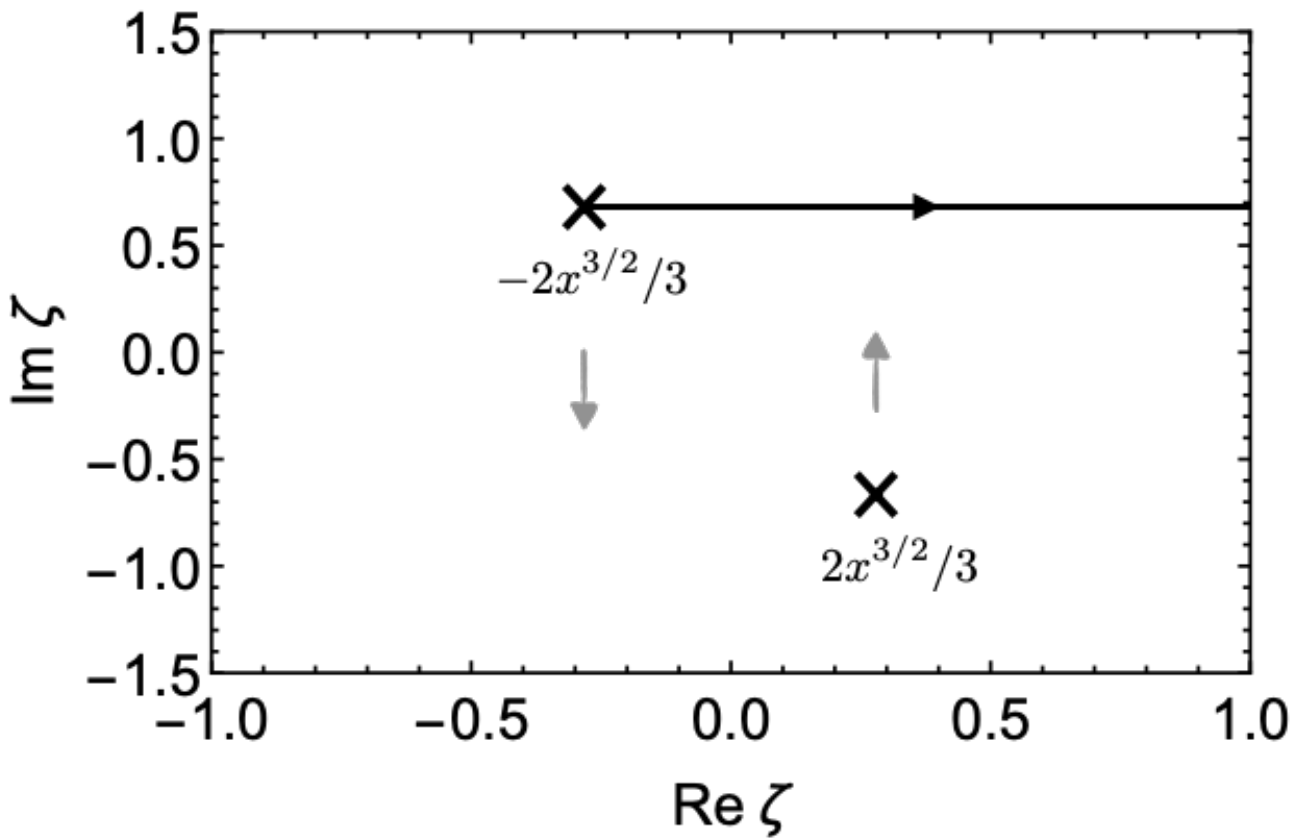}
   ~~~~~~~~~~~
    \includegraphics[width=0.4\linewidth]{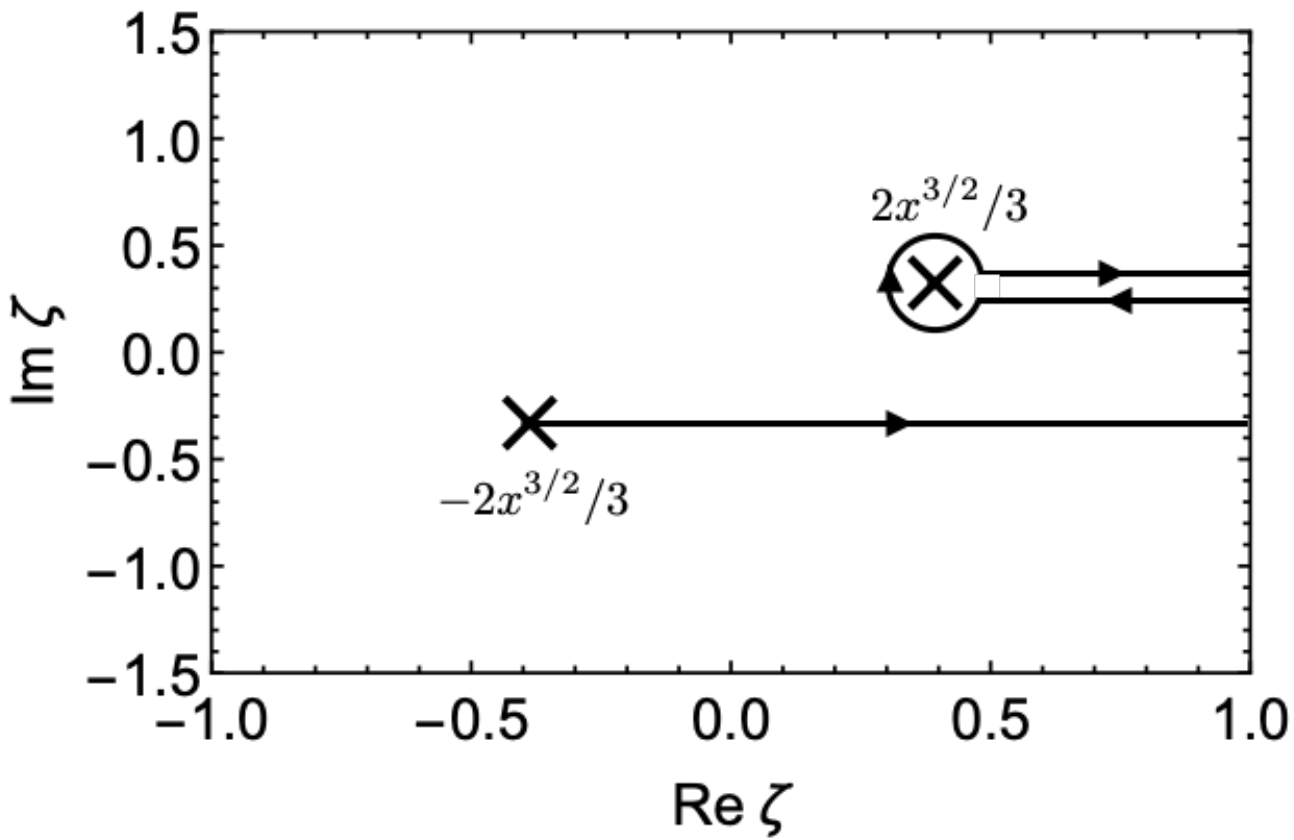}
   \vspace{0.2cm}
   \caption{
   Left: The integration path in the $\zeta$-plane for a point $x$ in Stokes region III.  
The crosses indicate the singularities at $\zeta = \pm \frac{2}{3} x^{3/2}$, and the solid lines with arrows show the integration path.  
The gray arrows indicate the motion of the singularities when $x$ moves along the path $P$ in Fig.~\ref{fig:x_stokes_lines_x}.  
Right: The integration path for the analytic continuation from Stokes region III to Stokes region I, with $x$ taken in Stokes region I.  
The solid lines with arrows denote the deformed integration path, which avoids the singularity at $\zeta = \frac{2}{3} x^{3/2}$.
} 
\label{fig:zeta_plane}
\end{figure}
%%%% 

Let us now discuss the analytic continuation of the Borel-summed WKB solution from one Stokes region to another. Further details can be found in the textbook~\cite{Kawai:1998book}.  
Suppose we consider the solution $\Psi_+^{(III)}$ in Stokes region III, as shown in Fig.~\ref{fig:x_stokes_lines_x}.  
When $x$ moves along the path denoted by $P$ in the figure, the integration contour in the $\zeta$-plane must be deformed, as illustrated in Fig.~\ref{fig:zeta_plane} (right), in order to avoid the singularity at $\zeta = \frac{2}{3} x^{3/2}$.  
The lower path, taken parallel to the real axis, gives the Borel-summed WKB solution in Stokes region I, denoted $\Psi_+^{(I)}$.  
In addition, there is an extra integration path, denoted by $\gamma$. This leads to the analytic continuation relation
\begin{align}
    \Psi_+^{(III)}(x,\eta) = \Psi_+^{(I)}(x,\eta) + \int_\gamma \ee^{-\eta \zeta} \psi_{+,B}(x,\zeta) \, d\zeta\ .
\end{align}
A similar discussion applies to $\Psi_-$.  
The second term on the right-hand side can be evaluated using the analytic continuation of the hypergeometric function.  
Applying Gauss' formula,
\begin{align*}
    F\Big(\frac{1}{6}, \frac{5}{6}, \frac{1}{2}; s\Big)
    &= \frac{\Gamma(1/2)^2}{\Gamma(1/6)\Gamma(5/6)} (1-s)^{-1/2} F\Big(\frac{1}{3}, -\frac{1}{3}, \frac{1}{2}; 1-s\Big) \\
    &\quad + \frac{\Gamma(1/2)\Gamma(-1/2)}{\Gamma(1/3)\Gamma(-1/3)} F\Big(\frac{1}{6}, \frac{5}{6}, \frac{3}{2}; 1-s\Big),
\end{align*}
together with Kummer's classical formula,
\(
    F(a,b,c;z) = (1-z)^{c-a-b} F(c-a, c-b, c; z),
\)
one finds
\begin{align}
    \int_\gamma \ee^{-\eta \zeta} \psi_{+,B}(x,\zeta) \, d\zeta = i \, \Psi_-^{(I)}(x,\eta)\ .
\end{align}

The connection formula is then given by
\begin{align}
\label{appeq:connection_III_I_+}
    \Psi_+^{(III)}(x,\eta) = \Psi_+^{(I)}(x,\eta) + i \, \Psi_-^{(I)}(x,\eta)\ .
\end{align}
This formula describes how the WKB solution for $V=-x$ in Stokes region III is analytically continued to Stokes region I.%
\footnote{In the Borel plane, the connection formula \eqref{appeq:connection_III_I_+} corresponds to the discontinuity of $\psi_{+,B}$, which can be written as
\begin{align}
    \Delta_{\zeta = \frac{2}{3}x^{3/2}} \psi_{+,B}(x,\zeta) = i \, \psi_{-,B}(x,\zeta)\ ,
\end{align}
where $\Delta_{\zeta = \frac{2}{3}x^{3/2}}$ is the so-called alien derivative, defined by
\begin{align}
    (\gamma_+ - \gamma_-) \psi_{+,B}(x,\zeta)\ .
\end{align}
Here, $\gamma_+$ denotes the analytic continuation of $\psi_{+,B}$ along a path that avoids the singularity at $\zeta = \frac{2}{3}x^{3/2}$ from above, while $\gamma_-$ avoids the singularity from below.  
The alien derivative is a powerful tool for capturing the Stokes phenomenon.
}
On the right-hand side of \eqref{appeq:connection_III_I_+}, in addition to the original WKB solution $\Psi_+^{(I)}(x,\eta)$, there is a contribution from $\Psi_-^{(I)}(x,\eta)$.  
Thus, the positive-mode solution in a given Stokes region can be expressed as a mixture of positive- and negative-mode solutions in another Stokes region.

\begin{figure}[th]
\centering
\hspace{1cm}
   \includegraphics[width=0.4\linewidth]{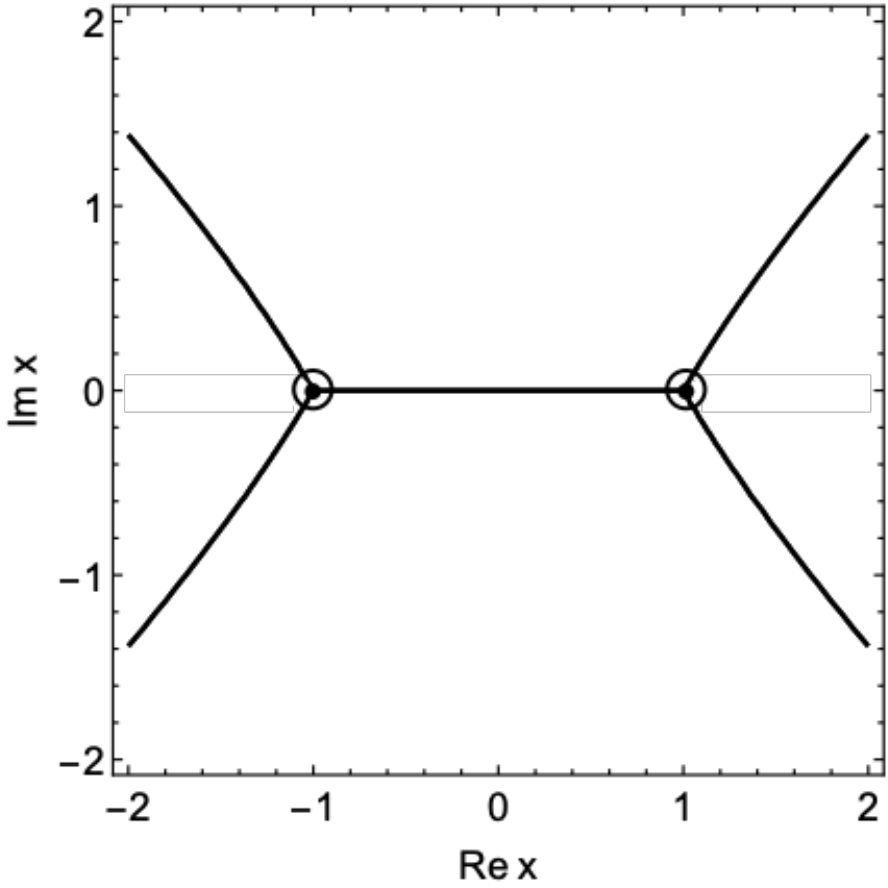}
   ~~~~~~~~~~~~~~~~~
    \includegraphics[width=0.4\linewidth]{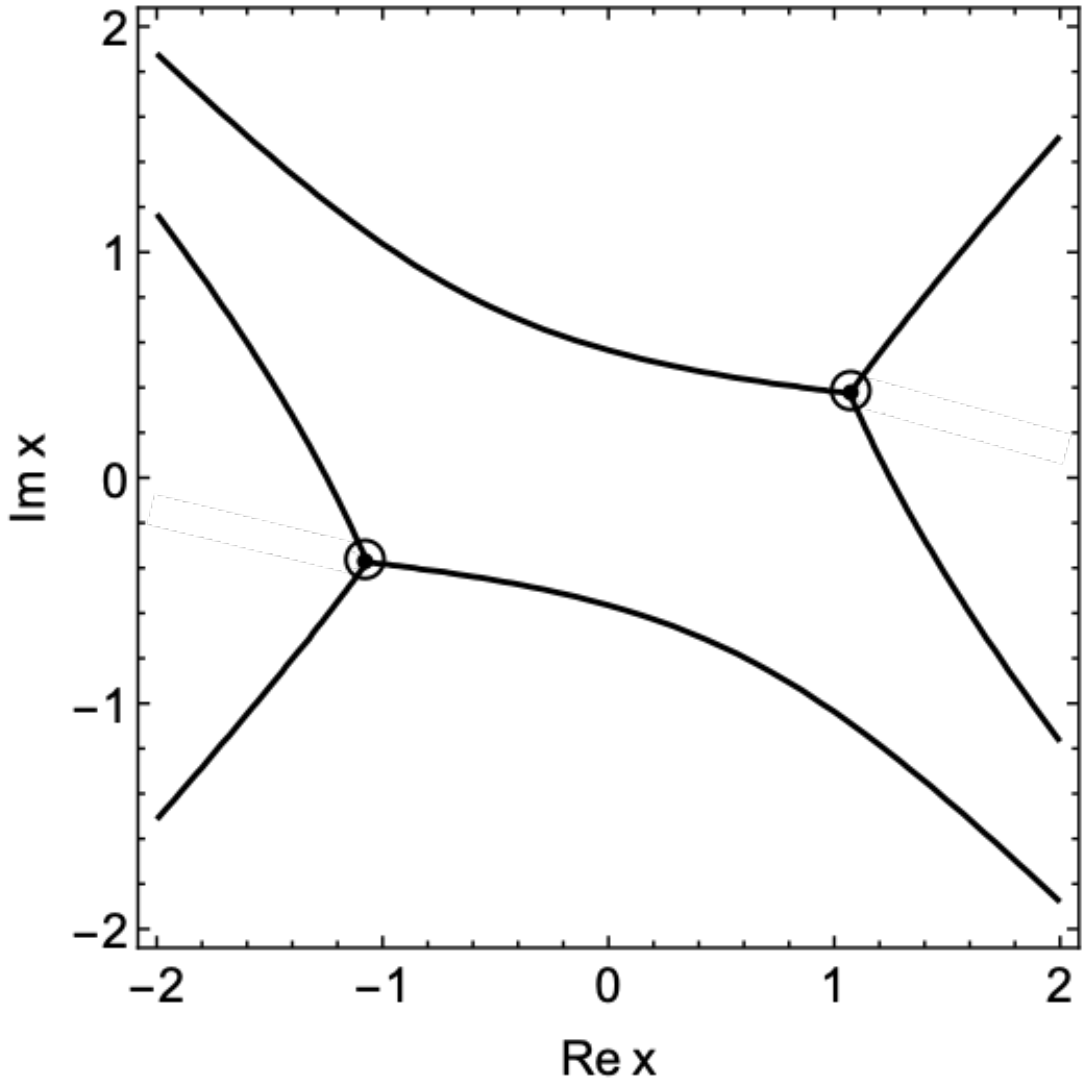}
   \vspace{0.2cm}
   \caption{The Stokes lines and Stokes regions for $V=-E+x^2/4~(E\in \mathbb{R})$ (left figure). The turning points are indicated by $\odot$.
   We also introduced small imaginary part into $E$ (right figure).
   }
\label{fig:stokes_lines}
\end{figure}

For other example potentials, we also considered $V=-E + x^2/4$ ($E \in \mathbb{R}$) in the main text.  
In this case, both endpoints of a Stokes line are turning points, as shown in Fig.~\ref{fig:stokes_lines} (left), which appears to contradict the above theorem on the Borel summability of the WKB solution.  
A simple way to resolve this issue is to introduce a small imaginary part to the potential, i.e., $V = -E + i \epsilon + x^2/4$ ($\epsilon \in \mathbb{R}$), and then take the limit $\epsilon \to 0$.  
The resulting Stokes lines for nonzero $\epsilon$ are illustrated in Fig.~\ref{fig:stokes_lines} (right), providing a consistent prescription for the Borel summation.

%%%%
%%%% 

\section{Particle production:$V=-E+x^2/4$, ${\rm Im}\,E<0$}
\label{app:negative_epsilon}

%%%%
\begin{figure}
\centering
\hspace{1cm}
   \includegraphics[width=0.4\linewidth]{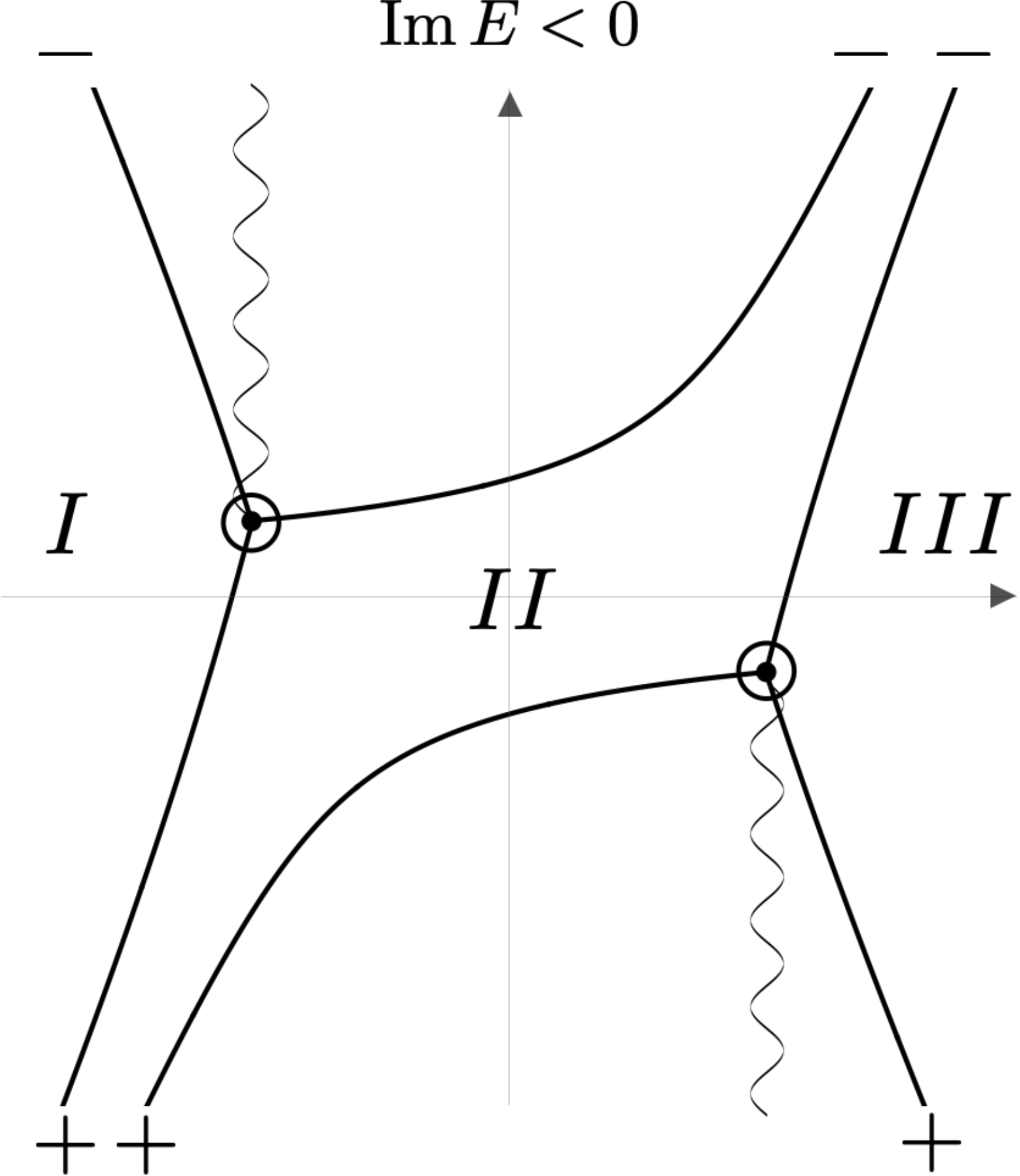}
   \caption{The Stokes curves and regions for $V=-E+x^2/4$ with ${\rm Im}(E)<0$. Three (relevant) Stokes regions are labeled by I, II, III. The branch cuts are emanating from the turning points (dot surrounded by a circle). The plus and minus sings denote the sign for the dominance relation.
   } 
\label{fig:stokes_curves_negative_imaginary}
\end{figure}
%%%% 

We compute the connection matrix for $V=-E+x^2/4$ with ${\rm Im} E<0$.
The connection matrix is
\begin{align}
    \left(
\begin{array}{c}
u^{(-\infty)}_{+}\\
u^{(-\infty)}_{-}
\end{array}
\right)
=&
\begin{pmatrix}
    0 & 1\\
    i & 0
\end{pmatrix}
\left(
\begin{array}{cc}
\exp(-V^{(-\infty)<}_{\rm voros}) & 0\\
0 & \exp(V^{(-\infty)<}_{\rm voros})
\end{array}
\right)
\left(
\begin{array}{cc}
1 & i\\
0 & 1
\end{array}
\right)
\left(
\begin{array}{cc}
\ee^{-\pi E\eta} & 0\\
0 & \ee^{\pi E\eta} 
\end{array}
\right)
\left(
\begin{array}{cc}
1 & 0\\
-i & 1
\end{array}
\right)\\
&\left(
\begin{array}{cc}
\exp(V^{(\infty)<}_{\rm voros}) & 0\\
0 & \exp(-V^{(\infty)<}_{\rm voros})
\end{array}
\right)
\begin{pmatrix}
    0 & -i\\
    1 & 0
\end{pmatrix}
 \left(
\begin{array}{c}
u^{(\infty)}_{+}\\
u^{(\infty)}_{-}
\end{array}
\right)\ .
\end{align}
Here, $V^{(\infty)<}_{\rm voros}$ denotes the Voros coefficient for ${\rm Im}E<0$.
Denoting the Voros coefficient for ${\rm Im}E>0$ as  $V^{(\infty)>}_{\rm voros}$, we have a relation,
\begin{align}
   -\overline{(V^{(\infty)<}_{\rm voros})}=V^{(\infty)>}_{\rm voros}\ .
\end{align} 
The connection matrix is reduced to,
\begin{align}
 \left(
\begin{array}{c}
u^{(-\infty)}_{+}\\
u^{(-\infty)}_{-}
\end{array}
\right)&=   \begin{pmatrix}
        \ee^{- 2\mathcal{V}_{\rm voros}^{(\infty)<}} \ee^{\pi E \eta} & -\ee^{\pi E\eta}\\
        \ee^{\pi E\eta} &  \ee^{2\mathcal{V}_{\rm voros}^{(\infty)<} } (\ee^{\pi E\eta}+\ee^{-\pi E\eta} )
    \end{pmatrix}
  \left(
\begin{array}{c}
u^{(\infty)}_{+}\\
u^{(\infty)}_{-}
\end{array}
\right)\ .
\end{align}
This is equivalent to the result with ${\rm Im}E>0$.

\section{The Voros coefficient for $V(x) = \frac{1}{4} - \frac{\nu^2 - \eta^{-2} / 4}{x^2}$}
\label{app:caseII}

We discuss the Voros coefficient of
\begin{align}
  \left[ - \frac{\dd^2}{\dd x^2} - \eta^2 V(x,\eta) \right] \psi(x) = 0 \; , \qquad
  V(x,\eta) = \frac{1}{4}-\frac{q(\nu,\eta)}{x^2} \; .
\end{align}
Here let us first consider the parametrization of $q(\nu,\eta)$. In a way, $\eta$ is merely a book-keeping parameter for the WKB expansion, and when we actually solve the differential equation (numerically or whatever), we may consider setting $\eta=1$. Keeping this in mind, we know by experience of cosmological perturbations that a good parametrization is
\begin{align}
  \label{eq:50}
  q(\nu,1) = \nu^2 - \frac{1}{4} \; .
\end{align}
By solving the equation \eqref{eq:eom_case2} for the asymptotic behaviors, we find
\begin{subequations}
  \label{eq:asymptotic_behavior_case2}
  \begin{align}
    x \to \infty & \quad \implies \quad
                   V \simeq \frac{1}{4} \quad \implies \quad
                   \psi \sim {\rm e}^{\pm i x/2} \; ,  \\
    x \to 0 & \quad \implies \quad
              V \simeq -\frac{\nu^2 - 1/4}{x^2} \quad \implies \quad
              \psi \sim x^{\frac{1}{2} \pm \nu} \; ,
  \end{align}
\end{subequations}
while in order for the WKB solutions to reproduce these asymptotic behaviors we are advised to set
\begin{align}
  \label{eq:52}
  q(\nu,\eta) = \nu^2 - \frac{\eta^{-2}}{4} \; .
\end{align}
Then we see the correct asymptotic solutions in \eqref{eq:asymptotic_behavior_case2} can be obtained via the WKB method (and taking $\eta=1$ at the end).
Thus the system of equation we would like to consider here is
\begin{align}
  \label{eq:eom_case2}
  \left[ - \frac{\dd^2}{\dd x^2} - \eta^2 V(x,\eta) \right] \psi(x) = 0 \; , \qquad
  V(x,\eta) = \frac{1}{4} - \frac{\nu^2}{x^2} + \frac{\eta^{-2}}{4 x^2} \; .
\end{align}

In this system, we have
\begin{align}
  \mbox{Turning points: } & \; x = \pm 2 \nu \equiv x_\pm \; , \\
  \mbox{Regular singuluarity: } & \; x = 0 \; , \\
  \mbox{Irregular singularity: } & \; x = \infty \; .
\end{align}
The branch cut runs from $x_+$ along the positive real axis, to $+\infty$, which is identified as $-\infty$, and then to $x_-$ along the negative real axis.
Let us take the branch of $S_{-1}(x)$ such that the first Riemann sheet takes
\begin{align}
  \label{eq:branch_case2}
  {\rm e}^{-i \pi/2} \sqrt{-\frac{1}{4} + \frac{\nu^2}{x^2}}
  > 0 \; , \qquad
  \mbox{ for } \;
  \nu > 0 \; , \;
  x > 2\nu \; .
\end{align}

The WKB solution normalized at the turning point is thus
\begin{align}
  \psi_{\pm,x_\pm} (x)
  = \exp\left[ \int_{x_\pm}^x S^{(\pm)}(x') \, \dd x' \right]
  = \frac{1}{\sqrt{S_{\rm odd}(x)}} \, \exp\left[ \pm \int_{x_\pm}^x S_{\rm odd}(x') \, \dd x' \right] \; ,
\end{align}
where
\begin{align}
  S_{\rm odd}(x)
  & = \eta \, Q_0^{1/2}
    + \eta^{-1} \, \frac{- x^2 - 16 \nu^2}{2^7 x^4 Q_0^{5/2}}
    + \eta^{-3} \, \frac{-25 x^6 - 1824 \nu^2 x^4 - 8960 \nu^4 x^2 - 4096 \nu^6}{2^{15} x^{10} Q_0^{11/2}} + \dots \; , \\
  S_{\rm even}(x)
  & = \frac{\nu^2}{2 x^3 Q_0}
    + \eta^{-2} \, \frac{x^4 + 40 \nu^2 x^2 + 64 \nu^4}{2^9 x^7 Q_0^4}
    + \eta^{-4} \, \frac{13 x^8 + 1496 \nu^2 x^6 + 14784 \nu^4 x^4 + 23552 \nu^6 x^2 + 4096 \nu^8 }{2^{15} x^{13} Q_0^7} \dots \; ,
\end{align}
and
\begin{align}
  S_{\rm odd}(x) \, \dd x/ \dd y
  & = \eta \, \frac{- Q_0^{1/2}}{y^2}
    + \eta^{-1} \, \frac{ 1 + 16 \nu^2 y^2}{2^7 Q_0^{5/2}}
    + \eta^{-3} \, \frac{y^2 \left( 25 + 1824 \nu^2 y^2 + 8960 \nu^4 x^4 + 4096 \nu^6 y^6 \right)}{2^{15} Q_0^{11/2}} + \dots \; , \\
  S_{\rm even}(x) \, \dd x/ \dd y
  & = \frac{- \nu^2 y}{2 Q_0}
    + \eta^{-2} \, \frac{- y \left( 1 + 40 \nu^2 y^2 + 64 \nu^4 y^4 \right)}{2^9 Q_0^{4}}\nonumber\\
  &  + \eta^{-4} \, \frac{- y^3 \left( 13 + 1496 \nu^2 y^2 + 14784 \nu^4 y^4 + 23552 \nu^6 y^6 + 4096 \nu^8 y^8 \right)}{2^{15} Q_0^7} \dots \; ,
\end{align}
with with $Q_0 \equiv - \frac{1}{4} + \frac{\nu^2}{x^2} = - \frac{1}{4} + \nu^2 y^2$ and $y = 1/x$.
Note that there is a relation
\begin{align}
  \label{eq:Sj_symm_case2}
  S_j(- x) =
  \begin{cases}
    S_j(x) \; , & \qquad \mbox{for  odd } \, j \; , \vspace{1mm}\\
    - S_j(x) \; , & \qquad \mbox{for even} \, j \; ,
  \end{cases}
\end{align}
and thanks to this property, we have~
\begin{align}
  \label{eq:89}
  & \psi_{\pm , x_{+}}(-x)
    = \exp \left[ \int^x_{x_-} S^{(\mp)}(x') \, \dd x' \right]
    = \frac{1}{\sqrt{S_{\rm odd}(x)}} \exp \left[ \mp \int^x_{x_-} S_{\rm odd}(x') \, \dd x' \right]
    = \psi_{\mp , x_-}(x) \; ,
\end{align}
and similarly $\psi_{\pm , x_-}(-x) = \psi_{\mp , x_+}(x)$.
The diverging terms at the singular points are,
\begin{align}
  \label{eq:51}
  S_{-1} \; , \;\; S_0 \qquad & @ \quad x \to 0 \; , \nonumber\\
  S_{-1} \qquad & @ \quad  x \to \infty \; .
\end{align}
Hence we define the WKB solution normalized at $x=0, \infty$ as
\begin{align}
  \psi_{\pm, x_\pm}^{(0)}(x,\nu,\eta)
  & \equiv \frac{\exp\left[ \pm \int_{x_\pm}^x \eta S_{-1}(x') \, \dd x' \right]}{\sqrt{S_{\rm odd}(x)}} \, \exp\left[ \pm \int_{0}^x \left( S_{\rm odd}(x') - \eta S_{-1}(x') \right) \dd x' \right] \; , \\
  \psi_{\pm,x_\pm}^{(\infty)}(x,\nu,\eta)
  & \equiv \frac{\exp\left[ \pm \int_{x_\pm}^x \eta S_{-1}(x') \, \dd x' \right]}{\sqrt{S_{\rm odd}(x)}} \, \exp\left[ \pm \int_{\infty}^x \left( S_{\rm odd}(x') - \eta S_{-1}(x') \right) \dd x' \right] \; ,
\end{align}
where in each case the integral path approaches to the corresponding singular points such that it does not cross the Stokes curves.
Then we define the Voros coefficients as
\begin{align}
  \label{eq:voroscoeff_case2}
  \mathcal{V}^{(0)}_{x_\pm}(\nu, \eta) & \equiv \int_{x_\pm}^0 \left[ S_{\rm odd}(x', \nu^2, \eta) - \eta S_{-1}(x',\nu^2) \right] \dd x' \; , \\
  \mathcal{V}^{(\infty)}_{x_\pm}(\nu, \eta) & \equiv \int_{x_\pm}^\infty \left[ S_{\rm odd}(x', \nu^2, \eta) - \eta S_{-1}(x',\nu^2) \right] \dd x' \; .
\end{align}
Thanks to the symmetry in \eqref{eq:Sj_symm_case2}, we have the relation
\begin{align}
  \label{eq:53}
  \mathcal{V}^{(0)}_{x_+}(\nu,\eta) & = - \mathcal{V}^{(0)}_{x_-}(\nu,\eta) \equiv \mathcal{V}^{(0)}(\nu,\eta) \; , \\
  \mathcal{V}^{(\infty)}_{x_+}(\nu,\eta) & = - \mathcal{V}^{(\infty)}_{x_-}(\nu,\eta) \equiv \mathcal{V}^{(\infty)}(\nu,\eta)
\end{align}
Then the relations of $\psi^{(0,\infty)}$ to the turning-point-normalized $\psi$ are
\begin{align}
  \label{eq:55}
  \psi_{+, x_\pm}(x,\nu,\eta) & = {\rm e}^{\pm \mathcal{V}^{(0)}(\nu,\eta)} \, \psi^{(0)}_{+, x_\pm}(x,\nu,\eta) \; , \qquad
  \psi_{-, x_\pm}(x,\nu,\eta) = {\rm e}^{\mp \mathcal{V}^{(0)}(\nu,\eta)} \, \psi^{(0)}_{-, x_\pm}(x,\nu,\eta) \; , \\
  \psi_{+, x_\pm}(x,\nu,\eta) & = {\rm e}^{\pm \mathcal{V}^{(\infty)}(\nu,\eta)} \, \psi^{(\infty)}_{+, x_\pm}(x,\nu,\eta) \; , \qquad
  \psi_{-, x_\pm}(x,\nu,\eta) = {\rm e}^{\mp \mathcal{V}^{(\infty)}(\nu,\eta)} \, \psi^{(\infty)}_{-, x_\pm}(x,\nu,\eta) \; .
\end{align}

We can write the Voros coefficients as
\begin{subequations}
  \label{eq:voroscoeff_all_case2}
  \begin{align}
    \mathcal{V}^{(0)}(\nu,\eta)
    & = \int_{x_+}^0 \left[ S^{(+)}(x,\nu^2,\eta) - \eta S_{-1}(x,\nu^2) - S_0(x,\nu^2) \right] \dd x
      = \int_{x_-}^0 \left[ S^{(-)}(x,\nu^2,\eta) + \eta S_{-1}(x,\nu^2) - S_0(x,\nu^2) \right] \dd x \; , \\
    \mathcal{V}^{(\infty)}(\nu,\eta)
    & = \int_{x_+}^\infty \left[ S^{+}(x,\nu^2,\eta) - \eta S_{-1}(x,\nu^2) - S_0(x,\nu^2) \right] \dd x
      = \int_{x_-}^\infty \left[ S^{-}(x,\nu^2,\eta) + \eta S_{-1}(x,\nu^2) - S_0(x,\nu^2) \right] \dd x \; .
  \end{align}
\end{subequations}

Now let us construct a ladder operator $\mathcal{L}$ that does the action that, for $\psi(x,\nu,\eta)$ satisfying
\begin{align}
  \label{eq:57}
  \left[ - \frac{\dd^2}{\dd x^2} - \eta^2 V(x, \nu, \eta) \right] \psi(x,\nu,\eta) = 0 \; ,
\end{align}
the following equation holds
\begin{align}
  \label{eq:58}
  \left[ - \frac{\dd^2}{\dd x^2} - \eta^2 V(x, \nu + c, \eta) \right] \left[ \mathcal{L} \, \psi(x,\nu,\eta) \right] = 0 \; ,
\end{align}
where $c$ is a constant to be determined.
Note that the shift is on $\nu$, not on $\nu^2$, which happens to be crucial to have a proper action of the ladder operator.
This implies the operator commutation
\begin{align}
  \label{eq:ladder_commutation_case2}
  \mathcal{L} \, H(x,\nu,\eta) - H(x, \nu + c, \eta) \, \mathcal{L} = 0 \; , \qquad
  H(x,\nu,\eta) \equiv - \frac{\dd^2}{\dd x^2} + \eta^2 Q(x, \nu, \eta) \; .
\end{align}
We make an ansatz
\begin{align}
  \label{eq:60}
  \mathcal{L} = \frac{\dd}{\dd x} + \tilde{c} \, x^\beta \; ,
\end{align}
where $\tilde{c}$ and $\beta$ are constants to be fixed.
The raising/lowering operators $\mathcal{L}_\pm$ are obtained as
\begin{align}
  \label{eq:64}
  & \mathcal{L}_\pm \equiv \frac{\dd}{\dd x} - \frac{\frac{1}{2} \pm \eta \nu}{x} \; , \\
  & H(x, \nu \pm \eta^{-1}, \eta) \left[ \mathcal{L}_\pm \psi(x,\nu,\eta) \right] = 0 \; .
\end{align}
Since the operators $\mathcal{L}_\pm$ do not change the dominant relation of $\psi_\pm$, the following relation must hold
\begin{align}
  \label{eq:65}
  \mathcal{L}_\pm \psi(x,\nu,\eta) = \left[ S^{(+)}(x,\nu,\eta) - \frac{\frac{1}{2} \pm \eta \nu}{x} \right] \psi(x,\nu,\eta) = C(\nu,\eta) \, \psi(x,\nu \pm \eta^{-1}, \eta) \; .
\end{align}
Then taking a logarithmic derivative of both sides, we obtain
\begin{align}
  \label{eq:differenceeq_S_case2}
  \Delta_{\nu}^{(\pm)} S^{(+)}(x,\nu,\eta) = \frac{\dd}{\dd x} \log \left[ S^{(+)}(x,\nu,\eta) - \frac{\frac{1}{2} \pm \eta \nu}{x} \right] \; ,
\end{align}
where
\begin{align}
  \label{eq:67}
  \Delta_{\nu}^{(\pm)}S^{(+)}(x,\nu,\eta) \equiv  S^{(+)}(x,\nu \pm \eta^{-1},\eta) - S^{(+)}(x, \nu, \eta)
  = \left[ \exp\left( \pm \eta^{-1} \partial_{\nu} \right) - 1 \right] S^{(+)}(x,\nu,\eta) \; .
\end{align}

Define
\begin{subequations}
  \label{eq:def_Ix_case2}
  \begin{align}
    I^x_{x_\pm}(x,\nu,\eta)
    & \equiv \int_{x_\pm}^x S^{(+)}(x',\nu,\eta) \, \dd x'
      = \frac{1}{2} \int_{\gamma_\pm^x} S^{(+)}(x',\nu,\eta) \, \dd x' \; , \\
    I^x_{j, x_\pm}(x,\nu)
    & \equiv \int_{x_\pm}^x S_j(x',\nu) \, \dd x'
      = \frac{1}{2} \int_{\gamma_\pm^x} S_j(x',\nu) \, \dd x'\; ,
  \end{align}
\end{subequations}
where the paths $\gamma_\pm^x$ of the integral are taken as usual on the first and second Riemann sheets.
Then from \eqref{eq:voroscoeff_all_case2}, the Voros coefficients can be written as
\begin{align}
  \label{eq:54}
  \mathcal{V}^{(0)}(\nu,\eta) & = \lim_{x \to 0} \left[ I^x_{x_+}(x,\nu,\eta) - \eta I^x_{-1,x_+}(x,\nu) - I^x_{0,x_+}(x,\nu) \right] \; , \\
  \mathcal{V}^{(\infty)}(\nu,\eta) & = \lim_{x \to \infty} \left[ I^x_{x_+}(x,\nu,\eta) - \eta I^x_{-1,x_+}(x,\nu) - I^x_{0,x_+}(x,\nu) \right] \; ,
\end{align}
where the limit should be taken along the path that does not cross the Stokes curves.
Now we aim to compute the difference equations
\begin{align}
  \label{eq:69}
  \Delta_{\nu}^{(+)} \mathcal{V}^{(0)}(\nu,\eta) & = \lim_{x \to 0} \left[ \Delta_{\nu}^{(+)}  I^x_{x_+}(x,\nu,\eta) - \eta \Delta_{\nu}^{(+)} I^x_{-1,x_+}(x,\nu) - \Delta_{\nu}^{(+)} I^x_{0,x_+}(x,\nu) \right] \; , \\
  \Delta_{\nu}^{(+)} \mathcal{V}^{(\infty)}(\nu,\eta) & = \lim_{x \to \infty} \left[ \Delta_{\nu}^{(+)} I^x_{x_+}(x,\nu,\eta) - \eta \Delta_{\nu}^{(+)} I^x_{-1,x_+}(x,\nu) - \Delta_{\nu}^{(+)} I^x_{0,x_+}(x,\nu) \right] \; .
\end{align}
Using the difference equation for $S^{(+)}$ in \eqref{eq:differenceeq_S_case2}, we have
\begin{align}
  \label{eq:66}
  \Delta_{\nu}^{(+)} I^x_{x_\pm}(x,\nu,\eta)
  & = \int^x_{x_\pm} \Delta_{\nu}^{(+)} S^{(+)}(x',\nu,\eta) \, \dd x' \nonumber\\
  & = \frac{1}{2} \log \left[ \frac{S^{(+)}(x,\nu,\eta)  - \frac{\frac{1}{2} + \eta \nu}{x}}{S^{(-)}(x,\nu,\eta)  - \frac{\frac{1}{2} + \eta \nu}{x}} \right] \; ,
\end{align}
where $\tilde x$ denotes the point in the second Riemann sheet.
Keeping in mind our choice of the branch in \eqref{eq:branch_case2}, around the singular points, we have
\begin{align}
  \label{eq:68}
  \begin{cases}
    \displaystyle
    S_{-1} \simeq 
    \left( \frac{\nu}{x} - \frac{x}{8\nu} - \frac{x^3}{128 \nu^3} \right) + \mathcal{O}(x^5) \vspace{1mm} \\
    \displaystyle
    S_0 \simeq \frac{1}{2x} + \frac{x}{8\nu^2} + \frac{x^3}{32 \nu^4} + \mathcal{O}(x^5)
    \vspace{1mm} \\
    \displaystyle
    S_1 \simeq 
    \left( - \frac{x}{8 \nu^3} - \frac{11 x^3}{128 \nu^5} \right) + \mathcal{O}(x^5)
    \vspace{1mm} \\
    \displaystyle
    S_2 \simeq \frac{x}{8 \nu^4} + \frac{13 x^3}{64 \nu^6} + \mathcal{O}(x^5)
    \vspace{1mm} \\
    \displaystyle
    S_3 \simeq 
    \left( - \frac{x}{8 \nu^5} - \frac{57 x^3}{128 \nu^7} \right) + \mathcal{O}(x^5)
    \vspace{1mm} \\
    \displaystyle
    S_4 \simeq \frac{x}{8 \nu^6} + \frac{15 x^3}{16 \nu^8} + \mathcal{O}(x^5)
  \end{cases}
  \quad , \qquad
  x \to 0 \; ,
\end{align}
and 
\begin{align}
  \label{eq:68}
  \begin{cases}
    \displaystyle
    S_{-1} \simeq i \left( \frac{1}{2} - \frac{\nu^2}{x^2} - \frac{\nu^4}{x^4} \right) + \mathcal{O}(x^{-6})
    \vspace{1mm} \\
    \displaystyle
    S_0 \simeq - \frac{2\nu^2}{x^3} - \frac{8 \nu^4}{x^5} + \mathcal{O}(x^{-7})
    \vspace{1mm} \\
    \displaystyle
    S_1 \simeq i \left( \frac{1}{4x^2} + \frac{13 \nu^2}{2 x^4} \right) + \mathcal{O}(x^{-6})
    \vspace{1mm} \\
    \displaystyle
    S_2 \simeq \frac{1}{2x^3} + \frac{28 \nu^2}{x^5} + \mathcal{O}(x^{-7})
    \vspace{1mm} \\
    \displaystyle
    S_3 \simeq - i \, \frac{25}{16 x^4} + \mathcal{O}(x^{-6})
    \vspace{1mm} \\
    \displaystyle
    S_4 \simeq - \frac{13}{2 x^5} + \mathcal{O}(x^{-7})
  \end{cases}
  \quad , \qquad
  x \to \infty \; .
\end{align}
Then, assuming $\nu \in \Re$ and $\nu > 0$,
we have 
\begin{align}
  \label{eq:70}
  S^{(+)}(x,\nu,\eta) - \frac{\frac{1}{2} + \eta\nu}{x}
  & \simeq - \frac{\eta x}{8 \nu} \left[1 - \left( \eta \nu \right)^{-1} + \left( \eta \nu \right)^{-2} - \left( \eta \nu \right)^{-3} + \left( \eta \nu \right)^{-4} - \left( \eta \nu \right)^{-5} + \dots \right] \nonumber\\
  & = - \frac{\eta x}{8 \nu} \sum_{n=0}^\infty \left( - \eta \nu \right)^{-n}
    = - \frac{\eta x}{8 \nu} \, \frac{\eta \nu}{1 + \eta \nu} \nonumber\\
  & = - \frac{\eta^2 x}{8 \left( 1 + \eta \nu \right)} \; ,
    \qquad
    x \to 0 \; ,\\
  S^{(-)}(x,\nu,\eta) - \frac{\frac{1}{2} + \eta\nu}{x}
  & \simeq - \frac{2 \eta \nu}{x} \; , \qquad
    x \to 0 \; .
\end{align}
On the other hand,
\begin{align}
  \label{eq:72}
  S^{(+)}(x,\nu,\eta) - \frac{\frac{1}{2} + \eta\nu}{x}
  & \simeq \frac{i \eta}{2} - \frac{\frac{1}{2} + \eta\nu}{x} \; ,
    \qquad
    x \to \infty \; ,\\
  S^{(-)}(x,\nu,\eta) - \frac{\frac{1}{2} + \eta\nu}{x}
  & \simeq - \frac{i \eta}{2} - \frac{\frac{1}{2} + \eta\nu}{x} \; , \qquad
    x \to \infty \; .
\end{align}
Therefore
\begin{align}
  \label{eq:73}
  \Delta_{\nu}^{(+)} I_{x_\pm}^x(x,\nu,\eta)
  & \simeq \frac{1}{2} \log \left[ \frac{\eta x^2}{16 \nu \left( 1 + \eta \nu \right)} \right] \; , \qquad
    x \to 0 \; , \\
  \Delta_{\nu}^{(+)} I_{x_\pm}^x(x,\nu,\eta)
  & \simeq \frac{1}{2} \log \left( \frac{ 2\eta\nu + 1 - i \eta x}{2\eta\nu + 1 + i \eta x} \right) \; , \qquad
    x \to \infty \; .
\end{align}

For $\Delta_{\nu}^{(+)} I^x_{0,x_\pm}$, since $S_{\rm even}$ is a single-valued function, the integration paths $\gamma_\pm^x$  are in fact closed contours. Among the terms in $S_{\rm even}$, only $S_0$ has poles at $x=0,x_\pm$, and in fact
\begin{align}
  \label{eq:74}
  \mathop{\rm Res}_{x=0} S_0 = \frac{1}{2} \; , \qquad
  \mathop{\rm Res}_{x=x_\pm} S_0 = \frac{- 2 \nu^2}{\left( \pm 2 \nu \right)^2 \pm 2 \left( \pm 2 \nu \right) \nu} = - \frac{1}{4} \; ,
\end{align}
and they do not depend on $\nu$. Therefore,
\begin{align}
  \label{eq:75}
  \Delta_{\nu}^{(+)} I^x_{0,x_\pm} = 0 \; .
\end{align}

Now for $\Delta_{\nu}^{(+)} I^x_{-1,x_\pm}$, we perform the direct evaluation of the integral
\begin{align}
  \label{eq:76}
  I^x_{-1,x_\pm}(x,\nu,\eta)
  & = \int_{x_\pm}^x \sqrt{- \frac{1}{4} + \frac{\nu^2}{x'{}^2}} \, \dd x'
    \nonumber\\
      \nonumber\\
  & = x \sqrt{- \frac{1}{4} + \frac{\nu^2}{x^2}}
    - \frac{\nu}{2} \left[ \log \left( 1 + \frac{x}{\nu} \sqrt{- \frac{1}{4} + \frac{\nu^2}{x^2}} \right)
    - \log \left( 1 - \frac{x}{\nu} \sqrt{- \frac{1}{4} + \frac{\nu^2}{x^2}} \right) \right] \; .
\end{align}
Since
\begin{align}
  \label{eq:71}
  \sqrt{- \frac{1}{4} + \frac{\nu^2}{x^2}}
  \simeq
  \begin{cases}
    \displaystyle
    \sqrt{\frac{\nu^2}{x^2}}
    - \frac{1}{8} \left( \frac{\nu^2}{x^2} \right)^{-1/2}
    - \frac{1}{128} \left( \frac{\nu^2}{x^2} \right)^{-3/2}
    + \dots
    = \frac{\nu}{x} - \frac{x}{8\nu} - \frac{x^3}{128 \nu^3} 
    + \dots \; , \qquad
    & x \to 0 \; ,
    \vspace{2mm}\\
    \displaystyle
    i \left( \frac{1}{2} - \frac{\nu^2}{x^2} - \frac{\nu^4}{x^4} \right)
    + \dots \; , \qquad
    & x \to \infty \; ,
  \end{cases}
\end{align}
where we have assumed $\nu > 0$, we have
\begin{align}
  \label{eq:77}
  I^x_{-1,x_\pm}(x,\nu,\eta)
  & \simeq
    \begin{cases}
      \displaystyle
      \nu - \frac{\nu}{2} \log \left( \frac{16 \nu^2}{x^2} \right) \; , \qquad
      & x \to 0 \; ,
        \vspace{2mm}\\
      \displaystyle
      \frac{i x}{2} - \frac{\nu}{2} \log \left( \frac{i x + 2\nu}{- i x + 2\nu} \right) \; , \qquad
      & x \to \infty \; ,
    \end{cases}
\end{align}
and hence
\begin{align}
  \label{eq:78}
  \Delta_\nu^{(+)} I^x_{-1,x_\pm}(x,\nu,\eta)
  & \simeq
    \begin{cases}
      \displaystyle
      \frac{1}{\eta}
      - \nu \log \left(1 + \frac{1}{\eta\nu} \right)
      - \frac{1}{2\eta} \log \left( \frac{16 \left( 1 + \eta \nu \right)^2}{\eta^2 x^2} \right) \; , \qquad
      & x \to 0 \; ,
        \vspace{2mm}\\
      \displaystyle
      - \frac{1}{2\eta} \log \left( \frac{i \eta x + 2 \left( \eta \nu + 1 \right)}{- i \eta x + 2 \left( \eta \nu + 1 \right)} \right) \; , \qquad
      & x \to \infty \; .
    \end{cases}
\end{align}

Combining the above results, we obtain
\begin{align}
  \label{eq:79}
  \Delta_\nu^{(+)} \mathcal{V}^{(0)}(\nu,\eta)
  & = \lim_{x \to 0} \left[
    \frac{1}{2} \log \left[ \frac{\eta x^2}{16 \nu \left( 1 + \eta \nu \right)} \right]
    - 1
    + \eta\nu \log \left(1 + \frac{1}{\eta\nu} \right)
    + \frac{1}{2} \log \left( \frac{16 \left( 1 + \eta \nu \right)^2}{\eta^2 x^2} \right)
    \right]
  \\
  & = \frac{1}{2} \log \left( 1 + \frac{1}{\eta\nu} \right) + \eta\nu \log \left( 1 + \frac{1}{\eta\nu} \right) - 1 \; ,
\end{align}
and
\begin{align}
  \label{eq:79}
  \Delta_\nu^{(+)} \mathcal{V}^{(\infty)}(\nu,\eta)
  & = \lim_{x \to \infty} \left[
    \frac{1}{2} \log \left( \frac{ 2\eta\nu + 1 - i \eta x}{2\eta\nu + 1 + i \eta x} \right)
    + \frac{1}{2} \log \left( \frac{i \eta x + 2 \left( \eta \nu + 1 \right)}{-i \eta x + 2 \left( \eta \nu + 1 \right)} \right)
    \right]    
  \\
  & = 0 \; .
\end{align}

Now let's consider Borel-transforming the Voros coefficient $\mathcal{V}^{(0,\infty)}$.
Notice that there is a scaling relation
\begin{align}
  \label{eq:80}
  S_j (\nu x, \nu) = \nu^{-j - 1} S_j(x,1) \; ,
\end{align}
and this tells us
\begin{align}
  \label{eq:81}
 \int_{\pm 2 \nu}^{0,\infty} S_j(x,\nu) \, \dd x
  = \nu^{-j} \int_{\pm 2}^{0,\infty} S_j(t,1) \, \dd t \; ,
\end{align}
where $t = \nu^{-1} x$, which results in
\begin{align}
  \mathcal{V}^{(0,\infty)}(\nu,\eta)
  & = \int_{2 \nu}^{0,\infty} \left[ S_{\rm odd}(x,\nu,\eta) - \eta S_{-1}(x,\nu) \right] \dd x \nonumber\\
  & = \sum_{j=1}^\infty \mathcal{V}_{2j-1}^{(0,\infty)}(\nu) \, \eta^{-2j+1} \nonumber\\
  & = \sum_{j=1}^\infty \mathcal{V}_{2j-1}^{(0,\infty)}(1) \, \left( \eta \nu \right)^{-2j+1} \; ,
    \label{eq:Voros_series_case2}
\end{align}
where $\mathcal{V}_j^{(0,\infty)}(\nu) \equiv \int_{\pm 2\nu}^{0,\infty} S_j(x,\nu) \, \dd x$.
Hence,
\begin{align}
  \label{eq:83}
  \mathcal{V}^{(0,\infty)}(\nu+\eta^{-1},\eta)
  & = \sum_{j=1}^\infty \mathcal{V}_{2j-1}^{(0,\infty)}(1) \left[ \eta \left( \nu + \eta^{-1} \right) \right]^{-2j+1} \nonumber\\
  & = \sum_{j=1}^\infty \mathcal{V}_{2j-1}^{(0,\infty)}(1) \left[ \left( \eta + \nu^{-1} \right) \nu \right]^{-2j+1}
    \nonumber\\
  & = \mathcal{V}^{(0,\infty)}(\nu, \eta + \nu^{-1}) \; .
\end{align}
Therefore, we have
\begin{align}
  \label{eq:84}
  \Delta_\eta^{(+)} \mathcal{V}^{(0)}(\nu, \eta)
  & = \Delta_\nu^{(+)} \mathcal{V}^{(0)}(\nu, \eta)
    = \frac{1}{2} \log \left( 1 + \frac{1}{\eta\nu} \right) + \eta\nu \log \left( 1 + \frac{1}{\eta\nu} \right) - 1 \; , \\
  \Delta_\eta^{(+)} \mathcal{V}^{(\infty)}(\nu, \eta)
  & = 0
\end{align}
where
\begin{align}
  \label{eq:85}
  \Delta_\eta^{(+)} \mathcal{V}^{(0,\infty)}(\nu, \eta)
  \equiv \mathcal{V}^{(0,\infty)}(\nu, \eta + \nu^{-1}) - \mathcal{V}^{(0,\infty)}(\nu, \eta)
  = \left( {\rm e}^{\nu^{-1} \partial_\eta} - 1 \right)  \mathcal{V}^{(0,\infty)}(\nu, \eta)
\end{align}
As in a similar manner with the one in the main text, the Borel transformation of \eqref{eq:84} leads to
\begin{align}
  \label{eq:42}
  \left( {\rm e}^{- y / \nu} - 1 \right)
  & \, \mathcal{V}_B^{(0)}(\nu,y)
    = - \frac{1}{2 y} \left( {\rm e}^{-y/\nu} - 1 \right)
    + \frac{\nu}{y^2} \left( {\rm e}^{-y/\nu} - 1 \right) + \frac{{\rm e}^{-y/\nu}}{y} \; , \nonumber \\
  \implies
  & \mathcal{V}_B^{(0)}(\nu,y) = - \frac{1}{y} \left( \frac{1}{{\rm e}^{y/\nu} - 1} - \frac{\nu}{y} + \frac{1}{2} \right) \; , 
\end{align}
also we have
\begin{align}
      & \mathcal{V}_B^{(\infty)}(\nu,y) = 0 \; .
\end{align}
Finally, using (5) of Section 1.9 in~\cite{Erdelyi:HTF1}, we can compute the Laplace transform to get the Borel sum of the Voros coefficients, $\mathscr{V}^{(0)}$, giving
\begin{align}
  \label{eq:47}
  \mathscr{V}^{(0)}(\nu,\eta)
  & = \int_0^\infty \mathcal{V}_B^{(0)}(\nu,y) \, {\rm e}^{- \eta y} \, \dd y
    = - \int_0^\infty \frac{1}{t} \left( \frac{1}{{\rm e}^{t} - 1} - \frac{1}{t} + \frac{1}{2} \right) \, {\rm e}^{-  \eta \nu t} \, \dd t \nonumber\\
  & = - \log \frac{\Gamma(\eta \nu)}{\sqrt{2\pi}} + \left( \eta \nu - \frac{1}{2} \right) \log \eta\nu - \eta\nu \; , \qquad
    {\rm Re} \left( \eta\nu \right) > 0 \; , \\
  \mathscr{V}^{(\infty)}(\nu,\eta)
  & = 0 \; .
\end{align}
From \eqref{eq:Voros_series_case2}, we see the parity relation
\begin{align}
  \label{eq:82}
  \mathcal{V}^{(0,\infty)}(-\eta\nu) = - \mathcal{V}^{(0,\infty)}(\eta\nu) \; ,
\end{align}
and we have the result in the domain of ${\rm Re} \left( \eta\nu \right) < 0$:
\begin{align}
  \label{eq:47}
  \mathscr{V}^{(0)}(\nu,\eta)
  & =
    \begin{cases}
      \displaystyle
       - \log \frac{\Gamma(\eta \nu)}{\sqrt{2\pi}} + \left( \eta \nu - \frac{1}{2} \right) \log \eta\nu - \eta\nu \; , \qquad
      & {\rm Re} \left( \eta\nu \right) > 0 \; , \vspace{2mm} \\
      \displaystyle
      \log \frac{\Gamma(- \eta \nu)}{\sqrt{2\pi}} + \left( \eta \nu + \frac{1}{2} \right) \log ( - \eta\nu ) - \eta\nu \; , \qquad
      & {\rm Re} \left( \eta\nu \right) < 0 \; ,
    \end{cases}
  \\
  \mathscr{V}^{(\infty)}(\nu,\eta)
  & = 0 \; .
\end{align}
This concludes the calculation of the Voros coefficients in the case of $Q = - \frac{1}{4} + \frac{\nu^2 - \eta^{-2}/4}{x^2}$.

\section{The Voros coefficient for $V(x) = 1 - \frac{2\xi}{x}
  $}
\label{app:caseIII}
Now consider
\begin{align}
  \label{eq:eom_case3}
  \left[ - \frac{\dd^2}{\dd x^2} - \eta^2 V(x) \right] \psi(x) = 0 \; , \qquad
  V(x) = 1 - \frac{2\xi}{x} \; .
\end{align}
The leading-order WKB solutions match the exact solution in the asymptotic region \( x \to \infty \). However, they fail to do so near \( x \to 0 \), where the WKB approximation breaks down due to 
\( S_{-1} \ll S_1 \). 
As a result, the WKB approximation is valid in the intermediate region \( \frac{3}{64|\xi|} \ll |x| \ll 2|\xi| \).

In this system, we have
\begin{align}
  \mbox{Turning point: } & \; x = 2 \xi \equiv x_t \; , \\
  \mbox{Pole of order 1: } & \; x = 0 \; , \\
  \mbox{Irregular singularity: } & \; x = \infty \; .
\end{align}
The branch cut runs between $x_t$ and $0$.
Let us take the branch of $S_{-1}(x)$ such that
\begin{align}
  \label{eq:branch_case3}
  {\rm e}^{-i \pi/2} \sqrt{-1 + \frac{2\xi}{x}}
  = {\rm e}^{-i \pi/2} \frac{\sqrt{2\xi - x}}{\sqrt{x}} > 0 \; , \qquad
  \mbox{ for } \;
  \xi > 0 \; , \;
  x > 2 \xi \; .
\end{align}
Note that the pole $x=0$ behaves more like a turning point than a singular point, and three Stokes curves (which can be degenerate depending on parameter values, though) are attached to it.
The connection formula for crossing a Stokes curve coming out of $x=0$ has a different form compared to the usual one~\cite{Ko2}.

The WKB solution normalized at the turning point is 
\begin{align}
  \psi_{\pm} (x)
  = \exp\left[ \int_{2\xi}^x S^{(\pm)}(x') \, \dd x' \right]
  = \frac{1}{\sqrt{S_{\rm odd}(x)}} \, \exp\left[ \pm \int_{2\xi}^x S_{\rm odd}(x') \, \dd x' \right] \; .
\end{align}
Since
\begin{align}
  S_{\rm odd}(x)
  & = \eta \frac{\sqrt{2\xi - x}}{\sqrt{x}}
    + \eta^{-1} \frac{\xi \left( -4 x + 3 \xi \right)}{8 x^{3/2} \left( 2 \xi - x \right)^{5/2}}
    + \eta^{-3} \frac{- \xi \left( 192 x^3 - 176 \xi x^2 + 168 \xi^2 x - 63 \xi^3 \right)}{128 x^{5/2} \left( 2\xi - x \right)^{11/2}}
    + \dots \; , \\
  S_{\rm even}(x)
  & = \frac{\xi}{2 x \left( 2\xi - x \right)}
    + \eta^{-2} \frac{3 \xi \left( 2x^2 - 2 \xi x + \xi^2 \right)}{8 x^2 \left( 2\xi - x \right)^4}
    + \eta^{-4} \frac{3\xi \left( 40 x^4 - 24 \xi x^3 + 42 \xi^2 x^2 - 30 \xi^3 x + 9 \xi^4 \right)}{32 x^3 \left( 2\xi - x \right)^7} \dots \; ,
\end{align}
we have
\begin{align}
  S_{\rm odd}(x) \, \dd x/\dd y
  & = \eta \frac{- \sqrt{2\xi y - 1}}{y^2}
    + \eta^{-1} \frac{\xi y \left( 4 - 3 \xi y \right)}{8 \left( 2 \xi y - 1 \right)^{5/2}}
    + \eta^{-3} \frac{\xi y^3 \left( 192 - 176 \xi y + 168 \xi^2 y^2 - 63 \xi^3 y^3 \right)}{128 \left( 2\xi y - 1 \right)^{11/2}}
    + \dots \; , \\
  S_{\rm even}(x) \, \dd x/\dd y
  & = \frac{- \xi}{2 \left( 2\xi y - 1 \right)}
    + \eta^{-2} \frac{- 3 \xi y^2 \left( 2 - 2 \xi y + \xi^2 y^2 \right)}{8 \left( 2\xi y - 1 \right)^4}
    + \eta^{-4} \frac{- 3\xi y^4 \left( 40 - 24 \xi y + 42 \xi^2 y^2 - 30 \xi^3 y^3 + 9 \xi^4 y^4 \right)}{32 \left( 2\xi y - 1 \right)^7} \dots \; ,
\end{align}
where $y = 1/x$.
The diverging term at the singular point $x = \infty$ is only,
\begin{align}
  \label{eq:51}
  S_{-1} \qquad & @ \quad  x \to \infty \; .
\end{align}
We define the WKB solution normalized at $x=0, \infty$ as
\begin{align}
  \psi_{\pm}^{(0)}(x,\xi,\eta)
  & \equiv \frac{1}{\sqrt{S_{\rm odd}(x)}} \, \exp\left[ \pm \int_{0}^x S_{\rm odd}(x') \, \dd x' \right] \; , \\
  \psi_{\pm}^{(\infty)}(x,\xi,\eta)
  & \equiv \frac{\exp\left[ \pm \int_{2\xi}^x \eta S_{-1}(x') \, \dd x' \right]}{\sqrt{S_{\rm odd}(x)}} \, \exp\left[ \pm \int_{\infty}^x \left( S_{\rm odd}(x') - \eta S_{-1}(x') \right) \dd x' \right] \; ,
\end{align}
where in each case the integral path approaches to the corresponding singular points such that it does not cross the Stokes curves.
While the pole $x=0$ behaves like a turning point as mentioned earlier, we define the Voros coefficients as
\begin{align}
  \label{eq:voroscoeff_case3}
  \mathcal{V}^{(0)}(\xi, \eta) & \equiv \int_{2\xi}^0 S_{\rm odd}(x', \xi, \eta) \, \dd x' \; , \\
  \mathcal{V}^{(\infty)}(\xi, \eta) & \equiv \int_{2\xi}^\infty \left[ S_{\rm odd}(x', \xi, \eta) - \eta S_{-1}(x',\xi) \right] \dd x' \; .
\end{align}
Then the relations of $\psi^{(0,\infty)}$ to the turning-point-normalized $\psi$ are
\begin{align}
  \label{eq:55}
  \psi_{\pm}(x,\xi,\eta) & = {\rm e}^{\pm \mathcal{V}^{(0)}(\xi,\eta)} \, \psi^{(0)}_{\pm}(x,\xi,\eta) \; , \\
  \psi_{\pm}(x,\xi,\eta) & = {\rm e}^{\pm \mathcal{V}^{(\infty)}(\xi,\eta)} \, \psi^{(\infty)}_{\pm}(x,\xi,\eta) \; .
\end{align}
Recall that the integral path for $2 \int_{2\xi}^x \dd x$ is defined such that it starts at the point in $2$nd Riemann sheet corresponding to $x$, goes on the same sheet to $2\xi$, crosses the branch cut to the $1$st Riemann sheet, circles around $2\xi$, and then goes to $x$ on the $1$st Riemann sheet. Let us call this path by $\gamma_x$.
Since each term in $S_{\rm even}(x)$ is a single-valued function in $x$, the path is in fact a closed contour. Since, among the terms in $S_{\rm even}(x)$, only $S_0$ has poles, which are at $x=0, 2\xi$.
In particular, $S_0$ does not have a pole in the region encircled by the same contour but outside.
Therefore, we observe
\begin{align}
  \int^0_{2\xi} S_{\rm even}(x) \, \dd x
  & = \frac{1}{2} \oint_{\gamma_0} S_{\rm even}(x) \, \dd x
    = \frac{1}{2} \oint_{\gamma_0} S_{0}(x) \, \dd x
    = 0 \; , \\
  \int^\infty_{2\xi} S_{\rm even}(x) \, \dd x
  & = \frac{1}{2} \oint_{\gamma_\infty} S_{\rm even}(x) \, \dd x
    = \frac{1}{2} \oint_{\gamma_\infty} S_{0}(x) \, \dd x\ .
\end{align}
Using these we can write the Voros coefficients as
\begin{subequations}
  \label{eq:voroscoeff_all_case3}
  \begin{align}
    \mathcal{V}^{(0)}(\xi,\eta) & = \int_{2\xi}^0 S^{+}(x,\xi,\eta) \, \dd x \; , \\
    \mathcal{V}^{(\infty)}(\xi,\eta) & = \int_{2\xi}^\infty \left[ S^{+}(x,\xi,\eta) - \eta S_{-1}(x,\xi) - S_0(x,\xi) \right] \dd x \; .
  \end{align}
\end{subequations}
In fact $\mathcal{V}^{(0)}$ can be computed immediately. Since the cut is between $0$ and $2\xi$, the path $\gamma_0$ is a closed contour not only for $S_{\rm even}$ but also for $S_{\rm odd}$. Moreover, only $S_{-1}$ has a residue at $x=\infty$, we obtain
\begin{align}
  \label{eq:88}
  \mathcal{V}^{(0)}(\xi,\eta)
  = \frac{1}{2} \oint_{\gamma_0} S^{+}(x,\xi,\eta) \, \dd x
  = \pm \pi i \left[ - \eta \mathop{\rm Res}_{x=\infty} S_{-1}(x,\xi,\eta) \right]
  = \pm \pi \eta \xi \; ,
\end{align}
where the square root is taken with respect to the branch choice \eqref{eq:branch_case3}.
Here the plus and minus signs correspond to the direction of $\gamma_0$ being counterclockwise and clockwise, respectively.

Now we only want to compute $\mathcal{V}^{(\infty)}$. First let us construct a ladder operator $\mathcal{L}$ that does the action that, for $\psi(x,\xi,\eta)$ satisfying
\begin{align}
  \left[ - \frac{\dd^2}{\dd x^2} + \eta^2 Q(x, \xi, \eta) \right] \psi(x,\xi,\eta) = 0 \; ,
\end{align}
the following equation holds
\begin{align}
  \left[ - \frac{\dd^2}{\dd x^2} + \eta^2 Q(x, \xi + c, \eta) \right] \left[ \mathcal{L} \, \psi(x,\xi,\eta) \right] = 0 \; ,
\end{align}
where $c$ is a constant to be determined.
This implies the operator commutation
\begin{align}
  \label{eq:ladder_commutation_case3}
  \mathcal{L} \, H(x,\xi,\eta) - H(x, \xi + c, \eta) \, \mathcal{L} = 0 \; , \qquad
  H(x,\xi,\eta) \equiv - \frac{\dd^2}{\dd x^2} + \eta^2 Q(x, \xi, \eta) \; .
\end{align}
We make an ansatz
\begin{align}
  \mathcal{L} = x^\alpha \, \frac{\dd}{\dd x} + \tilde{c} \, x^\beta + b \; .
\end{align}
The
raising/lowering operators $\mathcal{L}_\pm$ are obtained as
\begin{align}
  & \mathcal{L}_\pm \equiv x \, \frac{\dd}{\dd x} \pm i \eta \left( x - \xi \right) \; , \\
  & H(x, \xi \pm i \eta^{-1}, \eta) \left[ \mathcal{L}_\pm \psi(x,\xi,\eta) \right] = 0 \; .
\end{align}
Since the operators $\mathcal{L}_\pm$ do not change the dominant relation of $\psi_\pm$, the following relation must hold
\begin{align}
  \label{eq:65}
  \mathcal{L}_\pm \psi(x,\xi,\eta) = \left[ x S^{(+)}(x,\xi,\eta) \pm i \eta \left( x - \xi \right) \right] \psi(x,\xi,\eta) = C(\xi,\eta) \, \psi(x,\xi \pm i \eta^{-1}, \eta) \; .
\end{align}
Then taking a logarithmic derivative of both sides, we obtain
\begin{align}
  \label{eq:differenceeq_S_case3}
  \Delta_{\xi}^{(\pm)} S^{(+)}(x,\xi,\eta) = \frac{\dd}{\dd x} \log \left[ x S^{(+)}(x,\xi,\eta) \pm i \eta \left( x - \xi \right) \right] \; ,
\end{align}
where
\begin{align}
  \Delta_{\xi}^{(\pm)}S^{(+)}(x,\xi,\eta) \equiv  S^{(+)}(x,\xi \pm i \eta^{-1},\eta) - S^{(+)}(x, \xi, \eta)
  = \left[ \exp\left( \pm i \eta^{-1} \partial_{\xi} \right) - 1 \right] S^{(+)}(x,\xi,\eta) \; ,
\end{align}
in the same sense as the Taylor expansion.

Define
\begin{subequations}
  \label{eq:def_Ix_case3}
  \begin{align}
    I^x(x,\xi,\eta)
    & \equiv \int_{2\xi}^x S^{(+)}(x',\xi,\eta) \, \dd x'
      = \frac{1}{2} \int_{\gamma_x} S^{(+)}(x',\xi,\eta) \, \dd x' \; , \\
    I^x_{j}(x,\xi)
    & \equiv \int_{2\xi}^x S_j(x',\xi) \, \dd x'
      = \frac{1}{2} \int_{\gamma_x} S_j(x',\xi) \, \dd x'\; ,
  \end{align}
\end{subequations}
where the paths $\gamma_x$ of the integral are taken as usual on the first and second Riemann sheets.
Then from \eqref{eq:voroscoeff_all_case3}, the Voros coefficient $\mathcal{V}^{(\infty)}$ can be written as~%
\begin{align}
  \label{eq:54}
  \mathcal{V}^{(\infty)}(\xi,\eta) & = \lim_{x \to \infty} \left[ I^x(x,\xi,\eta) - \eta I^x_{-1}(x,\xi) - I^x_{0}(x,\xi) \right] \; ,
\end{align}
where the limit should be taken along the path that does not cross the Stokes curves.
Now we aim to compute the difference equation
\begin{align}
  \label{eq:69}
  \Delta_{\xi}^{(+)} \mathcal{V}^{(\infty)}(\xi,\eta) & = \lim_{x \to \infty} \left[ \Delta_{\xi}^{(+)} I^x(x,\xi,\eta) - \eta \Delta_{\xi}^{(+)} I^x_{-1}(x,\xi) - \Delta_{\xi}^{(+)} I^x_{0}(x,\xi) \right] \; .
\end{align}
Using the difference equation for $S^{(+)}$ in \eqref{eq:differenceeq_S_case3}, we have
\begin{align}
  \label{eq:66}
  \Delta_{\xi}^{(+)} I^x(x,\xi,\eta)
  & = \frac{1}{2} \int_{\gamma_x} \Delta_{\xi}^{(+)} S^{(+)}(x',\xi,\eta) \, \dd x' \nonumber\\
  & = \frac{1}{2} \log \left[ x S^{(+)}(x,\xi,\eta) + i\eta \left( x - \xi \right) \right]
    - \frac{1}{2} \log \left[ \tilde x S^{(+)}(\tilde x,\xi,\eta) + i\eta \left( \tilde x - \xi \right) \right] \nonumber\\
  & = \frac{1}{2} \log \left[ \frac{x S^{(+)}(x,\xi,\eta) + i\eta \left( x - \xi \right)}{x S^{(-)}(x,\xi,\eta) + i\eta \left( x - \xi \right)} \right] \; ,
\end{align}
where $\tilde x$ is the point on the $2$nd Riemann sheet that corresponds to $x$ on the $1$st Riemann sheet.
Keeping in mind our choice of the branch in \eqref{eq:branch_case3}, around the singular point at $x=\infty$, we have
\begin{align}
  \begin{cases}
    \displaystyle
    S_{-1} \simeq i \left( 1 - \frac{\xi}{x} - \frac{\xi^2}{2x^2} - \frac{\xi^3}{2x^3} - \frac{5 \xi^4}{8x^4} \right) + \mathcal{O}(x^{-5})
    \vspace{1mm} \\
    \displaystyle
    S_0 \simeq - \frac{\xi}{2x^2} - \frac{\xi^2}{x^3} - \frac{2\xi^3}{x^4} + \mathcal{O}(x^{-5})
    \vspace{1mm} \\
    \displaystyle
    S_1 \simeq i \left( \frac{\xi}{2x^3} + \frac{17 \xi^2}{x^4} \right) + \mathcal{O}(x^{-5})
    \vspace{1mm} \\
    \displaystyle
    S_2 \simeq \frac{3\xi}{4x^4} + \mathcal{O}(x^{-5})
  \end{cases}
  \quad , \qquad
  x \to \infty \; .
\end{align}
Then we have
\begin{align}
  x S^{(+)}(x,\nu,\eta) + i \eta \left( x - \xi \right)
  & \simeq 2 i \eta \left( x - \xi \right)
    \simeq 2 i \eta x \; ,
    \qquad
    x \to \infty \; ,\\
  x S^{(-)}(x,\nu,\eta) + i \eta \left( x - \xi \right)
  & \simeq \frac{\xi}{2x} \left( i \eta \xi - 1 \right) \; , \qquad
    x \to \infty \; .
\end{align}
Therefore
\begin{align}
  \Delta_{\xi}^{(+)} I^x(x,\xi,\eta)
  & \simeq \frac{1}{2} \log \left[ \frac{4 i \eta x^2}{\xi \left( i \eta \xi - 1 \right)} \right] \; , \qquad
    x \to \infty \; .
\end{align}

For $\Delta_{\xi}^{(+)} I^x_{0}$, since $S_{\rm even}$ is a single-valued function, the integration path $\gamma_x$ is in fact closed contours. Among the terms in $S_{\rm even}$, only $S_0$ has a pole at $x=0$, and in fact
\begin{align}
  \mathop{\rm Res}_{x=0} S_0(x) = \frac{1}{4} \; , \qquad
\end{align}
and it does not depend on $\xi$. Therefore,
\begin{align}
  \label{eq:75}
  \Delta_{\xi}^{(+)} I^x_{0} = 0 \; .
\end{align}

Now for $\Delta_{\xi}^{(+)} I^x_{-1}$, we perform the direct evaluation of the integral
\begin{align}
  \label{eq:76}
  I^x_{-1}(x,\xi,\eta)
  & = \int_{2\xi}^x \sqrt{-1 + \frac{\xi}{x'}} \, \dd x'
    \nonumber\\
  & = i x \sqrt{1 - \frac{2\xi}{x}} - i \xi \left[ \log \left( 1 + \sqrt{1 - \frac{2\xi}{x}} \right) - \log \left( 1 - \sqrt{1 - \frac{2\xi}{x}} \right) \right]
    \nonumber\\
  & \simeq i \left( x - \xi \right) - i \xi \left[ \log \left( 2 - \frac{\xi}{x} \right) - \log \frac{\xi}{x} \right]
    \nonumber\\
  & \simeq i \left( x - \xi \right) - i \xi \log \frac{2x}{\xi} \; , \qquad
    x \to \infty \; ,
\end{align}
and hence
\begin{align}
  \label{eq:78}
  \Delta_\xi^{(+)} I^x_{-1}(x,\xi,\eta)
  & \simeq \eta^{-1} + \eta^{-1} \log \frac{2x}{\xi + i \eta^{-1}} + i \xi \log \frac{\xi + i \eta^{-1}}{\xi}
    \; , \qquad x \to \infty
\end{align}

Combining the above results, we obtain
\begin{align}
  \label{eq:79}
  \Delta_\xi^{(+)} \mathcal{V}^{(\infty)}(\xi,\eta)
  & = \lim_{x \to \infty} \left[
    \frac{1}{2} \log \left[ \frac{4 \eta x^2}{\xi \left( \eta \xi + i \right)} \right]
    - 1
    - \log \left( \frac{2 \eta x}{\eta \xi + i} \right)
    - i \eta \xi \log \left( 1 + \frac{i}{\eta \xi} \right)
    \right]
  \\
  & = \frac{1}{2} \log \left( 1 - \frac{1}{i \eta \xi} \right) - i \eta \xi \log \left( 1 - \frac{1}{i \eta \xi} \right) - 1 \; .
\end{align}

Now let's consider Borel-transforming the Voros coefficient $\mathcal{V}^{(\infty)}$.
Notice that there is a scaling relation
\begin{align}
  S_j (\xi x, \xi) = \xi^{-j - 1} S_j(x,1) \; ,
\end{align}
and this tells us
\begin{align}
  \int_{2\xi}^{\infty} S_j(x,\xi) \, \dd x
  = \xi^{-j} \int_{2}^{\infty} S_j(t,1) \, \dd t \; ,
\end{align}
where $t = \xi^{-1} x$, which results in
\begin{align}
  \mathcal{V}^{(\infty)}(\xi,\eta)
  & = \int_{2\xi}^{\infty} \left[ S_{\rm odd}(x,\xi,\eta) - \eta S_{-1}(x,\xi) \right] \dd x \nonumber\\
  & = \sum_{j=1}^\infty \mathcal{V}_{2j-1}^{(\infty)}(\xi) \, \eta^{-2j+1} \nonumber\\
  & = \sum_{j=1}^\infty \mathcal{V}_{2j-1}^{(\infty)}(1) \, \left( \eta \xi \right)^{-2j+1} \; ,
    \label{eq:Voros_series_case3}
\end{align}
where $\mathcal{V}_j^{(\infty)}(\xi) \equiv \int_{2\xi}^{\infty} S_j(x,\xi) \, \dd x$.
Hence,
\begin{align}
  \mathcal{V}^{(\infty)}(\xi+ i \eta^{-1},\eta)
  & = \sum_{j=1}^\infty \mathcal{V}_{2j-1}^{(\infty)}(1) \left[ \eta \left( \xi + i \eta^{-1} \right) \right]^{-2j+1} \nonumber\\
  & = \sum_{j=1}^\infty \mathcal{V}_{2j-1}^{(\infty)}(1) \left[ \left( \eta + i \xi^{-1} \right) \xi \right]^{-2j+1}
    \nonumber\\
  & = \mathcal{V}^{(\infty)}(\xi, \eta + i \xi^{-1}) \; .
\end{align}
Therefore, we have
\begin{align}
  \label{eq:84}
  \Delta_\eta^{(+)} \mathcal{V}^{(\infty)}(\xi, \eta)
  & = \Delta_\xi^{(+)} \mathcal{V}^{(\infty)}(\xi, \eta)
    = \frac{1}{2} \log \left( 1 - \frac{1}{i \eta\xi} \right) - i \eta\xi \log \left( 1 - \frac{1}{i \eta\xi} \right) - 1 \; .
\end{align}
where
\begin{align}
  \label{eq:85}
  \Delta_\eta^{(+)} \mathcal{V}^{(\infty)}(\xi, \eta)
  \equiv \mathcal{V}^{(\infty)}(\xi, \eta + i \xi^{-1}) - \mathcal{V}^{(\infty)}(\xi, \eta)
  = \left( {\rm e}^{i \xi^{-1} \partial_\eta} - 1 \right)  \mathcal{V}^{(\infty)}(\xi, \eta)
\end{align}
We now consider Borel-transforming the above expression. 
In a similar manner with previous calculations,
we immediately obtain
\begin{align}
  \label{eq:42}
  \left( {\rm e}^{- i y / \xi} - 1 \right)
  & \, \mathcal{V}_B^{(\infty)}(\xi,y)
    = - \frac{1}{2 y} \left( {\rm e}^{- i y/\xi} - 1 \right)
    + \frac{- i \xi}{y^2} \left( {\rm e}^{- i y/\xi} - 1 \right) + \frac{{\rm e}^{-i y/\xi}}{y} \; , \nonumber \\
  \implies
  & \mathcal{V}_B^{(\infty)}(\nu,y) = - \frac{1}{y} \left( \frac{1}{{\rm e}^{i y/\xi} - 1} - \frac{- i \xi}{y} + \frac{1}{2} \right) \; .
\end{align}
Finally, 
we can compute the Laplace transform to get the Borel sum of the Voros coefficients, $\mathscr{V}^{(\infty)}$, giving
\begin{align}
  \mathscr{V}^{(\infty)}(\xi,\eta)
  & = \int_0^\infty \mathcal{V}_B^{(\infty)}(\xi,y) \, {\rm e}^{- \eta y} \, \dd y
    = - \int_0^\infty \frac{1}{t} \left( \frac{1}{{\rm e}^{t} - 1} - \frac{1}{t} + \frac{1}{2} \right) \, {\rm e}^{i \eta \xi t} \, \dd t \nonumber\\
  & = - \log \frac{\Gamma(- i \eta \xi)}{\sqrt{2\pi}} + \left( - i \eta \xi - \frac{1}{2} \right) \log \left( - i \eta \xi \right) + i \eta\xi \; , \qquad
    {\rm Re} \left( - i \eta\xi \right) > 0 \; . 
\end{align}
From \eqref{eq:Voros_series_case3}, we see the parity relation
\begin{align}
  \mathcal{V}^{(\infty)}(-\eta\xi) = - \mathcal{V}^{(\infty)}(\eta\xi) \; ,
\end{align}
and we have the result in the domain of ${\rm Re} \left( - i \eta\xi \right) < 0$:
\begin{align}
  \mathscr{V}^{(\infty)}(\xi,\eta)
  & =
    \begin{cases}
      \displaystyle
      = - \log \frac{\Gamma(- i \eta \xi)}{\sqrt{2\pi}} - \left( i \eta \xi + \frac{1}{2} \right) \log \left( - i \eta\xi \right) + i \eta\xi \; , \qquad
      & {\rm Re} \left( - i \eta\xi \right) > 0 \; , \vspace{2mm} \\
      \displaystyle
      = \log \frac{\Gamma(i \eta \xi)}{\sqrt{2\pi}} - \left( i \eta \xi - \frac{1}{2} \right) \log ( i \eta\xi ) + i \eta\xi \; , \qquad
      & {\rm Re} \left( - i \eta\xi \right) < 0 \; ,
    \end{cases}
\end{align}
This concludes the calculation of the Voros coefficients in the case of $Q = -1 + \frac{2\xi}{x}$.

\bibliography{bib}
\end{document}